\documentclass{article}
\usepackage{geometry}
\usepackage{authblk}
\usepackage[titletoc]{appendix}
\usepackage{enumitem}
\usepackage{booktabs,diagbox}
\usepackage{xcolor}
\usepackage{tablefootnote}

\usepackage{tikz} 
\usetikzlibrary{cd}
\usetikzlibrary{arrows}

\usepackage{amssymb,amsmath,amsthm,bbold,mathtools,slashed,bm} 
\def\-{\raisebox{.75pt}{-}}

\numberwithin{table}{section}
\numberwithin{equation}{section}

\newtheorem{defn}{Definition}[section]

\newtheorem{exmp}{Example}[section]
\newtheorem{rmk}{Remark}[section]
\theoremstyle{plain}
\newtheorem{lem}{Lemma}[section]
\newtheorem{prop}{Proposition}[section]
\newtheorem{thm}{Theorem}[section]
\newtheorem*{thm*}{Theorem}
\newtheorem{cor}{Corollary}[section]

\usepackage[alphabetic,initials]{amsrefs}

\usepackage{hyperref,cleveref}

\usepackage{fixltx2e}

\title{Cohomological Lagrangian field theory}
\author{Shuhan Jiang}
\date{}
\affil[1]{Max Planck Institute for Mathematics in the Sciences}

\begin{document}
	
	\maketitle

	\begin{abstract}
		This paper introduces a geometric framework for classical cohomological field theories based on $G^{\star}$-algebras and gauge natural field theories. A BV-BFV extension of the framework is provided, which incorporates the cotangent lift of the Donaldson-Witten theory as an illustrative example.
	\end{abstract}

	\tableofcontents
	
	\section{Introduction}
	
	Cohomological field theories over a Riemannian manifold $(M,g)$ are typically characterized by the following property \cite{Witten1988,Witten1991}: 
	\begin{align}\label{cohprp}
		\frac{\delta}{\delta g_{\mu\nu}} \left(\mathcal{O} \exp\left(-S\right)\right)  \text{ is } Q\text{-exact},
	\end{align}
	where $Q$ is the scalar supersymmetry, $S$ is the action, and $\mathcal{O}$ is an observable of the theory. Since $S$ and $\mathcal{O}$ are $Q$-closed, \eqref{cohprp} holds when both $\frac{\delta}{\delta g_{\mu\nu}} \mathcal{O}$ and the Einstein-Hilbert energy-momentum tensor $T^{\mu\nu}:=\frac{\delta}{\delta g_{\mu\nu}} S$ of the theory are $Q$-exact.
	
	Let's restate everything in a more mathematically rigorous manner. A field theory over $M$ should be specified by the space of sections $\Gamma(Y)$  of a bundle $Y$ over $M$ and a local form $\mathcal{L}$ of degree $(n,0)$ in the variational bicomplex $\Omega_{loc}(M \times \Gamma(Y))$ of $Y$, known as the Lagrangian of the theory \cite{Zuckerman1987, Deligne1999, Blohmann14}. The action $S$ of the theory should then be defined as the integration $\int_M \mathcal{L}$. For our purposes, it is essential to assume that $Y$ is graded, and $\Gamma(Y)$ is equipped with an evolutionary cohomological vector field $Q$. The space of local forms then becomes a tricomplex $\Omega_{loc} = \bigoplus_{r,s,t} \Omega_{loc} ^{r,s,t}$, with the extra differential given by the Lie derivative $\mathrm{Lie}_Q$ along $Q$. The condition $S$ being $Q$-closed is equivalent to $\mathrm{Lie}_Q \mathcal{L}$ being $d_h$-exact, where $d_h$ is the horizontal differential of $\Omega_{loc}$. In fact, it is often the case that
	\begin{align}\label{brstdecomp}
		H_{\mathrm{Lie}_Q}^{p,0,\bullet}(\Omega_{loc}(M \times \Gamma(Y))/d_h \Omega_{loc}(M \times \Gamma(Y))) \cong \Omega^p(M)/d \Omega^{p-1}(M)
	\end{align}
	for $p=1,\cdots,n$, where $d$ is the de Rham differential of the de Rham complex $\Omega(M)$ of $M$. In such cases, $\mathcal{L}$ can be decomposed into two parts: $\mathcal{L}=\mathcal{L}_{0} + \mathrm{Lie}_Q \mathcal{V}$, where $\mathcal{L}_{0}$ corresponds to the pullback of an $n$-form over $M$. From a physical point of view, the decomposition tells us that $Q$ and $\mathcal{V}$ should be interpreted as a BRST operator and a gauge fixing fermion. In the context of cohomological field theories, the condition $T^{\mu\nu}$ being $Q$-exact is equivalent to the condition $\mathcal{L}_{0}$ being topological, meaning it does not depend on the Riemannian metric $g$. Such $\mathcal{L}_{0}$ usually possesses a very large symmetry group. It becomes necessary for $\mathcal{V}$ to be dependent on $g$ to gauge fix these symmetries. This formulation represents the essence of the BRST approach to cohomological field theories \cite{Baulieu1988,Baulieu1989}.
	
	There exists a systematic way to construct cohomological vector fields $Q$ such that \eqref{brstdecomp} holds true \cite{Jiang2023a}. The idea is to consider the shifted vertical bundle $Y=V[1]Y'$ of a graded bundle $Y'$ over $M$. The vertical differential $d_v$ of $\Omega_{loc}(M \times \Gamma(Y'))$ can then be viewed a cohomological vector field $Q$ over $\Gamma(Y)$. When $Y'$ is affine, one can show that $H_{d_v}^{p,0,\bullet}(\Omega_{loc}(M \times \Gamma(Y'))/d_h \Omega_{loc}(M \times \Gamma(Y'))) \cong \Omega^p(M)/d \Omega^{p-1}(M)$ \cite[Proposition A.1]{Sharapov2015}, which then implies \eqref{brstdecomp}. Such $Q$ alone is not the right BRST operator because it does not encode any information about the symmetries of the theory. However, this problem can be overcome by ``deforming" $Q$ using the Mathai-Quillen map introduced in \cite{Mathai1986} to bridge the Weil and Cartan models of equivariant cohomology.

	Let $\mathfrak{X}_Q(\Gamma(Y))$ denote the ideal of the graded Lie superalgebra $\mathfrak{X}(\Gamma(Y))$ of evolutionary vector fields over $\Gamma(Y)$ generated by $Q$. For $[Q,\Xi] \in \mathfrak{X}_Q$, we have
	\begin{align}\label{qs}
		[Q,\Xi] S = Q (\Xi S)\pm \Xi(Q(S)) = Q(\Xi S).
	\end{align}
	In other words, the $Q$-cohomology class of $S$ is preserved under the action of $\mathfrak{X}_Q(\Gamma(Y))$. The contractions and Lie derivatives on $\Omega_{loc}(M \times \Gamma(Y'))$ can be also viewed as evolutionary vector fields over $\Gamma(Y)$ of degrees $-1$ and $0$, respectively. Together with $Q$, they span a graded Lie superalgebra $\widetilde{\mathfrak{X}(\Gamma(Y'))} \subset \mathfrak{X}_Q(\Gamma(Y))$. If the bundle $Y'$ is natural, the diffeomorphism group $\mathrm{Diff}(M)$ of $M$ acts canonically on $\Gamma(Y')$ and every vector field $X \in \mathfrak{X}(M)$ determines canonically two vector fields $K_X$ in $\widetilde{\mathfrak{X}(\Gamma(Y'))}_{-1}$ and $\xi_{X} \in \widetilde{\mathfrak{X}(\Gamma(Y'))}_0$ such that $\xi_X = [Q,K_X]$. Combining with \eqref{qs}, this in particular implies that the $Q$-cohomology class of $S$ is preserved under the $\mathrm{Diff}(M)$-action.
	The evolutionary vector field $K_X$ is, in general, not a Noether symmetry of the Lagrangian. In the case of $M=\mathbb{R}^n$, one can choose $X$ to be $\frac{\partial}{\partial x^{\mu}}$ and ``deform'' $K_{\mu}:=K_{\frac{\partial}{\partial x^{\mu}}}$ properly such that it becomes a Noether symmetry of the theory. Such $K_{\mu}$ are known as the vector supersymmetries in the physics literature \cite{Baulieu2005,Ouvry1989,Sorella1998}. Their existence guarantees the $Q$-exactness of the canonical energy-momentum tensor of the theory.
	
	An observable $\mathcal{O}$ in a cohomological field theory (or any BRST theory) is usually obtained by integrating a local form $\mathcal{O}^{(p)} \in \Omega_{loc}^{n-p,0,p}(M \times \Gamma(Y))$ over a $p$-cycle in $M$. $\mathcal{O}^{(p)}$ are solutions to the so-called descent equations
	\begin{align}\label{0deq}
		\mathrm{Lie}_Q \mathcal{O}^{(p)} = d_h \mathcal{O}^{(p-1)}
	\end{align}
	for $p=1,\cdots,n$ with $\mathrm{Lie}_Q \mathcal{O}^{(0)}=0$.  The existence of solutions to \eqref{0deq} is guaranteed by the isomorphism \eqref{brstdecomp}. It is shown in \cite{Jiang2023a} that every solution $\sum_{p=0}^{n} \mathcal{O}^{p}$ to \eqref{0deq} is locally equivalent to $\exp(-K) \mathcal{W}$, where $\mathcal{W}$ is any $Q$-closed local form of total degree $n$, and $K:= dx^{\mu} \wedge \mathrm{Lie}_{K_{\mu}}$ is a locally defined ``homotopy operator''.
	
	Our main achievement here is the construction of a rigorous geometric framework for cohomological Lagrangian field theories, which solidifies the mathematical basis of the above BRST picture. The framework is built upon the variational bicomplex of gauge natural bundles \cite{Fatibene2003} and the concept of a $G^{\star}$-algebra, which was introduced in \cite{Guillemin2013} to provide an axiomatic treatment of algebraic equivariant cohomology. Our framework also admits a direct extension to the extended BV-BFV formalism \cite{Cattaneo2014,Mnev2019} via the standard cotangent lift procedure of a BRST theory \cite{Zinn1975}. In both the BRST and BV world, we use the Donaldson-Witten theory \cite{Witten1988} as a primary example. Interestingly, our approach aligns with the AKSZ construction of the Donaldson-Witten theory \cite{Bonechi2020} when the auxiliary fields are integrated out.
	
	
	\section{\texorpdfstring{$G^{\star}$}{TEXT}-algebras and equivariant cohomology}
	
	\subsection{\texorpdfstring{$G^{\star}$}{TEXT}-algebras}
	
	A graded Lie supergroup is a triple $(G,\widetilde{\mathfrak{g}},\tau)$ where 
	\begin{enumerate}
		\item $G$ is a Lie group with Lie algebra $\mathfrak{g}$;
		\item $\widetilde{\mathfrak{g}}=\bigoplus_{i \in \mathbb{Z}} \widetilde{\mathfrak{g}}_i$ is a graded Lie superalgebra with $\widetilde{\mathfrak{g}}_{0}=\mathfrak{g}$;
		\item $\tau: G \rightarrow \mathrm{Aut}(\widetilde{\mathfrak{g}})$ is an action of $G$ on $\widetilde{\mathfrak{g}}$ by graded algebra automorphisms such that its restriction to $\widetilde{\mathfrak{g}}_0$ is the adjoint action of $G$ on $\mathfrak{g}$.
	\end{enumerate}
	
	\begin{exmp}
		Let $A$ be a graded commutative algebra. The automorphism group $\mathrm{Aut}(A)$, the graded Lie superalgebra $\mathrm{Der}(A)$ of derivations of $A$, and the adjoint action of $\mathrm{Aut}(A)$ on $\mathrm{Der}(A)$ determines a graded Lie supergroup which we denote by $G(A)$.
	\end{exmp}
	
	Given two graded Lie supergroups $(G,\widetilde{\mathfrak{g}},\tau)$ and $(G',\widetilde{\mathfrak{g}}',\tau')$. A homomorphism between them is a pair $(\phi, \varphi)$ where
	\begin{enumerate}
		\item $\phi: G \rightarrow G'$ is a Lie group homomorphism;
		\item $\varphi: \widetilde{\mathfrak{g}} \rightarrow \widetilde{\mathfrak{g}}'$ is a graded Lie superalgebra homomorphism;
		\item $\phi$ and $\varphi$ are compatible in the sense that
		\begin{align}\label{shcm}
			\varphi|_{\widetilde{\mathfrak{g}}_0} = d \phi|_{\mathrm{Id}}, \quad \tau'(\phi(g)) \circ \varphi = \varphi \circ \tau(g), ~\forall g \in G. 
		\end{align}
	\end{enumerate}
	Let $X$ be a $G$-manifold. To each $\xi \in \mathfrak{g}$ we can associate a vector field $v_{\xi}$ on $X$, which again induces a contraction $\iota_{\xi}$ and a Lie derivative $\mathrm{Lie}_{\xi}$ on the de Rham complex $\Omega(X)$ of $X$. Let $d$ denote the de Rham differential. Fix a basis $\{\xi_a\}$ of $\mathfrak{g}$. Let $f^c_{ab}$ denote the structure constants of $\mathfrak{g}$ with respect to $\{\xi_a\}$. Let $\iota_a$ and $\mathrm{Lie}_a$ denote the contraction and Lie derivative associated to $\xi_a$. $d$, $\iota_a$, $\mathrm{Lie}_a$ satisfy the following relations
	\begin{align}
		&\mathrm{Lie}_a\mathrm{Lie}_b -\mathrm{Lie}_b  \mathrm{Lie}_a = f^c_{ab} \mathrm{Lie}_c, \quad \mathrm{Lie}_a \iota_b - \iota_b \mathrm{Lie}_a = f^c_{ab} \iota_c, \quad
		\mathrm{Lie}_a d - d \mathrm{Lie}_a = 0, \label{e_5_1a} \\
		&d^2=0, \quad 
		\iota_a \iota_b + \iota_b \iota_a=0, \quad 
		d \iota_a + \iota_a d =\mathrm{Lie}_a, \label{e_5_1b}    
	\end{align}     
	which are known as the Cartan calculus.  \eqref{e_5_1a} and \eqref{e_5_1b} define a differential graded Lie superalgebra $\widetilde{\mathfrak{g}}=\widetilde{\mathfrak{g}}_{-1} \oplus \widetilde{\mathfrak{g}}_{0} \oplus \widetilde{\mathfrak{g}}_1$ where $\widetilde{\mathfrak{g}}_0$ is spanned by $\mathrm{Lie}_a$, $\widetilde{\mathfrak{g}}_{-1}$ is spanned by $\iota_a$, and $\widetilde{\mathfrak{g}}_1$ is spanned by $d$. $G$, $\widetilde{\mathfrak{g}}$, and the adjoint action of $G$ on $\widetilde{\mathfrak{g}}$ determines a graded Lie supergroup which we follow \cite{Guillemin2013} to denote by $G^{\star}$.
	
	\begin{defn}
		We call $G^{\star}$ the Cartan graded Lie supergroup associated to $G$.
	\end{defn}
	Let $G^{\star}$ and $H^{\star}$ be the Cartan graded Lie supergroups associated to the Lie groups $G$ and $H$, respectively. A homomorphism $\phi: G \rightarrow H$ naturally defines a homomorphism $\phi^{\star}=(\phi, \varphi)$ from $G^{\star}$ to $H^{\star}$ where $\varphi$ is defined by setting
	\begin{align*}
		\varphi(d) = d, \quad \varphi(\mathrm{Lie}_{\xi}) = \mathrm{Lie}_{d\phi|_{\mathrm{Id}}(\xi)}, \quad \varphi(\iota_{\xi}) = \iota_{d\phi|_{\mathrm{Id}}(\xi)}
	\end{align*}
	for all $\xi \in \mathfrak{g}$. One can easily verify that $\phi$ and $\varphi$ satisfy \eqref{shcm}. It follows that a subgroup $L$ of $G$ defines naturally a sub-supergroup $L^{\star}$ of $G^{\star}$.
	
	\begin{defn}
		A $G^{\star}$-algebra is a graded commutative algebra $A$ together with a $G^{\star}$-action, i.e., a morphism $\rho: G^{\star} \rightarrow G(A)$ of graded Lie supergroups. A morphism between two $G^{\star}$-algebras is just a graded algebra homomorphism which is compatible with the $G^{\star}$-action.
	\end{defn} 
	An element $\alpha \in A$ is called horizontal if $\iota_{\xi} \alpha =0$ for all $\xi \in \mathfrak{g}$. A horizontal element $\alpha$ is called basic if in addition $\mathrm{Lie}_{\xi} \alpha =0$ for all $\xi \in \mathfrak{g}$. Let $A_{hor}$ and $A_{bas}$ denote the subalgebras of horizontal and basic elements in $A$, respectively. (They are subalgebras because $\iota_a$ and $\mathrm{Lie}_a$ are derivations.) It is easy to see that for $\alpha \in A_{bas}$, $d\alpha$ is also in $A_{bas}$. We use $H(A)$ to denote the cohomology of $(A,d)$ and $H_{bas}(A)$ to denote the cohomology of $(A_{bas},d)$.
	\begin{defn}
		Let $A$ and $B$ be two $G^{\star}$-algebras. A semi-homotopy is a linear map $K: A \rightarrow B$ of degree $-1$ which satisfies
		\begin{align}
			\iota_{\xi} K + K \iota_{\xi}=0, ~ \forall \xi \in \mathfrak{g}		
		\end{align}
		and
		\begin{align}
			B_{hor} \subset \ker(\mathrm{Lie}_{\xi} K - K \mathrm{Lie}_{\xi}), ~ \forall \xi \in \mathfrak{g}.
		\end{align}
		A semi-homotopy $K$ is said to be a homotopy if 
		$
		(\mathrm{Lie}_{\xi} K - K \mathrm{Lie}_{\xi})=0
		$
		for all $\xi \in \mathfrak{g}$. Two morphisms $\alpha_0$ and $\alpha_1: A \rightarrow B$ are (semi-)homotopic if they are equal up to a (semi-)homotopy, i.e., if
		\begin{align*}
			\alpha_1 - \alpha_2 = d K + K d.
		\end{align*}
	\end{defn}
	\begin{rmk}
		Let $A$ and $B$ be two $G^{\star}$-algebras with a morphism $\alpha: A \rightarrow B$.  A (semi-)homotopy $K : A \rightarrow B$ is said to be a (semi-)homotopy relative to $\alpha$ if
		\begin{align*}
			K(xy) = K(x)\alpha(y) + (-1)^{d(x)}\alpha(x)K(y)
		\end{align*}
		for all $x, y \in A$. For $B=A$, one can take $\alpha$ to be the identity morphism $\mathrm{Id}: A \rightarrow A$ and such $K$ becomes a derivation.
	\end{rmk}
	\begin{prop}
		Let $\alpha_0$ and $\alpha_1: A \rightarrow B$ be two morphisms between $G^{\star}$-algebras. They induce the same morphism $H(A) \rightarrow H(B)$ if they are homotopic. They induce the same morphism $H_{bas}(A) \rightarrow H_{bas}(B)$ if they are semi-homotopic.
	\end{prop}
	\begin{proof}
		Let $L=d K + K d$ and $P_{\xi}=\mathrm{Lie}_{\xi} K - K \mathrm{Lie}_{\xi}$. It is not hard to show that $\iota_{\xi} L - L \iota_{\xi} = P_{\xi}$ and $\mathrm{Lie}_{\xi} L - L \mathrm{Lie}_{\xi} = d P_{\xi} - P_{\xi}d$. If $K$ is a homotopy, then $P_{\xi}=0$ and $L$ is a morphism of $G^{\star}$-algebras, hence induces a graded commutative algebra morphism $H(A) \rightarrow H(B)$. If $K$ is a semi-homotopy, then $L$ still commutes with $\mathrm{Lie}_{\xi}$ when restricted to the basic part of $B$, hence induces a graded commutative algebra morphism $H_{bas}(A) \rightarrow H_{bas}(B)$. The rest of the proof follows directly from the standard arguments of homological algebras. 
	\end{proof}
	
	\subsection{Algebraic equivariant cohomology}
	
	\begin{defn}
		A $G^{\star}$-algebra $E$ is said to be of type (C) if there exists a $G$-invariant free submodule $C$ of the $A_0$-module $A_1$ such that the contractions 
		\begin{align*}
			\iota_a: A_1 \rightarrow A_0
		\end{align*}
		form a basis of $C^*$, the dual module of $C$ over $A_0$.
	\end{defn}
	\begin{rmk}
		$C$ be can be seen as an algebraic analogue of the dual bundle of the vertical bundle $VP$ of a principal $G$-bundle $P$. 
	\end{rmk}
	\begin{exmp}
		The de Rham complex of a principal $G$-bundle is of type (C).
	\end{exmp}
	A $G^{\star}$-algebra $E$ is of type (C) if and only if there are elements $\theta^a \in E_1$ such that \cite{Guillemin2013}
	\begin{align}
		&\iota_a \theta^b = \delta_a^b, \label{conncontr}\\
		&\mathrm{Lie}_a \theta^b = -f^b_{ac} \theta^c. \label{connlie}
	\end{align}
	Using Cartan's magic formula, one can show that there exists elements $\phi^a \in E_2$ satisfying
	\begin{align}
		d \theta^a = \phi^a - \frac{1}{2}f^a_{bc} \theta^b \theta^c. \label{connd}
	\end{align}
	The actions of $d$, $\iota_a$ and $\mathrm{Lie}_a$ on $\phi^b$ are uniquely determined by \eqref{conncontr} to \eqref{connd}.
	
	There is a universal object in the category of $G^{\star}$-algebras of type (C).
	\begin{defn}\label{d_5_1}
		The Weil Algebra of a Lie algebra $\mathfrak{g}$ is a $G^{\star}$-algebra of type (C) with underlying graded commutative algebra
		\begin{align*}
			W(\mathfrak{g})= \Lambda(\mathfrak{g}^*) \otimes \mathrm{S}(\mathfrak{g}^*).
		\end{align*}
		$W(\mathfrak{g})$ is graded by assigning degree $1$ to elements of $\mathfrak{g}^* \subset \Lambda(\mathfrak{g}^*)$ and degree $2$ to elements of $\mathfrak{g}^* \subset \mathrm{S}(\mathfrak{g}^*)$. The action of $G$ on $W(\mathfrak{g})$ is induced by its coadjoint action on $\mathfrak{g}^*$. The action of $\widetilde{\mathfrak{g}}$ on $W(\mathfrak{g})$ is specified by \eqref{conncontr} to \eqref{connd} and
		\begin{align}
			&\iota_a \phi^b=0, \label{e_5_5} \\
			&d\phi^a=f^a_{bc}\phi^b\theta^c, \label{e_5_6} \\
			&\mathrm{Lie}_a \phi^b=-f^b_{ac}\phi^c,
		\end{align}
		where $\theta^a = \xi^a \otimes 1$, $\phi^a = 1 \otimes \xi^a$, $\{\xi^a\}$ is a basis of $\mathfrak{g}^*$. 
	\end{defn}
	\begin{rmk}
		$(W(\mathfrak{g}),d)$ is acyclic \cite[Theorem 3.2.1]{Guillemin2013}. 
	\end{rmk}
	It is easy to show that the tensor product $A \otimes B$ of two $G^{\star}$-algebra is again a $G^{\star}$ algebra, and that $A \otimes B$ is of type (C) if $B$ is of type (C).
	\begin{defn}\label{algeequivcoh}
		The algebraic equivariant cohomology of a $G^{\star}$-algebra $A$, denoted by $H_G(A)$, is defined as $H_{bas}(A \otimes E)$, where $E$ is an acyclic $G^{\star}$-algebra of type (C).
	\end{defn}
	\begin{rmk}
		$E$ should be thought of as an algebraic analogue of the universal $G$-space $EG$ of a Lie group $G$. Just like in the topological case, Definition \ref{algeequivcoh} does not depend on the choice of $E$ \cite[Section 4.4]{Guillemin2013}. 
	\end{rmk}
	
	\subsubsection{Weil model}
	
	Let's consider the tensor product $W(\mathfrak{g})\otimes \Omega(X)$. It has a canonical $G^{\star}$-algebra structure where the contractions, the Lie derivatives, and the differential are
	\begin{align*}
		\iota_a \otimes 1 + 1 \otimes \iota_a,\quad \mathrm{Lie}_a \otimes 1 + 1 \otimes \mathrm{Lie}_a,\quad d \otimes 1 + 1 \otimes d =: d_W.
	\end{align*}
	The equivariant cohomology of $X$ is defined as $H_G(X):=H_G(\Omega(X))=H_{bas}(W(\mathfrak{g})\otimes \Omega(X))$.
	\begin{defn}
		$(W(\mathfrak{g})\otimes \Omega(X),d_W)$ is called the Weil Model for the equivariant cohomology of $X$. $d_W$ is called the Weil differential. 
	\end{defn}
	
	\subsubsection{BRST model and Cartan model}
	
	Let's consider the automorphism map $j$ of $W(\mathrm{g})\otimes \Omega(X)$ defined by $j = \exp{(-\theta^a \otimes \iota_a)}$. $j$ is known as the Mathai-Quillen map. Let $d_K := j \circ d_W \circ j^{-1}$. One can show that \cite{Kalkman1993}
	\begin{align*}
		&d_K = d_W + \theta^a \otimes \mathrm{Lie}_a - \phi^a \otimes \iota_a, \\
		&\iota_a \otimes 1 = j \circ (\iota_a \otimes 1 + 1 \otimes \iota_a) \circ j^{-1},\\
		&\mathrm{Lie}_a \otimes 1 + 1 \otimes \mathrm{Lie}_a = j \circ (\mathrm{Lie}_a \otimes 1 + 1 \otimes \mathrm{Lie}_a) \circ j^{-1}.
	\end{align*}
	\begin{defn}
		$(W(\mathrm{g})\otimes \Omega(X), d_K)$ is called the BRST (or Kalkman) model of the equivariant cohomology of $X$. $d_K$ is called the BRST (or Kalkman) differential. 
	\end{defn}	
	The basic part of $W(\mathrm{g})\otimes \Omega(X)$ in the Kalkman model of $X$ can be identified as $(S(\mathfrak{g}^*) \otimes \Omega(X))^G$, and the restriction of $d_K$ to $(S(\mathfrak{g}^*) \otimes \Omega(X))^G$, denoted by $d_C$, takes the form $d_C = d \otimes 1 - \phi^a \otimes \iota_a$. $((S(\mathfrak{g}^*) \otimes \Omega(X))^G, d_C)$ is called the Cartan Model of the equivariant cohomology of $X$. $d_C$ is called the Cartan differential.

	
	\subsection{BV extensions of \texorpdfstring{$G^{\star}$}{TEXT}-algebras and gauge fixings}
	
	
	\subsubsection{BRST and BV systems}
	
	Let $\mathcal{M}$ be a graded manifold equipped with a $G^{\star}$-action. The graded commutative algebra $C^{\infty}(\mathcal{M})$ of functions over $\mathcal{M}$ is a $G^{\star}$-algebra. The differential of $\widetilde{\mathfrak{g}}$ induces a cohomological vector field $Q$ over $\mathcal{M}$.  With a slight abuse of notation, we denote the vector fields generated by $\iota_{\xi} \in \widetilde{\mathfrak{g}}_{-1}$ and $\mathrm{Lie}_{\xi} \in \widetilde{\mathfrak{g}}_{0}$ again as $\iota_{\xi}$ and $\mathrm{Lie}_{\xi}$, respectively. Let $S$ be a $Q$-closed basic function over $\mathcal{M}$ of degree $0$. 
	\begin{defn}
		We call the triple $(\mathcal{M}, Q, S)$ a ($G^{\star}$-equivariant)  BRST system.
	\end{defn}
	\begin{defn}
		A gauge fixing procedure of a BRST system $(\mathcal{M}, Q, S)$ is referred to as the modification of $S$ as $S + Q(\Psi)$, where $\Psi$ is a function of degree $-1$. $\Psi$ is called a gauge fixing fermion. Such a procedure is called $G^{\star}$-invariant if $\Psi$ is a basic function.
	\end{defn}
	\begin{rmk}
		In particular, one can choose $\mathcal{M}$ to be the graded manifold associated to the graded vector bundle $\underline{\mathfrak{g}}[1] \oplus \underline{\mathfrak{g}}[2] \oplus T[1] M$. We have $C^{\infty}(\mathcal{M}) \cong W(\mathfrak{g}) \otimes \Omega(M)$ as graded commutative algebras. Let the $G^{\star}$-action on $\mathcal{M}$ be induced from the $G^{\star}$-algebraic structure on $W(\mathfrak{g}) \otimes \Omega(M)$ through the Weil (or BRST) model. Then the cohomology $H_{bas}(\mathcal{M})$ of basic functions over $\mathcal{M}$ is nothing but the equivariant cohomology $H_G(M)$ of its underlying $G$-manifold $M$. Moreover, one can see that a $G^{\star}$-invariant gauge fixing of $S$ does not change the cohomology class of $S$ in $H_{bas}(\mathcal{M})$.
	\end{rmk}
	Let $(\mathcal{M},\omega)$ be an odd symplectic graded manifold equipped with a Hamiltonian $G^{\star}$-action, where $\omega$ is an odd symplectic form over $\mathcal{M}$ of degree $-1$.\footnote{We choose the sign convention such that $\iota_{X_f} \omega - df = 0$, where $X_f$ is the Hamiltonian vector field of $f$.} Let $S$ be a Hamiltonian function associated to the cohomological vector field $Q$. Let $I_{\xi}$ and $L_{\xi}$ be Hamiltonian functions associated to $\iota_{\xi}$ and $\mathrm{Lie}_{\xi}$. It follows that
	\begin{align*}
		Q(S)=\{S,S\}=0, \quad \iota_{\xi}(S) = \{I_{\xi},S\} = L_{\xi}, \quad \mathrm{Lie}_{\xi}(S) = \{L_{\xi}, S\}=0,
	\end{align*}
	where $\{\cdot,\cdot\}$ is the graded Poisson bracket associated to $\omega$ defined by setting $\{f,g\}=\iota_{X_f}\iota_{X_g} \omega$.
	\begin{defn}
		We call the quadruple $(\mathcal{M}, \omega, Q, S)$ a ($G^{\star}$-equivariant) BV system.
	\end{defn}
	Obviously, $S$ is not basic with respect to a $G^{\star}$-action with nontrivial underlying $G$-action. 
	However, one can introduce a ``$G^{\star}$-invariant'' gauge fixing procedure to fix this problem. Recall that a gauge fixing procedure of a BV system is the restriction of $S$ to a Lagrangian submanifold $\mathcal{L}$ of $\mathcal{M}$. There is also an implicit assumption that the cohomological vector field $Q$ is tangential to $\mathcal{L}$ so that the restriction of $Q$ to $\mathcal{L}$ is well-defined and we still have $Q(S|_{\mathcal{L}})=0$. Likewise, we can require $\mathcal{L}$ to be $G^{\star}$-invariant so that the restriction of the $G^{\star}$-action to $\mathcal{L}$ is well-defined.
	\begin{defn}
		A gauge fixing procedure of a $BV$ system is called $G^{\star}$-invariant if the corresponding Lagrangian submanifold $\mathcal{L}$ is $G^{\star}$-invariant and is contained in the zero locus of $L_{\xi}$ for all $\xi \in \mathfrak{g}$.
	\end{defn}
	For a $G^{\star}$-invariant gauge fixing, we have
	\begin{align*}
		Q(S|_{\mathcal{L}}) = Q(S)|_{\mathcal{L}}=0, \quad \iota_\xi(S|_{\mathcal{L}}) = \iota_{\xi}(S)|_{\mathcal{L}} = L_{\xi}|_{\mathcal{L}}=0, \quad \mathrm{Lie}_{\xi}(S|_{\mathcal{L}}) = \mathrm{Lie}_{\xi}(S)|_{\mathcal{L}}=0.
	\end{align*}
	Therefore, $S|_{\mathcal{L}}$ is a $Q$-closed basic function over the $G^{\star}$-manifold $\mathcal{L}$. In other words, we obtain a (gauge fixed) BRST system $(\mathcal{L}, Q, S|_{\mathcal{L}})$ from the BV system $(\mathcal{M}, \omega, Q, S)$.
	
	\subsubsection{Cotangent lift of a BRST system}
	
	One can also obtain a BV system out of a BRST system using a trick called cotangent lift. Let $(\mathcal{M},Q,S)$ be a BRST system. Let $T^*[-1] \mathcal{M}$ be the cotangent bundle of $\mathcal{M}$ shifted by degree $-1$ equipped with the canonical odd symplectic form $\omega$. Let $(x^{\mu},\theta^a)$ be a coordinate system of $\mathcal{M}$ and $(x^{\mu},\theta^a,x^+_{\mu},\theta^+_{a})$ be the induced coordinate system on $T^*[-1]\mathcal{M}$. $\omega$ can be locally expressed as
	\begin{align*}
		\omega = dx^+_{\mu} \wedge dx^{\mu} + d\theta^+_{a} \wedge d \theta^a.
	\end{align*}
	Every vector field $X = X^{\mu} \frac{\partial}{\partial x^{\mu}} + X^{a} \frac{\partial}{\partial \theta^{a}}$ over $\mathcal{M}$ can be lifted to a function $\widetilde{X}$ over $T^*[-1]\mathcal{M}$ defined by the formula
	\begin{align*}
		\widetilde{X} = x^+_{\mu} X^{\mu} + \theta^+_{a} X^a.
	\end{align*}
	Let $X_{cl}$ denote the Hamiltonian vector field associated to $\widetilde{X}$ over $T^*[-1] \mathcal{M}$. One can easily check that
	\begin{align*}
		CL: \mathfrak{X}(\mathcal{M}) &\rightarrow \mathfrak{X}(T^*[-1] \mathcal{M}) \\
		X &\mapsto X_{cl}
	\end{align*}
	is a graded Lie superalgebra homomorphism.\footnote{One can check that both the map $(\mathfrak{X}(M),[\cdot,\cdot]) \rightarrow (C^{\infty}(T^*[-1]\mathcal{M}), \{\cdot,\cdot\}), \quad  X \mapsto \widetilde{X}$ and the map $(C^{\infty}(T^*[-1]\mathcal{M}), \{\cdot,\cdot\}) \rightarrow (\mathfrak{X}(T^*[-1]\mathcal{M}),[\cdot,\cdot]), \quad  f \mapsto X_f$ are anti-homomorphisms. Therefore, their composition $CL$ is a homomorphism.} In local coordinates, we have
	\begin{align*}
		X_{cl} = X^{i} \frac{\partial}{\partial u^i} + (-1)^{(|u^i|+1)|u^j|}u^+_i \frac{\partial X^i}{\partial u^j} \frac{\partial}{\partial u^+_{j}},
	\end{align*}
	for $X$ odd and
	\begin{align*}
		X_{cl} = (-1)^{|u^i|}X^{i} \frac{\partial}{\partial u^i} + (-1)^{|u^i||u^j|+1}u^+_i \frac{\partial X^i}{\partial u^j} \frac{\partial}{\partial u^+_{j}},
	\end{align*}
	for $X$ even, where $(u^j)=(x^{\mu},\theta^a)$. On the other hand, any function over $\mathcal{M}$ can be canonically viewed as a function over $T^*[-1]\mathcal{M}$. Note that for an odd vector field $X$ over $\mathcal{M}$, $X_{cl}(f) = X(f)$ for $f \in C^{\infty}(\mathcal{M})$. Let $X_S$ be the Hamiltonian vector field of $S$.  $X_S$ locally takes the form
	\begin{align*}
		X_S=\frac{\partial S}{\partial u^j} \frac{\partial}{\partial u^+_j}.
	\end{align*}
	Since $S$ is $Q$-closed and basic, we have
	\begin{align*}
		[Q_{cl}, X_S] = [(\iota_{\xi})_{cl}, X_S] = 0.
	\end{align*}
	It follows from $[Q,\iota_{\xi}]=\mathrm{Lie}_{\xi}$ that $[(\mathrm{Lie}_{\xi})_{cl}, X_S]=0$. One can then deform the $G^{\star}$-action on $T^*[-1]\mathcal{M}$ induced by the cotangent lift by introducing the new cohomological vector field
	\begin{align*}
		Q_{BV} := Q_{cl} + X_S.
	\end{align*}
	The Hamiltonian function of $Q_{BV}$ is $S_{BV}:= S + \widetilde{Q}$. We have proved that
	\begin{prop}
		$(T^*[-1]\mathcal{M}, \omega, Q_{BV}, S_{BV})$ is a $BV$-system.
	\end{prop}
	It was shown by Schwarz that any Lagrangian submanifold of $T^*[-1]\mathcal{M}$ can be obtained by combining the following two types of examples \cite{Schwarz1993}.
	\begin{enumerate}[label=\emph\bfseries\arabic*),wide=0pt]
		\item[``Graph Lagrangian'':] Let $\Psi$ be a function over $\mathcal{M}$ of degree $-1$. The graph $\mathrm{Graph}(d\Psi)=: \mathcal{L}_{\Psi}$ of the $1$-form $d \Psi$ is a Lagrangian submanifold of $T^*[-1]\mathcal{M}$.
		\item[``Conormal Lagrangian'':] Let $\mathcal{N}$ be a submanifold of $\mathcal{M}$. The conormal bundle $N^*[-1] \mathcal{N}$ of $\mathcal{N}$ shifted by degree $-1$ is a Lagrangian submanifold of $T^*[-1]\mathcal{M}$.
	\end{enumerate}
	
	For the first type of Lagrangian submanifolds, one has a natural isomorphism $\mathcal{L}_0 \cong \mathcal{L}_{\Psi}$ where $0$ is the zero function over $\mathcal{M}$.  Under this isomorphism, we have
	\begin{align*}
		S_{BV}|_{\mathcal{L}_0} = S + Q(\Psi).
	\end{align*} 
	Obviously, this gauge fixing procedure is $G^{\star}$-invariant if $\Psi$ is a basic function. We have
	\begin{align*}
		&\iota_{\xi} (S_{BV}|_{\mathcal{L}_0}) = \iota_{\xi}(S)+ \iota_{\xi}Q(\Psi) = 0 + \mathrm{Lie}_{\xi}(\Psi) = 0, \\
		&\mathrm{Lie}_{\xi}(S_{BV}|_{\mathcal{L}_0} ) = \mathrm{Lie}_{\xi}(S) + Q(\mathrm{Lie}_{\xi}(\Psi))=0.
	\end{align*}
	
	For the second type of Lagrangian submanifolds, the gauge fixing is $G^{\star}$-invariant if $\mathcal{N}$ is a $G^{\star}$-invariant submanifold in $\mathcal{M}$. In fact, one can choose a local coordinate system $(x^{\mu},y^{\nu},\theta^a,\eta^b)$ such that $\mathcal{N}$ is determined locally by the equations $y^{\nu}=0$ and $\eta^b = 0$. Let $(x^{\mu},y^{\nu},\theta^a,\eta^b,x^+_{\mu},y^+_{\nu},\theta^+_a,\eta^+_b)$ be the induced local coordinate system on $T^*[-1]\mathcal{M}$. The conormal Lagrangian $N^*[-1]\mathcal{N}$ is then determined locally by $y^{\nu}=0, \eta^b = 0, x^+_{\mu}=0, \theta^+_a = 0$. We have
	\begin{align*}
		\mathrm{Lie}_{\xi}|_{N^*[-1]\mathcal{N}} = y^+_{\nu} \mathrm{Lie}_{\xi}^{\nu} + \eta^+_b \mathrm{Lie}_{\xi}^{b}=0
	\end{align*}
	since $\mathcal{N}$ is invariant under the $G^{\star}$-action, i.e., $\mathrm{Lie}_{\xi}^{\nu} =0,  \mathrm{Lie}_{\xi}^{b} =0$.	
	\begin{rmk}
		The preceding discussion makes it evident that one of the key advantages of BV systems over BRST systems is the increased flexibility in selecting a gauge-fixing procedure.
	\end{rmk}
	
	
	\section{Gauge natural field theories}
	
	\subsection{Gauge natural bundle}
	
	Let $M$ be a $n$-dimensional manifold and $s$ a positive integer. Consider the set
	\begin{align*}
		L^s(M) = \{j^s (\epsilon) (0)| \epsilon: \mathbb{R}^n \rightarrow M, \mathrm{locally}~\mathrm{invertible}~\mathrm{around}~ 0 \in \mathbb{R}^n\},
	\end{align*}
	where $j^s(\epsilon)$ is the $s$-th order jet prolongation of $\epsilon$. $L^s(M)$ is a fiber bundle over $M$ via the canonical projection
	\begin{align*}
		\pi: L^s(M) &\rightarrow M \\
		j^s (\epsilon) (0) &\mapsto  \epsilon(0).
	\end{align*}
	Moreover, it is a principal bundle with the standard fiber
	\begin{align*}
		\mathrm{GL}^s(n) = \{j^s (\alpha) (0)| \alpha: \mathbb{R}^n \rightarrow \mathbb{R}^n, \mathrm{locally}~\mathrm{invertible}~\mathrm{around}~ 0 \in \mathbb{R}^n, \alpha(0)=0\}.
	\end{align*}
	The group structure of $\mathrm{GL}^s(n)$ is specified by $j^s(\alpha)(0)j^s(\beta)(0):=j^s(\alpha \circ \beta)(0)$. The right action of $\mathrm{GL}^s(n)$ on $L^s(M)$ is specified by $j^s(\epsilon)(0)j^s(\alpha)(0):=j^s(\epsilon \circ \alpha)(0)$.	
	For $s=1$, one has the identification $\mathrm{GL}^1(n) \cong \mathrm{GL}(n)$ and $L^1(M) \cong \mathrm{Fr}(M)$.
	
	Let $G$ be a Lie group and $P$ be a principal $G$-bundle. The set 
	\begin{align*}
		J^r_n G:=\{j^r(a)(0)|a: \mathbb{R}^n \rightarrow G\}
	\end{align*}
	is also a Lie group with group multiplication defined by $j^r(a)(0)j^r(b)(0):=j^r(ab)(0)$. For $r \leq s$, one has a canonical right action of $\mathrm{GL}^s(n)$ on $J^r_n G$  defined by setting $j^s(a)(0)j^s(\alpha)(0):=j^s(a \circ \alpha)(0)$.	Consider the right semi-direct product
	\begin{align*}
		W^{r,s}_n(G):= J^r_n G \rtimes \mathrm{GL}^s(n)
	\end{align*}
	and the fiber product 
	\begin{align*}
		W^{r,s}_n P := J^r P \times_M L^s(M),
	\end{align*}
	where $J^r P$ denote the $r$-th order jet prolongation of $P$. $W^{r,s} P$ is again a principal bundle with structure group $W^{r,s}_n(G)$. The $W^{r,s}_n(G)$-action on $W^{r,s}_n P$ is defined by setting
	\begin{align*}
		(j^r(\sigma)(x),j^s(\epsilon)(0))(j^r(a)(0),j^s(\epsilon)(0)):=(j^r(\sigma \cdot (a \circ \alpha^{-1} \circ \epsilon^{-1}))(x),j^s(\epsilon \circ \alpha)(0)),
	\end{align*}
	where `$\cdot$' denote the $G$-action on $P$.
	
	\begin{defn}
		A gauge natural bundle of finite order $(r,s)$ is a fiber bundle associated to $W^{r,s}_n P$ so that $(r,s)$ is minimal.
	\end{defn}
	\begin{exmp}
		Recall that a connection $1$-form on a principal $G$-bundle $P$ is a $G$-equivariant $1$-form $A$ with values in the Lie algebra $\mathfrak{g}$ such that
		$
		A(K_{\xi})=\xi,~ \xi \in \mathfrak{g},
		$
		where $K_{\xi}$ is the fundamental vector field generated by $\xi$ on $P$. The space of all connections $\mathcal{A}$ is an affine space modeled on $\Omega^1(\mathrm{ad}P)$, where $\mathrm{ad}P$ is the adjoint bundle of $P$. 
		
		On the other hand, let $J^1 P$ be the first jet bundle of $P$. $J^1P$ is an affine bundle modeled on the vector bundle $T^*M \otimes_M VP$, where $VP$ is the vertical bundle over $P$ and the tensor product is taken over $M$. Let $j^1 \Phi: J^1 P \rightarrow J^1 P$ denote the jet prolongation of a bundle automorphism $\Phi$ of $P$, i.e., a $G$-equivariant diffeomorphism $\Phi: P \rightarrow P$. Such operations satisfy the chain rules
		\begin{align*}
			&j^1(\Phi_1 \circ \Phi_2) = j^1(\Phi_1)\circ j^1(\Phi_2), \\
			&j^1(\mathrm{id}_P) = \mathrm{Id}_{J^1 P}.
		\end{align*}
		Thus, $J^1P$ also has a principal $G$-action. The quotient space $C=J^1 P/G$ is then an affine bundle modeled on the vector bundle $(T^*M \otimes_M VP)/G \cong T^*M \otimes \mathrm{ad}P$ over $M$. One can show that there exists an identification $\mathcal{A} \cong \Gamma(C)$. Moreover, $C$ is a gauge natural bundle of order $(1,1)$.
	\end{exmp}
	
	Let $\mathrm{Aut}(P)$ denote the group of bundle automorphisms of $P$. Each $\Phi \in \mathrm{Aut}(P)$ determines uniquely a diffeomorphism $\phi$ of $M$. In other words, we have a group homomorphism $\mathrm{Aut}(P) \rightarrow \mathrm{Diff}(M)$. Let $\mathrm{Diff}_P(M)$ denote the image of $\mathrm{Aut}(P)$ under this homomorphism. Recall that there exists a bijection between the set of isomorphism classes of principal $G$-bundles over $M$ and the set of  of homotopy classes of maps from $M$ to the classifying space $BG$ of $G$. Let $f: M \rightarrow BG$ represent the isomorphism class of $P$. $\mathrm{Diff}_P(M)$ can be identified as the group of diffeomorphisms $\phi$ such that $f \circ \phi$ is homotopy equivalent to $f$. 
	
	We have a short exact sequence of groups
	\begin{align}\label{sespgrp}
		\mathrm{Id}  \rightarrow \mathrm{Gau}(P) \rightarrow \mathrm{Aut}(P) \rightarrow \mathrm{Diff}_P(M) \rightarrow \mathrm{Id},
	\end{align}
	where $\mathrm{Gau}(P)$ is the kernel of $\mathrm{Aut}(P) \rightarrow \mathrm{Diff}(M)$, known as the gauge group of $P$. The Lie algebra $\mathfrak{gau}(P)$ of $\mathrm{Gau}(P)$ can be identified with the space of sections of $\mathrm{ad} P$. \eqref{sespgrp} induces infinitesimally
	the short exact sequences of Lie algebras
	\begin{align}\label{sespalg}
		0 \rightarrow \mathfrak{gau}(P) \rightarrow \mathfrak{X}_{inv}(P) \rightarrow \mathfrak{X}_P(M) \rightarrow 0,
	\end{align}
	where $\mathfrak{X}_{inv}(P)$ denotes the set of $G$-invariant vector fields over $P$ and $\mathfrak{X}_P(M)$ denotes the set of vector fields over $M$ which can be lifted to $P$.
	
	There exists a canonical action of $\mathrm{Aut}(P)$ on $W^{r,s}_n P$ defined by
	\begin{align*}
		W^{r,s}_n \Phi: W^{r,s}_n P &\rightarrow W^{r,s}_n P \\
		(j^r(\sigma)(x),j^s(\epsilon)(0)) &\mapsto (j^r(\Phi \circ \sigma \circ \phi^{-1})(x), j^s(\epsilon \circ \phi)(0)).
	\end{align*}
	Therefore, $\mathrm{Aut}(P)$ also acts naturally on any gauge natural bundle over $M$. It follows that any $G$-invariant vector field $\Xi$ over $P$ which projects on a vector field $\xi$ over $M$ determines naturally a vector field $\Xi_Y$ over a gauge natural bundle $Y$ which projects on the same vector field $\xi$.
	\begin{defn}
		Let $\sigma$ be a section of $Y$. The Lie derivative of $\sigma$ with respect to $\Xi$ is defined to be the map
		\begin{align*}
			\mathrm{Lie}_{\Xi}\sigma: M \rightarrow VY,  \quad \mathrm{Lie}_{\Xi}\sigma:= T \sigma \circ \xi - \Xi_Y \circ \sigma,
		\end{align*}
		where $VY$ is the vertical bundle of $Y$.
	\end{defn}
	Let $\{(\Phi_t,\phi_t)\}$ denote the one-parameter group generated by $(\Xi_Y,\xi)$. Let $\sigma_t:= \Phi_t \circ \sigma \circ \phi_t^{-1}$. 
	\begin{prop}\cite[Proposition 2.6.5]{Fatibene2003}\label{liedergloloc}
		$\mathrm{Lie}_{\Xi} \sigma = -\frac{d}{dt} \sigma_t |_{t=0}$.
	\end{prop}
	
	\begin{exmp}
		Let $\Xi$ be a $G$-invariant vector field over $P$. Let $\{U, P|_U \cong U \times G\}$ be a local trivialization of $P$ equipped with a coordinate system $(x^{\mu})$. Let $\{\xi_a\}$ be a basis of $\mathfrak{g}$.  Every $G$-invariant vector field $\Xi$ over $P$ can be written as $\Xi = \Xi^{\mu}(x) \frac{\partial}{\partial x^{\mu}} + \Xi^a(x) \rho_a$, where $\rho_a$ is the right invariant vector field generated by $\xi_a$ over $G$. The vertical part $\Xi_v = \Xi^a(x) \rho_a$ of $\Xi$ can be identified with a section $\lambda$ of the adjoint bundle $\mathrm{ad} P$. Locally, $\lambda$ is given by $\lambda = \Xi^a(x)\xi_a$.	Recall that the infinitesimal (left) action of $\mathfrak{gau}(P)$ on $\mathcal{A}$ is given by
		\begin{align*}
			\mathfrak{gau}(P) \times \mathcal{A}  &\rightarrow T \mathcal{A} \\
			(\lambda, A) &\mapsto (A, d_A \lambda).
		\end{align*}
		where we use identification $T_A(\mathcal{A}) \cong \Omega^1(\mathrm{ad}P)$. This agrees with the Lie derivative of $A = A_{\mu}^a dx^{\mu}\otimes \xi_a \in \Gamma(C)$ with respect to $\Xi_v = \Xi^a(x) \rho_a$, which is locally given by
		\begin{align*}
			\mathrm{Lie}_{\Xi_v} A^a_{\mu} = (A^a_{\mu}, \partial_{\mu}\Xi^a + f^a_{bc}A^b_{\mu}\Xi^c),
		\end{align*}
		where $f^a_{bc}$ are the structure constants of $\mathfrak{g}$ with respect to $\xi_a$, and we identify $VC$ as $C \times_M (T^*M \otimes \mathrm{ad} P)$.
	\end{exmp}
	
	Let's consider the special case $G = \mathrm{Id}$ and $r=0$.
	
	\begin{defn}
		A natural bundle of finite order $s$ is a fiber bundle associated to $L^{s}(M)$ so that $s$ is minimal.
	\end{defn}
	
	Unlike the case of natural gauge bundles, the Lie derivative of a section of a natural bundle can be defined with respect to any vector field over the base manifold.
	
	\begin{exmp}
		Let $\mathrm{Met}(M)$ denote the bundle of (Riemannian) metrics over $M$. It is naturally associated to $\mathrm{Fr}(M)$, therefore a natural bundle of order $1$. The vertical bundle of $\mathrm{Met}(M)$ can be identified with $\mathrm{Met}(M) \times_M S^2 T^*M$, where $S^2 T^*M$ is the bundle of symmetric tensors of type $(0,2)$ over $M$. The Lie derivative of a metric $g = g_{\mu\nu}dx^{\mu} \otimes dx^{\nu}$ along a vector field $X=X^{\mu}\partial_{\mu}$ is locally given by
		\begin{align*}
			\mathrm{Lie}_{X} g_{\mu\nu} = (g_{\mu\nu}, \nabla_{\mu}X_{\nu}+\nabla_{\nu}X_{\mu}),
		\end{align*}
		where $\nabla_{\mu} X_{\nu} = \partial_{\mu} X_{\nu} - \Gamma_{\mu \nu}^{\rho} X_{\rho}$, $\Gamma_{\mu \nu}^{\rho}$ are the Christoffel symbols of $g$, and $X_{\mu}=g_{\mu\nu}X^{\nu}$.
	\end{exmp}
	
	\begin{exmp}
		Let $A(M)$ denote the bundle of affine connections over $M$. One can show that it is a natural bundle of order $2$. The vertical bundle of $A(M)$ can be identified with $A(M) \times_M (\mathrm{End}(M) \otimes T^*M)$, where $(\mathrm{End}(M) \otimes T^*M)$ is the bundle of tensors of type $(1,2)$ over $M$. The Lie derivative of an affine connection $\Gamma=\Gamma^{\rho}_{\mu \nu}\partial_{\rho}\otimes dx^{\mu} \otimes dx^{\nu}$ along a vector field $X=X^{\mu}\partial_{\mu}$ is locally given by \cite{Blohmann2023}
		\begin{align*}
			\mathrm{Lie}_X \Gamma^{\rho}_{\mu \nu} = (\Gamma^{\rho}_{\mu \nu}, \nabla_{\mu}\nabla_{\nu} X^{\rho} + X^{\lambda} R^{\rho}_{\mu\lambda\nu} - 2\nabla_{\mu}( \Gamma^{\rho}_{[\mu\lambda]} V^{\lambda})),
		\end{align*}
		where $\nabla_{\mu} X^{\rho}= \partial_{\mu} X^{\rho} + \Gamma^{\rho}_{\mu\nu} V^{\nu}$ and $\Gamma^{\rho}_{[\mu\nu]}=\frac{1}{2}(\Gamma^{\rho}_{\mu\nu} - \Gamma^{\rho}_{\nu\mu})$, i.e., the torsion of $\Gamma$. 
	\end{exmp}
	
	\subsection{Variational bicomplex}
	
	\subsubsection{Variational bicomplex of a gauge natural bundle}
	
	Let $M$ be an $n$-dimensional (compact) manifold. Let $P$ be a principal $G$-bundle over $M$. Let $\pi: Y \rightarrow M$ be a gauge natural bundle associated to $P$ with fiber $Z$ being an $m$-dimensional manifold.  Let $\Gamma(Y)$ be the space of sections of $Y$. Let's consider the de Rham complex $\Omega (M \times \Gamma(Y))$ of differential forms on $M \times \Gamma(Y)$. It is a bicomplex bigraded according to the product structure of $M \times \Gamma(Y)$. We can write
	\begin{align*}
		\Omega (M \times \Gamma(Y)) = \bigoplus_{p,q} \Omega^{p,q} (M \times \Gamma(Y)).
	\end{align*}
	Correspondingly, the de Rham differential $d_{tot}$ on $M \times \Gamma(Y)$ breaks into two parts $d_{tot}=d + \delta$, where $d$ is the de Rham differential on $M$ and $\delta$ is the de Rham differential on $\Gamma(Y)$.
	
	Let $J^{\infty}Y$ be the infinite jet bundle of $Y$ over $M$. Let $\mathrm{ev}$ be the evaluation map from $M \times \Gamma(Y)$ to $J^{\infty}Y$, i.e.,
	\begin{align*}
		\mathrm{ev}: M \times \Gamma(Y) &\rightarrow J^{\infty}Y \\
		(x,\Phi) &\mapsto j^{\infty}(\Phi)(x),
	\end{align*}
	where $j^{\infty}(\Phi)$ is the infinite jet prolongation of $\Phi$. The pull-back $\mathrm{ev}^* \Omega(J^{\infty}Y)$ is stable under both $d$ and $\delta$, hence a sub-bicomplex, which is denoted by $\Omega_{loc}(M \times \Gamma(Y))$ \cite{Zuckerman1987}. 
	\begin{defn}
		$\Omega_{loc}(M \times \Gamma(Y))$ is called the variational bicomplex of $Y$. Elements in $\Omega_{loc}(M \times \Gamma(Y))$ are called local forms. $d$ and $\delta$ restricted to $\Omega_{loc}(M \times \Gamma(Y))$ are called the horizontal differential, denoted by $d_h$, and the vertical differential, denoted by $d_v$, respectively.
	\end{defn}
	
	A vector field $\Xi$ over $M \times \Gamma(Y)$ is a map $\Xi: M \times \Gamma(Y) \rightarrow TM \times \Gamma(VY)$. $\Xi$ is called local if it projects on a vector field $\xi$ over $J^{\infty}Y$, i.e., if the following diagram commutes,
	\[ \begin{tikzcd}[row sep=large, column sep=large]
		M \times \Gamma(Y) \arrow{d}{\mathrm{ev}} \arrow{r}{\Xi} &TM \times \Gamma(VY) \arrow{d}{T\mathrm{ev}}\\
		J^{\infty}Y \arrow{r}{\xi} &  TJ^{\infty}(Y).
	\end{tikzcd} \] 
	A local vector field $\Xi$ is called (strictly) vertical if it is of the form $(0,\Xi')$ where $\Xi'$ is a vector field over $\Gamma(Y)$, and (strictly) horizontal if it is of the form $(X,0)$ where $X$ is a vector field over $M$.	
	\begin{rmk}
		Let $\iota_{\Xi}$ denote the contraction of the local vector field $\Xi$. Obviously, we have
		\begin{align*}
			[\iota_{\Xi},d_h] = 0
		\end{align*}
		if $\Xi$ is vertical and 
		\begin{align*}
			[\iota_{\Xi},d_v] = 0
		\end{align*}
		if $\Xi$ is horizontal. It follows that the Lie derivative $\mathrm{Lie}_{\Xi}$ is of the form $\mathrm{Lie}_{\Xi}=[d_v, \iota_{\Xi}]$ if $\Xi$ is vertical and $\mathrm{Lie}_{\Xi}=[d_h, \iota_{\Xi}]$ if $\Xi$ is horizontal.
	\end{rmk}
	
	\begin{exmp}\label{autpvarbi}
		Since $Y$ is gauge natural bundle, every $G$-invariant vector field over $P$ induces naturally a vertical local vector field over $\Gamma(Y)$. In particular, if $Y$ is a natural bundle, a vector field $X$ over $M$ induces a vertical local vector field $\xi_X$ over $\Gamma(Y)$.
	\end{exmp}
	
	On the other hand, every vector field $X$ over $M$ can be lifted to a vector field $\widehat{X}$ over $J^{\infty}(Y)$ via the Cartan connection, i.e., we have the following commutative diagram,
	\[ \begin{tikzcd}[row sep=large, column sep=large]
		J^{\infty}Y  \arrow{d} \arrow{r}{\widehat{X}} &TJ^{\infty}(Y) \arrow{d}\\
		M \arrow{r}{X}&  TM
	\end{tikzcd} \] 
	where the vertical morphisms are the canonical projections.  One can show that the lift of $\widehat{X}$ to $M \times \Gamma(Y)$ is exactly of the form $(X,0)$. With a slight abuse of notation, we denote $(X,0)$ as $\widehat{X}$.
	
	Let ${U}$ be an open neighborhood of $M$ such that $\pi^{-1}U \cong U \times Z$. Let ${V}$ be a coordinate chart of $Z$. $Y$ can then be covered by coordinate charts of the form $U \times V$ with coordinate functions $x^1,\dots,x^n,u^1,\dots,u^m$. Let $\mathcal{W}(U,V)$ be the set of pairs $(x,\Phi)$ such that $\Phi(x)$ is in $V$ for all $x \in U$, it is then an open neighborhood of $M \times \Gamma(Y)$. We can explicitly define functions $x^{\mu}$ and $u^j_I$ on $\mathcal{W}(U,V)$ by setting
	\begin{align*}
		x^{\mu}(x,\Phi)=x^{\mu}(x),\quad u^j_I(x,\Phi)=\partial_I(u^j(\Phi(x))),
	\end{align*}
	where $\partial_I$ is the partial derivative in $x^{\mu}$ with respect to the multi-index $I=(\mu_1,\dots,\mu_n)$. By definition, a local function on $\mathcal{W}(U,V)$ depends only on finitely many of $x^{\mu}$ and $u^j_I$. In particular, $x^{\mu}$ and $u^j_I$ themselves are local functions. Their derivatives $dx^{\mu}$ and $\delta u^j_I$ can be viewed as local forms of degree $(1,0)$ and $(0,1)$, respectively. One can write any local $(k,l)$-form $\omega$ as a finite sum
	\begin{align*}
		\omega = f_{\mu_l,\dots,\mu_k,j_1,\dots,j_l}^{I_1,\dots,I_l} dx^{\mu_1} \wedge \dots \wedge dx^{\mu_k} \wedge \delta u^{j_1}_{I_1} \wedge \dots \wedge \delta u^{j_l}_{I_l},
	\end{align*}
	where each $f_{\mu_l,\dots,\mu_k,j_1,\dots,j_l}^{I_1,\dots,I_l}$ is a local function. On the other hand, one can show that every horizontal local vector field is of the form
	\begin{align}
		\widehat{X} = X^{\mu} \partial_{\mu} + X^{\mu} u^j_{I \cup \{\mu\}} \frac{\partial}{\partial u^j_I}
	\end{align}
	and every vertical local vector field is of the form
	\begin{align}
		\Xi =  \Xi^j \frac{\partial}{\partial u^j} + \widehat{\partial_I}(\Xi^j) \frac{\partial}{\partial u^j_I}.
	\end{align}
	
	The differentials $d_h$ and $d_v$ can then be expressed as
	\begin{align}\label{difflielocal}
		d_h = dx^{\mu} \wedge \mathrm{Lie}_{\widehat{\partial_{\mu}}}, \quad
		d_v = \delta u^j_I \wedge \mathrm{Lie}_{\frac{\partial}{\partial u^j_I}}.
	\end{align}
	\begin{defn}
		A local form $\omega$ is called invariant with respect to a $G$-invariant vector field $\Xi \in \mathfrak{X}_{inv}(P)$ which covers a vector field $X \in \mathfrak{X}_P(M)$ if
		\begin{align*}
			\mathrm{Lie}_{\widehat{X} - \Xi} \omega = 0.
		\end{align*}
		$\omega$ is said to be invariant if it is invariant with respect to all $\Xi \in \mathfrak{X}_{inv}(P)$.
		In particular, if $Y$ is a natural bundle, $\omega$ is called invariant with respect to a vector field $X \in \mathfrak{X}(M)$ if
		\begin{align*}
			\mathrm{Lie}_{\widehat{X} - \xi_X} \omega = 0.
		\end{align*}
		where $\xi_X$ is defined as in Example \ref{autpvarbi}. $\omega$ is said to be invariant if it is invariant with respect to all $X \in \mathfrak{X}(M)$.
	\end{defn}
	
	By definition, the set $\Omega_{loc,inv}(M \times \Gamma(Y))$ of invariant local forms over $M \times \Gamma(Y)$ is a subbicomplex of the variational bicomplex of $Y$. 
	
	\begin{rmk}
		Let $(U, P|_U \cong U \times G)$ be a local trivialization of $P$ over an open neighborhood $U$ of $M$ with coordinates $x^{\mu}$. Obviously, every local vector field over $U$ can be lifted to a $G$-invariant vector field over $P|_U$. 
		Locally, the one-parameter group $(\Phi_t,\phi_t)$ generated by $\xi_{\partial_{\mu}}$ is nothing but the trivial transportation along $x \mapsto x + t x^{\mu}$. It follows from Proposition \ref{liedergloloc} that 
		\begin{align*}
			\mathrm{Lie}_{\xi_{\partial_{\mu}}} = \mathrm{Lie}_{u^j_{I \cup \{\mu\}} \frac{\partial}{\partial u^j_I}}.
		\end{align*}
		(Use $u^j_{I \cup \{\mu\}}=\lim_{t \rightarrow 0} \frac{u^j_I(x + t x^{\mu})}{t}$.) Therefore, we have \cite[Example 2.17]{Blohmann2023}
		\begin{align*}
			\mathrm{Lie}_{\widehat{\partial_{\mu}}-\xi_{\partial_{\mu}}} = \mathrm{Lie}_{\partial_{\mu}}.
		\end{align*}
		In other words, an invariant local form must be locally independent of the base coordinates $x^{\mu}$. The converse is not true. Let $\Omega_{\circ loc}(U \times \Gamma(Y|_U))$ denote the set of local forms  that are independent of $x^{\mu}$, we have
		\begin{align*}
			\Omega_{loc, inv}(U \times \Gamma(Y|_U)) \subsetneq \Omega_{\circ loc}(U \times \Gamma(Y|_U)) \subsetneq \Omega_{loc}(U \times \Gamma(Y|_U)).
		\end{align*}
	\end{rmk}
	
	Let $d_{h,inv}:= dx^{\mu} \wedge \mathrm{Lie}_{\xi_{\partial_{\mu}}}$ and $K_0:= dx^{\mu} \wedge \iota_{\xi_{\partial_{\mu}}}$. They preserves $\Omega_{\circ loc}(U \times \Gamma(Y|_U))$ because $[\xi_{\partial_{\mu}},\xi_{\partial_{\nu}}]=\xi_{[\partial_{\mu},\partial_{\nu}]}=0$. Note that the expressions for $d_{h,inv}$ and $K_0$ only make sense locally since the the map $\mathfrak{X}(U) \rightarrow \mathfrak{X}_{inv}(P|U), X \mapsto \xi_X$ is not $C^{\infty}(U)$-linear. 
	
	By definition, $d_v$, $d_{h,inv}$, and $K_0$ satisfy the following relations
	\begin{align}
		[d_v,d_v]=[d_v,d_{h,inv}]=[d_{h,inv},d_{h,inv}]=[K_0,K_0]=[K_0,d_{h,inv}]=0,\quad [d_v,K_0]=d_{h,inv}.
	\end{align}
	The last relation tells us the $K_0$ can be interpreted as a homotopy operator and $d_{h,inv}$ is locally homotopy equivalent to the zero differential.
	
	\begin{prop}
		$d_{h,inv}|_{\Omega_{\circ loc}(U \times \Gamma(Y|_U))} = d_h |_{\Omega_{\circ inv}(U \times \Gamma(Y|_U))}$.
	\end{prop}
	\begin{proof}
		This follows directly from \eqref{difflielocal} and the definition of $\Omega_{\circ loc}$.
	\end{proof}
	In particular, we have $d_{h,inv}|_{\Omega_{loc,inv}(U \times \Gamma(Y|_U))} = d_h |_{\Omega_{loc,inv}(U \times \Gamma(Y|_U))}$. It follows that $d_{h,inv}$ is globally well-defined when restricted to the subbicomplex of invariant local forms.
	
	\subsubsection{Variational bicomplex of a graded gauge natural bundle}
	
	The previous discussion can be easily generalized to the graded case.
	\begin{defn}
		A graded gauge natural bundle over $M$ is a composite fiber bundle $Y \rightarrow Y_0 \rightarrow M$ over $M$ where $Y_0 \rightarrow M$ is a ordinary gauge natural bundle over $M$ and $Y \rightarrow Y_0$ is a gauge natural graded vector bundle over $Y_0$.
	\end{defn}
	\begin{exmp}
		$Y=V[1]Y_0 \rightarrow Y_0 \rightarrow M$ where $V[1]Y_0$ is the vertical bundle of the gauge natural bundle $Y_0 \rightarrow M$ shifted by degree $1$.
	\end{exmp}
	The infinite jet bundle $J^{\infty}Y$ of $Y$ (as a bundle over $M$) is again a graded gauge natural bundle $J^{\infty}Y \rightarrow J^{\infty}(Y_0) \rightarrow M$ where $J^{\infty}Y \rightarrow J^{\infty}Y_0$ is induced from the bundle morphism $Y \rightarrow Y_0$ of fiber bundles over $M$, and the grading of the fibers of $J^{\infty}Y$ is induced from the grading of the fibers of $Y$. On the other hand,  $\Gamma(Y)$ can be viewed as a graded vector bundle over $\Gamma(Y_0)$ with fiber $\Gamma(\varphi^*Y)$ at the point $\varphi \in \Gamma(Y_0)$. It follows that $M \times \Gamma(Y)$ can be viewed as a graded vector bundle over $M \times \Gamma(Y_0)$. The evaluation map    
	\begin{align*}
		\mathrm{ev}: M \times \Gamma(Y) &\rightarrow J^{\infty}Y \\
		(x,\Phi) &\mapsto j^{\infty}(\Phi)(x)
	\end{align*}
	together with the evaluation map
	\begin{align*}
		\mathrm{ev}_0: M \times \Gamma(Y_0) &\rightarrow J^{\infty}Y_0 \\
		(x,\Phi_0) &\mapsto j^{\infty}(\Phi_0)(x)
	\end{align*}
	gives us a morphism of graded vector bundles, which induces a morphism of the corresponding graded manifolds.
	
	The evaluation map $\mathrm{ev}$ can be again used to define local forms and local vector fields over the graded manifold $M \times \Gamma(Y)$. The only subtleties of this generalization are  
	\begin{enumerate}
		\item $\Omega_{loc}(M \times \Gamma(Y))$ is trigraded. We write $\Omega_{loc}(M \times \Gamma(Y)) = \bigoplus_{p,q,r} \Omega_{loc}^{p,q,r}(M \times \Gamma(Y))$, where $p$ and $q$ are the horizontal and vertical form degree, respectively, and $r$ is the ghost number degree. $d_{h}$ and $d_v$ are of degrees $(1,0,0)$ and $(0,1,0)$, respectively. In local coordinates, we should assign degree $(0,0,d(u^j))$ to the local function $u^j_I$ and degree $(0,1,d(u^j))$ to the local form $\delta u^j_I$ where $d(u^j)$ is the degree of $u^j$ induced from the grading of $Y$.
		\item A local function should be a graded polynomial in $u^j_I$ when $d(u^j)\neq0$. We also need to fix a notion of commutativity for local forms. Our convention will be 
		\begin{align*}
			dx^{\mu} \wedge \delta u^j_I =\delta u^j_I \wedge dx^{\mu}, \quad \delta u^j_I \wedge \delta u^{j'}_{I'} =(-1)^{(d(u^j)+1)(d(u^{j'})+1)}\delta u^{j'}_{I'} \wedge \delta u^j_I.
		\end{align*}
		\item While horizontal local vector fields over $M \times \Gamma(Y)$ remain ungraded, vertical local vector fields are graded. In local coordinates, we should assign degree $(0,0,-d(u^j))$ to $\frac{\partial}{\partial u^j_I}$ and degree $(0,-1,-d(u^j))$ to the contraction $\iota_{\frac{\partial}{\partial u^j_I}}$.
		\item The Cartan calculus of the vertical part of $\Omega_{loc}(M \times \Gamma(Y))$ should also be modified properly. More precisely, we choose the bracket $[\cdot,\cdot]$ between $d_v$, $\iota_{\Xi}$, and $\mathrm{Lie}_{\Xi}$ such that
		\footnote{$\mathrm{Lie}_{\Xi}$ should be given by $[\iota_{\Xi}, d_v]$ instead of $[d_v, \iota_{\Xi}]$ because one needs $\mathrm{Lie}_{\Xi} f = \Xi(f)$ for a function $f$.}
		\begin{align*}
			&[\iota_{\Xi_1},\iota_{\Xi_2}]=\iota_{\Xi_1} \iota_{\Xi_2} - (-1)^{(-1+d(\Xi_1))(-1+d(\Xi_2))}\iota_{\Xi_2} \iota_{\Xi_1}=0, \\
			&\mathrm{Lie}_{\Xi}=[\iota_{\Xi}, d_v] = \iota_{\Xi} d_v - (-1)^{(-1+d(\Xi))} d_v \iota_{\Xi}  = \iota_{\Xi} d_v + (-1)^{d(\Xi)} d_v \iota_{\Xi}.
		\end{align*}
		Moreover, our sign convention also implies that $d_h$ should commute with $d_h$, $\iota_{\Xi}$, and $\mathrm{Lie}_{\Xi}$ without producing any sign factors.
	\end{enumerate}
	
	\begin{rmk}
		It is not hard to see that $\Omega_{loc}^{p,q}(M \times \Gamma(Y_0)) \cong \Omega_{loc}^{p,0,q}(M \times \Gamma(V[1]Y_0))$ for a ordinary gauge natural bundle $Y$ over $M$. And more generally, $\Omega_{loc}^{p,q,r}(M \times \Gamma(Y)) \cong \Omega_{loc}^{p,0,q+r}(M \times \Gamma(V[1]Y))$ for a graded gauge natural bundle $Y$ over $M$.
	\end{rmk}
	
	We also have a well-defined notion of invariant local forms in the graded case. Let again $\Omega_{loc,inv}(M\times \Gamma(Y))$ denote the bicomplex of invariant local forms. Every result we established for $\Omega_{loc,inv}(M\times \Gamma(Y))$ in the ungraded case remains valid in the graded case. In particular, $d_{h}$ is homotopic to $0$ on $\Omega_{loc,inv}(M\times \Gamma(Y))$, i.e., there exists a homotopy operator $K_0$ such that $K_0 d_v + d_v K_0 = d_{h,inv}$, where $d_{h,inv}$ and $K_0$ are given by the same definitions.
	
	\begin{rmk}
		From now on, we will frequently use the shorthand notation $\Omega_{loc}^{p,q,r}$ to refer to $\Omega_{loc}^{p,q,r}(M \times \Gamma(Y))$ and $\Omega_{loc}^{p,q}$ to denote $\bigoplus_r \Omega_{loc}^{p,q,r}$, provided there is no potential for confusion.
	\end{rmk}
	
	\subsection{Lagrangian field theory}
	
	Let $Y \rightarrow Y_0 \rightarrow M$ be a graded gauge natural bundle over $M$.
	
	\begin{defn}
		A (local) Lagrangian field theory (LFT) is a triple $(M,Y,\mathcal{L})$ where $\mathcal{L} \in \Omega_{loc}^{n,0,0}(M \times \Gamma(Y))$.\footnote{If $M$ is not compact, we require that $\mathcal{L}$ has compact support along $M$.} $M$ is called the spacetime manifold, $Y$ is called the configuration bundle, and $\mathcal{L}$ is called the Lagrangian. The theory is called generally covariant if $\mathcal{L}$ is an invariant local form.
	\end{defn}
	
	Recall that the interior Euler operator $\mathcal{I}: \Omega_{loc}^{n,s} \rightarrow \Omega_{loc}^{n,s}$, $s \geq 1$, is defined by setting
	$
	\mathcal{I} \omega = \delta u^j \wedge (\iota_{\frac{\partial}{\partial u^j}} \omega) + \delta u^j \wedge (\widehat{(-\partial_I)} (\iota_{\frac{\partial}{\partial u^j_I}} \omega)).
	$
	$\mathcal{I}$ has the properties
	\begin{align*}
		\mathcal{I}^2 = \mathcal{I}, \quad \mathcal{I} \circ d_h = 0.
	\end{align*}
	The subspace of $\Omega_{loc}^{n,s}$ that is invariant under $\mathcal{I}$ is denoted by $\mathcal{F}^s$. Elements in $\mathcal{F}^s$ for $s > 1$ are called functional form. Elements in $\mathcal{F}^1$ are called source forms. Note that every source form $\alpha$ can be locally written as $\alpha = \alpha_j \delta u^j \wedge \nu$, where $\alpha^j$ is any local function and $\nu$ is a volume form over $M$.
	\begin{lem}\cite[Theorem 6]{Sardanashvily2005}\label{flv}
		The interior rows of the augmented variational bicomplex 
		\begin{align*}
			0 \longrightarrow \Omega_{loc}^{0,s} \xlongrightarrow{d_h} \Omega_{loc}^{1,s} \xlongrightarrow{d_h} \cdots \xlongrightarrow{d_h} \Omega_{loc}^{n,s} \xlongrightarrow{\mathcal{I}} \mathcal{F}^s \longrightarrow 0,
		\end{align*}
		with $s \geq 1$, are globally exact. Moreover, the cohomology groups of the following cochain complex 
		\begin{align*}
			0 \longrightarrow \Omega_{loc}^{0,0} \xlongrightarrow{d_h} \Omega_{loc}^{1,0} \xlongrightarrow{d_h} \cdots \xlongrightarrow{d_h} \Omega_{loc}^{n,0} \xlongrightarrow{\mathcal{I} \circ d_v} \mathcal{F}^1  \xlongrightarrow{\mathcal{I} \circ d_v} \mathcal{F}^2 \xlongrightarrow{\mathcal{I} \circ d_v} \cdots
		\end{align*}
		is isomorphic to the de Rham cohomology of the fiber bundle $Y_0$.
	\end{lem}
	We also need an infinitesimal version the first part of Lemma \ref{flv}. A local form $\omega \in \Omega_{loc}^{r,s}$ is said to be $d_h$-closed at $\Phi \in \Gamma(Y)$ if $d_h \omega$ vanishes at $\Phi$. It is said to be $d_h$-exact at $\Phi$ if there exists another local form $\omega' \in \Omega_{loc}^{r-1,s}$ such that $\omega - d_h \omega'$ vanishes at $\Phi$. 
	\begin{lem}\cite[Proposition 6.3.23]{Blohmann14}\label{iflv}
		For $r < n$ and $s \geq 1$, $\omega$ is $d_h$-closed at $\Phi$ if and only if it is $d_h$-exact at $\Phi$.
	\end{lem}
	Let $\mathcal{E}:= \mathcal{I} \circ d_v: \Omega_{loc}^{n,0} \rightarrow \Omega_{loc}^{n,1}$. $\mathcal{E}$ is known as the Euler-Lagrange operator. Let $EL=\mathcal{E}(\mathcal{L})$. $EL$ is known as the Euler-Lagrange form of the LFT. 
	\begin{cor}\label{eleq}
		$d_v \mathcal{L} = EL + d_h \gamma$ for some $\gamma \in \Omega_{loc}^{n-1,1,0}$.
	\end{cor}
	$\gamma$ is known as the boundary form of the LFT. By Lemma \ref{flv}, $\gamma$ is unique up to a $d_h$-exact term for a fixed $\mathcal{L}$. 
	
	\begin{defn}
		The action functional $S \in C^{\infty}(\Gamma(Y))$ of a LFT $(M,Y,\mathcal{L})$ is defined as
		\begin{align*}
			S(\Phi) := \int_M \mathcal{L}(x,\Phi).
		\end{align*}
		A field $\Phi \in \Gamma(Y)$ is called on-shell if it is a critical point of $S$, i.e., if $\delta S(\Phi) = 0$. 
	\end{defn}
	
	Let $C_{\mathcal{L}} \subset  \Gamma(Y)$ denote the space of on-shell fields. We have
	\begin{align*}
		\delta S = \int_M d_v \mathcal{L} = \int_M EL.
	\end{align*}
	It follows that $\Phi \in C_{\mathcal{L}}$ if $\iota_{\delta_{\Phi}} EL(x,\Phi)$ vanishes identically over $M$, where $\delta_{\Phi}$ is a tangent vector in $T_{\Phi}\Gamma(Y)$. In fact, one can show that $\Phi$ is on-shell if and only if $(\iota_{\Xi} EL)(x,\Phi)$ vanishes identically over $M$ for all vertical local vector field $\Xi$ over $M \times \Gamma(Y)$ \cite{Blohmann2023}. 
	
	Note that every vertical vector field over $M \times \Gamma(Y)$ can be viewed canonically as a vector field over $\Gamma(Y)$ by definition. If $Y$ is a natural bundle and the LFT is generally covariant, we have
	\begin{align*}
		\mathrm{Lie}_{\xi_X} S = \int_M \mathrm{Lie}_{\xi_X} \mathcal{L} = \int_M \mathrm{Lie}_{\widehat{X}} \mathcal{L} = \int_M d(\iota_{\widehat{X}} \mathcal{L}) = 0
	\end{align*}
	for all $X \in \mathfrak{X}(M)$, i.e., $S$ is $\mathrm{Diff}(M)$-invariant.
	
	\subsubsection{Noether theorem}
	
	Let $(M,Y,\mathcal{L})$ be a LFT. A Noether current $j$ is an element in $\Omega_{loc}^{n-1,0}$ such that there exists a vertical local vector field $\Xi$, $d_h j = \iota_{\Xi} EL$. $(j,\Xi)$ is called a Noether pair \cite{Deligne1999}. Given two Noether pairs $(j_i, \Xi_i)$, $i=1,2$, one can define their bracket to be
	\begin{align*}
		\{(j_1,\Xi_1), (j_2, \Xi_2)\} = (\mathrm{Lie}_{\Xi_1} j_2 - (-1)^{d(\Xi_1)d(\Xi_2)}\mathrm{Lie}_{\Xi_2} j_1, [\Xi_1, \Xi_2]).
	\end{align*}
	Note that
	\begin{align*}
		\iota_{[\Xi_1,\Xi_2]} EL = [\mathrm{Lie}_{\Xi_1},\iota_{\Xi_2}] EL = d_h (\mathrm{Lie}_{\Xi_1} j_2 - (-1)^{d(\Xi_1)d(\Xi_2)} \mathrm{Lie}_{\Xi_2} j_1).
	\end{align*}
	$\{(j_1,\Xi_1), (j_2, \Xi_2)\} $ is again a Noether pair.  In other words, Noether pairs together with the bracket $\{\cdot,\cdot\}$ form a (graded) Lie (super)algebra and the map which sends $\Xi$ to $(j,\Xi)$ is a (graded) Lie (super)algebra homomorphism.
	
	A vertical local vector field $\Xi$ is said to be a (Noether) symmetry of the LFT if there exists an element $\alpha \in \Omega_{loc}^{n-1,0,0}$ such that $\mathrm{Lie}_{\Xi} \mathcal{L} = d_h \alpha$. $\alpha$ is unique up to a $d_h$-closed term for a fixed $\mathcal{L}$. Moreover, if the ghost number degree of $\Xi$ is non-zero, $\alpha$ is unique up to a $d_h$-exact term by Lemma \ref{flv}.
	\begin{thm}
		Let $j:= \alpha - \iota_{\Xi} \gamma$. $j$ is a Noether current.
	\end{thm}
	\begin{proof}
		$d_h j = d_h \alpha - \iota_{\Xi} d_h \gamma = \mathrm{Lie}_{\Xi} \mathcal{L} - \iota_{\Xi} (d_v \mathcal{L} - EL) = \iota_{\Xi} EL$.
	\end{proof}
	
	Let $\Xi$ be a Noether symmetry of the LFT with $\mathrm{Lie}_{\Xi} = d_h \alpha$. We have
	\begin{align*}
		\Xi(S) = \int_M \mathrm{Lie}_{\Xi} \mathcal{L} = \int_M d_h \alpha = 0,
	\end{align*}
	i.e., $S$ is invariant under $\Xi$. Let $\Sigma$ be a hyper-surface in $M$. The integral 
	\[
	\mathcal{J}_{\Sigma}(\Phi) := \int_\Sigma j(x, \Phi)
	\]
	is known as the Noether charge associated to $(j,\Xi)$. Note that for an on-shell $\Phi$, $d_h j(x, \Phi) = (\iota_{\Xi}EL)(x,\Phi)=0$. It follows that $\mathcal{J}_{\Sigma}(\Phi)=\mathcal{J}_{\Sigma'}(\Phi)$ when $\Sigma$ and $\Sigma'$ are two hypersurfaces having the same homology. Let $\omega_{\Sigma}(\Phi):= \int_{\Sigma} (d_v \gamma)(x,\Phi)$. $\delta \omega_{\Sigma} = 0$. $\omega_{\Sigma}$ defines a presymplectic form over $\Gamma(Y)$. 
	We have
	\begin{align*}
		\delta \mathcal{J}_{\Sigma} = \int_{\Sigma} d_v j = \int_{\Sigma} (d_v \alpha - \mathrm{Lie}_{\Xi} \gamma +\iota_{\Xi} d_v \gamma) = \iota_{\Xi} \omega_{\Sigma} + \int_{\Sigma}(d_v \alpha - \mathrm{Lie}_{\Xi} \gamma).
	\end{align*}
	By \eqref{eleq}, we have
	\begin{align*}
		d_h (d_v \alpha - \mathrm{Lie}_{\Xi} \gamma) = d_v \mathrm{Lie}_{\Xi} \mathcal{L} - d_h \mathrm{Lie}_{\Xi}\gamma = \mathrm{Lie}_{\Xi}EL.
	\end{align*}
	One can show that $(\mathrm{Lie}_{\Xi}EL)(x,\Phi)$ vanishes over $M$ for an on-shell $\Phi$.  
	By Lemma \ref{iflv}, $d_v \alpha - \mathrm{Lie}_{\Xi} \gamma$ is $d_h$-exact on-shell. Therefore, we have
	\[
	\left(\delta \mathcal{J}_{\Sigma} - \iota_{\Xi} \omega_{\Sigma}\right)|_{C_{\mathcal{L}}}=0.
	\]
	In other words, $\Xi|_{C_{\mathcal{L}}}$\footnote{This restriction makes sense because $\Xi(S)=0$.} is the Hamiltonian vector field associated to the presymplectic form $\omega_{\Sigma}|_{C_{\mathcal{L}}}$. 
	\begin{rmk}
		Let $\{\cdot,\cdot\}_{C_{\mathcal{L}}}$ denote the Poisson bracket associated to $\omega_{\Sigma}|_{C_{\mathcal{L}}}$. Let $\Xi_1$ and $\Xi_2$ be two Noether symmetries of the LFT. Let $\mathcal{J}_{\Sigma,i}$ be the Noether charges associated to the Noether pairs $(j_i,\Xi_i)$, $i=1,2$.  We have
		\begin{align*}
			\{\mathcal{J}_{\Sigma,1},\mathcal{J}_{\Sigma,2}\}_{C_{\mathcal{L}}} = \Xi_1(\mathcal{J}_{\Sigma,2}) = - (-1)^{d(\Xi_1)d(\Xi_2)} \Xi_2(\mathcal{J}_{\Sigma,1}) = \frac{1}{2} \int_{\Sigma} (\mathrm{Lie}_{\Xi_1} j_2 - (-1)^{d(\Xi_1)d(\Xi_2)}\mathrm{Lie}_{\Xi_2} j_1).
		\end{align*}
		In other words, up to a factor $2$, the bracket $\{\cdot,\cdot\}$ between Noether pairs can be viewed as an off-shell extension of the Poisson bracket between Noether charges. 
	\end{rmk}
	
	\section{Cohomological Lagrangian field theory}
	
	
	\subsection{\texorpdfstring{$QKG^{\star}$}{TEXT}-structures}
	
	Let $Y$ be a gauge natural bundle associated to the principal $G$-bundle $P$ over $M$. Let $\mathrm{Aut}(P)^{\star}$ denote the Cartan graded Lie supergroup associated to the automorphism group $\mathrm{Aut}(P)$ of $P$. There is a canonical sub-supergroup $\mathrm{Gau}(P)^{\star}$ of $\mathrm{Aut}(P)^{\star}$, which is the Cartan graded Lie supergroup associated to the gauge group $\mathrm{Gau}(P)$ of $P$. Noting that in the case of a trivial group $G=\mathrm{Id}$, $\mathrm{Gau}(P)^{\star}$ is trivial and $\mathrm{Aut}(P)^{\star} = \mathrm{Diff}(M)^{\star}$, the Cartan graded Lie supergroup associated to the diffeomorphism group $\mathrm{Diff}(M)$ of $M$.
	
	\begin{defn}
		A $QKG^{\star}$-structure on $M \times \Gamma(Y)$ is a vertical local $\mathrm{Aut}(P)^{\star}$-action on $M \times \Gamma(Y)$ whose underlying $\mathrm{Aut}(P)$-action is the canonical $\mathrm{Aut}(P)$-action on $M \times \Gamma(Y)$.
		A $QK^{\star}$-structure is a $QKG^{\star}$-structure with $G = \mathrm{Id}$. 
	\end{defn}
	\begin{rmk}
		$\bigoplus_r \Omega_{loc}^{p,q,r}(M \times \Gamma(Y))$ is an $\mathrm{Aut}(P)^{\star}$-algebra, hence also a $\mathrm{Gau}(P)^{\star}$-algebra
	\end{rmk}
	
	Equivalently, a $QKG^{\star}$-structure on $M \times \Gamma(Y)$ is specified by the following data:
	\begin{enumerate}
		\item A vertical local vector field $Q$ of degree $1$ over $M \times \Gamma(Y)$ satisfying $Q^2=0$;
		\item A family of vertical local vertical field $K_{\Xi}$ of degree $-1$ over $M \times \Gamma(Y)$, parameterized by $G$-invariant vector fields $\Xi$ over $P$, satisfying
		\begin{align*}
			[K_{\Xi_1}, K_{\Xi_2}] = 0, \quad [\Xi_1, K_{\Xi_2}]= K_{[\Xi_1, \Xi_2]}, \quad [Q, K_{\Xi}] = \Xi,
		\end{align*}
		where we identify $\Xi$ as a vertical local vector field of degree $0$ over $M \times \Gamma(Y)$ via taking Lie derivatives.
	\end{enumerate}
	
	Let $\Xi$ be a $G$-invariant vector field over $P$. Let $\{U, P|_U \cong U \times G\}$ be a local trivialization of $P$ equipped with a coordinate system $(x^{\mu})$. Let $\{\xi_a\}$ be a basis of $\mathfrak{g}$. Recall that $\Xi$ locally takes the form $\Xi = \Xi^{\mu}(x) \frac{\partial}{\partial x^{\mu}} + \Xi^a(x) \rho_a$, where $\rho_a$ is the right invariant vector field generated by $\xi_a$ over $G$. The second component of $\Xi$ can be identified with a local section $\lambda=\Xi^a(x) \xi_a$ of the adjoint bundle $\mathrm{ad} P$. Let $K_{\mu}:=K_{\frac{\partial}{\partial x^{\mu}}}$ and $I_{\lambda}:=K_{\Xi^a \rho_a}$. We have
	\begin{align*}
		&[Q,K_{\mu}] = \xi_{\partial_{\mu}}, \quad [K_{\mu}, K_{\nu}] = 0, \quad [\xi_{\partial_{\mu}}, K_{\nu}] = 0,\\
		&[Q, I_{\lambda}] = \delta_{\lambda}, \quad [I_{\lambda}, I_{\lambda'}] = 0, \quad [\delta_{\lambda},I_{\lambda'}]=I_{[\lambda,\lambda']}, \\
		&[K_{\mu},I_{\lambda}]=0, \quad [\xi_{\partial_{\mu}}, I_{\lambda}] = I_{\partial_{\mu}\lambda},
	\end{align*}
	where $\xi_{\partial_{\mu}}$ is the vertical lift of $\frac{\partial}{\partial x^{\mu}}$ on $U \times \Gamma(Y|_U)$ via taking Lie derivatives, $\delta_{\lambda}$ is the infinitesimal gauge transformation induced by $\lambda$, and $\partial_{\mu} \lambda := \partial_{\mu}\Xi^a \xi_a$. Note also that
	\begin{align*}
		[K_{\mu},\delta_{\lambda}] = [[K_{\mu},Q],I_{\lambda}] - [Q,[K_{\mu},I_{\lambda}]]=[\xi_{\partial_{\mu}}, I_{\lambda}] = I_{\partial_{\mu}\lambda}.
	\end{align*}
	Let $K:=dx^{\mu} \wedge \mathrm{Lie}_{K_{\mu}}: \Omega_{loc}^{p,q,r}(U \times \Gamma(Y|_U))\rightarrow \Omega_{loc}^{p+1,q,r-1}(U \times \Gamma(Y|_U))$. Just like $K_0$, $K$ is only locally well-defined.  
	\begin{prop}
		$Q$, $K$, and $d_{h,inv}$ satisfy the following relations. (With a slight abuse of notation, we often use $Q$ instead of $\mathrm{Lie}_Q$ to denote the Lie derivative along $Q$.)
		\begin{align}\label{QKd}
			Q^2=0, \quad QK+KQ = d_{h,inv}, \quad Kd_{h,inv} + d_{h,inv} K = 0.
		\end{align}
	\end{prop}
	\begin{proof}
		$[Q,K]=dx^{\mu} \wedge \mathrm{Lie}_{[Q,K_{\mu}]} = d_{h,inv}$, $[d_{h,inv},K]=dx^{\mu} \wedge dx^{\nu} \wedge \mathrm{Lie}_{[\xi_{\partial_{\mu}},K_{\nu}]}=0$.
	\end{proof}
	\begin{rmk}
		$Q$, $K$, and $d_{h,inv}$ together with the relations \eqref{QKd} define a $QK$-algebra which is studied in details in \cite{Jiang2023a}.
	\end{rmk}
	\begin{lem}\label{shk}
		$K: \bigoplus_r \Omega_{loc}^{p,q,r}(U \times \Gamma(Y|_U)) \rightarrow \bigoplus_r \Omega_{loc}^{p+1,q,r}(U \times \Gamma(Y|_U))$ is a semi-homotopy of $\mathrm{Gau}(P)^{\star}$-algebras.
	\end{lem}
	\begin{proof}
		We need to show that $[K,I_{\lambda}] = 0$ and $[K,\delta_{\lambda}]$ vanishes for local functions that are basic with respect to the $\mathrm{Gau}(P)^{\star}$-action. The first one follows directly from $[K_{\mu},I_{\lambda}]=0$ and the second one follows from $[K_{\mu},\delta_{\lambda}]=I_{\partial_{\mu} \lambda}$. 
	\end{proof}
	For the reader's convenience, we summarize the global and local derivations we have defined on $\Omega_{loc}$ in Table \ref{t_1}.
	\begin{table}[!htbp]
		\centering
		\begin{tabular}{c c c c c c c c}
			\hline
			Operation & $d_h$ & $d_{h,inv}$ & $d_v$ & $\mathrm{Lie}_Q$ & $\iota_Q$ & $K_0$ & $K$ \\
			\hline
			Degree & $(1,0,0)$ & $(1,0,0)$ & $(0,1,0)$ & $(0,0,1)$ & $(0,-1,1)$ & $(1,-1,0)$ & $(1,0,-1)$\\
			Global/local & Global & Global\tablefootnote{$d_{h,inv}$ is global only when it is restricted to $\Omega_{loc, inv}$.}  & Global & Global & Global & Local & Local \\
			\hline
		\end{tabular}    
		\caption{Global/local derivations on $\Omega_{loc}$ .}\label{t_1}
	\end{table}
	
	For later use, we compute that
	\begin{align}
	    &[K, \iota_Q] = dx^{\mu}\wedge [\mathrm{Lie}_{K_{\mu}}, \iota_Q] = dx^{\mu} \wedge \iota_{[K_{\mu}, Q]} = K_0, \label{KiQ} \\
		&[\mathrm{Lie}_Q,K_0] = [\mathrm{Lie}_Q,[K,\iota_Q]] = [d_{h,inv}, \iota_Q] -[K,[\mathrm{Lie}_Q,\iota_Q]] = 0, \label{KK0}
	\end{align}
	where we use $[d_h, \iota_Q]=0$ and $[\mathrm{Lie}_Q,\iota_Q]=\iota_{[Q,Q]}=0$.
	
	\begin{defn}
		A deformation of a $QKG^{\star}$-structure is a family of deformations
		\begin{align*}
			K_{\Xi} \mapsto K_{\Xi} + t K'_{\Xi}
		\end{align*}
		for $t \in \mathbb{R}$, where $K'_X$ is a vertical local vector field of degree $-1$ satisfying
		\begin{align*}
			[Q, K'_{\Xi}]=0, \quad [K'_{\Xi_1}, K'_{\Xi_2}]=0.
		\end{align*}
		A deformation is said to be vertically compatible with the original $QKG^{\star}$-structure if $[K'_{\Xi}, I_{\lambda}] = 0$ for all $\lambda \in \mathrm{ad} P$. It is said to be (fully) compatible with the original $QKG^{\star}$-structure if $[K'_{\Xi}, K_{\Xi'}] = 0$ for all $\Xi' \in \mathfrak{X}_{inv}(P)$. 
	\end{defn}
	
	Obviously, $Q$ and $K_{\Xi} + t K'_{\Xi}$ define a new $QKG^{\star}$-structure if the deformation is compatible with the original $QKG^{\star}$-structure.
	
	\subsection{Cohomological Lagrangian field theories and supersymmetries}
	
	Let $(M,g)$ be a Riemannian manifold. Let $Y$ be a graded natural bundle over $M$. 
	Let $\mathrm{Iso}(M)$ denote the isometry group of the Riemannian manifold $M$. Let $\mathfrak{iso}(M)$ denote the Lie algebra of $\mathrm{Iso}(M)$, whose elements are Killing vector fields over $M$.
	\begin{defn}\label{CohLFT}
		A cohomological Lagrangian field theory (CohLFT) is a LFT $(M,Y,\mathcal{L})$ such that
		\begin{enumerate}
			\item $M \times \Gamma(Y)$ is equipped with a (deformed) $QK^{\star}$-structure;
			\item $Q$ is a Noether symmetry of the LFT, i.e., there exists a local form $\alpha_Q$ of degree $(n-1,0,1)$ such that $Q \mathcal{L} = d_h \alpha_Q$.
		\end{enumerate}
		A CohLFT is called supersymmetric if $K_X$ is a Noether symmetry of the LFT for all $X$ in (a nontrivial subalgebra of) $\mathfrak{iso}(M)$.
	\end{defn}
	
	Supersymmetric CohLFTs are usually obtained by applying a trick called topological twisting to supersymmetric LFTs  with $R$-symmetries \cite{Witten1988}. The idea is to ``twist'' the structure of the super Poincar\'e algebra via a nontrivial group homomorphism from the spin group to the $R$-symmetry group of the LFT. For more details, we refer the reader to Section 5 of \cite{Jiang2023a}.
	
	\begin{defn}
		A preobservable $\mathcal{O}$ in the CohLFT $(M, Y,\mathcal{L})$ is an (invariant) local form over $M \times \Gamma(Y)$ of vertical form degree $0$ which is $Q$-closed up to a $d_h$-exact term.  
	\end{defn}
	
	Let $\mathcal{O}$ be a preobservable of horizontal form degree $p$. Let $\gamma$ be a $p$-dimensional submanifold in $M$. One can define
	\begin{align*}
		O[\gamma](\Phi) := \int_{\gamma}\mathcal{O}(x,\Phi).
	\end{align*}
	Obviously, $O[\gamma]$ is a $Q$-closed function over $\Gamma(Y)$. Moreover, $O[\gamma] = O[\gamma']$ if $\gamma$ and $\gamma'$ are two $p$-dimensional submanifolds having the same homology.	
	\begin{prop}\label{diffqexO}
		$O[\gamma]$ is $\mathrm{Diff}(M)$-invariant up to a $Q$-exact term.
	\end{prop} 
	\begin{proof}
		By definition, one can find a local form $\mathcal{O}'$ such that $Q \mathcal{O} = d_h \mathcal{O}'$. We have
		\begin{align*}
			\xi_{X}(O[\gamma]) = \int_{\gamma} \mathrm{Lie}_{\xi_X} \mathcal{O} = \int_{\gamma} (Q K_X + K_X Q) \mathcal{O} = Q (\int_{\gamma} K_X \mathcal{O}) + \int_{\gamma} d_h (K_X \mathcal{O}') = Q (\int_{\gamma} K_X \mathcal{O}),
		\end{align*}
		where we use $\mathrm{Lie}_{\xi_X} = [Q, K_{X}]$ and $[K_X, d_h]=0$.
	\end{proof}
	Using similar arguments as in the proof of Proposition \ref{diffqexO}, one can prove
	\begin{prop}\label{diffqexS}
		The action $S$ of a CohLFT is $\mathrm{Diff}(M)$-invariant up to a $Q$-exact term.
	\end{prop}
	Let $\mathrm{Loc}(\Gamma(Y))$ denote the subspace of $C^{\infty}(\Gamma(Y))$ spanned by functions $F_{\alpha,\gamma}$ of the form 
	\begin{align*}
		F_{\alpha,\gamma}:=\int_{\gamma} \alpha
	\end{align*}
	where $\alpha$ is a local form of horizontal degree $p$, $\gamma$ is a $p$-dimensional submanifold of $M$.
	\begin{defn}
		An observable of the CohLFT $(M, Y,\mathcal{L})$ is an element in $\mathrm{Loc}(\Gamma(Y))$ which is $Q$-closed and $\mathrm{Diff}(M)$-invariant up to a $Q$-exact term, i.e., a $\mathrm{Diff}(M)$-invariant element in the cohomology group $H_Q(\mathrm{Loc}(\Gamma(Y)))$.
	\end{defn}    
	\begin{rmk}
		By definition, the function $O[\gamma]$ over $\Gamma(Y)$ obtained by integrating the preobservable $\mathcal{O}$ over $\gamma$ is an observable of $(M, Y,\mathcal{L})$.
	\end{rmk}
	
	The expectation value of an observable in quantum field theory is given by the formula
	\begin{align*}
		\langle O \rangle = \int_{\Gamma(Y)} \mathcal{D}\Phi O(\Phi)\exp(-S(\Phi)),
	\end{align*}
	where $\mathcal{D}\Phi$ is the path integral measure on $\Gamma(Y)$. If this measure is invariant under the action of the Lie supergroup generated by $Q$, one can show that $\langle Q(O)\rangle =0$. Therefore, $\langle \cdot \rangle$ in a CohLFT can be viewed as a map
	\begin{align}
		\langle \cdot \rangle: H_Q(\mathrm{Loc}&(\Gamma(Y)))\rightarrow \mathbb{R} \\
		O &\mapsto \langle O \rangle. \notag
	\end{align}
	Let's assume that the path integral measure $\mathcal{D}\Phi$ is also $\mathrm{Diff}(M)$-invariant. Let $\{\phi_t\}$ be the one-parameter group generated by $X \in \mathfrak{X}(M)$.  Let $O$ be a observable of the CohLFT, i.e., a $\mathrm{Diff}(M)$-invariant element in $H_Q(\mathrm{Loc}(\Gamma(Y)))$. By Proposition \ref{diffqexS}, we have
	\begin{align*}
		\lim_{t \rightarrow 0} \frac{d}{dt}\left(\phi_t \langle O\rangle  \right) = - \int_{\Gamma(Y)} \mathcal{D}\Phi \mathrm{Lie}_{\xi_X}\left(O(\Phi)\exp(-S(\Phi))\right) 
		=  \int_{\Gamma(Y)} \mathcal{D}\Phi (O(\Phi)\exp(-S(\Phi)) \mathrm{Lie}_{\xi_X}(S)) 
		=0.
	\end{align*}
	Therefore, $\langle \cdot \rangle$ is constant when restricted to the $\mathrm{Diff}(M)$-invariant subspace of $H_Q(\mathrm{Loc}(\Gamma(Y)))$. 
	
	\subsubsection{Vector supersymmetries}
	
	Let $(M,g)$ be the $n$-dimensional Euclidean space $\mathbb{R}^n$ equipped with the canonical Euclidean metric. The Killing vector fields over $M$ are $\partial_{\mu}$, which generate the translations, and $x^{\mu} \partial_{\nu} - x^{\nu} \partial_{\mu}$, which generate the rotations. In this case, the vertical local vector field $K_{\mu}$ is globally well-defined and we have
	\begin{align}\label{qkeucli}
		Q K_{\mu} + K_{\mu} Q = \xi_{\partial_{\mu}}.
	\end{align} 
	If a CohLFT $(M,Y,\mathcal{L})$ is supersymmetric, $K_{\mu}$ is a Noether symmetry of $(M,Y,\mathcal{L})$. In the physics literature, $K_{\mu}$ is referred to as a vector supersymmetry \cite{Baulieu2005,Ouvry1989,Piguet2008}. It follows from \ref{qkeucli} that the infinitesimal translation $\xi_{\partial_{\mu}}$ is also a Noether symmetry of the theory. Let $\mathcal{Q}$, $\mathcal{G}_{\mu}$, and $\mathcal{T}_{\mu}$ denote the Noether currents associated to $Q$, $K_{\mu}$, and $\xi_{\partial_{\mu}}$, respectively. We then have
	\begin{align*}
		(\mathcal{T}_{\mu},\xi_{\partial_{\mu}}) = \{(\mathcal{Q},Q),(\mathcal{G}_{\mu},K_{\mu})\},
	\end{align*}
	where $\{\cdot,\cdot\}$ is the bracket between Noether pairs.
	Note that $\mathcal{K}_{\mu}$ and $\mathcal{G}_{\mu}$ can be written as
	\begin{align*}
		\mathcal{K}_{\mu} = \mathcal{G}_{\mu\nu} \star dx^{\nu}, \quad  \mathcal{T}_{\mu} = \mathcal{T}_{\mu\nu} \star dx^{\nu},
	\end{align*}
	where $\star$ is the Hodge star operator. $\mathcal{T}_{\mu\nu}$ is known as the canonical energy-momentum tensor. In the physics literature, one often writes
	\begin{align*}
		\mathcal{T}_{\mu\nu} = \{\mathcal{Q}, \mathcal{G}_{\mu\nu}\}
	\end{align*}
	to emphasize the $Q$-exact nature of $\mathcal{T}_{\mu\nu} $. 

	\subsubsection{Descendant sequences of preobservables}
	
	Let $\mathcal{O}^{(0)}$ be a $Q$-closed preobservable of degree $(0,0,n)$ of the CohLFT $(M, Y,\mathcal{L})$. 
	\begin{defn}
		A descendant sequence of $\mathcal{O}^{(0)}$ is a sequence $\{ \mathcal{O}^{(p)}\}_{p=0}^n$ of preobservables satisfying
		\begin{align}\label{deq}
			Q \mathcal{O}^{(p)} = d_h \mathcal{O}^{(p-1)}
		\end{align}
		for $p=1,\dots,n$.  (\ref{deq}) is called the (topological) descent equations of preobservables. 
	\end{defn}
	
	\begin{rmk}
		Descendant sequences of preobservables can be introduced for not only a CohLFT, but also any LFT with a scalar supersymmetry $Q$. In such a LFT, let's consider a preobservable $\mathcal{O}$ which is an invariant local form of horizontal form degree $p$. We then have
		\begin{align*}
			\xi_X(O[\gamma]) = \int_{\gamma} \mathrm{Lie}_{\xi_X} \mathcal{O} = \int_{\gamma} \mathrm{Lie}_{\widehat{X}} \mathcal{O},
		\end{align*}
		where $\xi_X$ is the vertical local vector field induced by the Lie derivatives along $X$. Now apply Cartan's formula $[d_h, \iota_{\widehat{X}}] = \mathrm{Lie}_{\widehat{X}}$. We obtain
		\begin{align*}
			\xi_X(O[\gamma]) = \int_{\gamma} d_h(\iota_{\widehat{X}} \mathcal{O}) + \iota_{\widehat{X}}(d_h \mathcal{O}) = Q( \int_{\gamma} \iota_{\widehat{X}} \mathcal{O}'),
		\end{align*}
		where $\mathcal{O}'$ is a preobservable of horizontal degree $p+1$ satisfying $Q \mathcal{O}' = d_h \mathcal{O}$. In other words, the observable $O[\gamma]$ associated to $\mathcal{O}$ is again $\mathrm{Diff}(M)$-invariant up to a $Q$-exact term despite the absence of a $QK^{\star}$-strucutre.
	\end{rmk}
	
	Recall that we have (locally) a operator $K$ on $\Omega_{\circ loc}$, the bicomplex consisting of those local forms that are locally independent of the base coordinates $x^{\mu}$, satisfying 
	\begin{align*}
		QK + KQ = d_{h,inv}, \quad K d_{h,inv} + d_{h,inv} K = 0.
	\end{align*}
	
	\begin{defn}
		The standard $K$-sequence of an invariant preobservable $\mathcal{O}^{(0)}$ is locally defined by
		\begin{align*}
			\mathcal{O}^{(p)}:= \frac{K^p}{p!} \mathcal{O}^{(0)}.
		\end{align*}
		for $p=1,\cdots,n$.
	\end{defn}
	By definition, $\mathcal{O}^{(p)}$ is again independent of $x^{\mu}$, i.e., an element in $\Omega_{\circ loc}$. 
	
	\begin{prop}\label{kseq}
		Every standard $K$-sequence $\{ \mathcal{O}^{(p)}\}_{p=0}^n$ of an invariant preobservable $\mathcal{O}^{(0)}$ is a descendant sequence.
	\end{prop}
	\begin{proof}
		We have $Q \mathcal{O}^{(p)}=\frac{1}{p!}Q K^p \mathcal{O}^{(0)}=\frac{1}{p!} [Q,K^p] \mathcal{O}^{(0)}=\frac{p}{p!}d_{h,inv} K^{p-1}\mathcal{O}^{(0)}=d_h \mathcal{O}^{(p-1)}$.
	\end{proof}
	
	\begin{defn}
		Let $\mathcal{W}^{(q)}$ be a $Q$-closed invariant local form of degree $(q,0,n-q)$, $1 \leq q \leq n$. A (general) $K$-sequence of an invariant preobservable $\mathcal{O}^{(0)}$ is a sequence $\{ \mathcal{O}^{(p)}\}_{p=0}^n$, where 
		\begin{align*}
			\mathcal{O}^{(p)}:= \frac{1}{p!}K^p \mathcal{O}^{(0)} + \sum_{q=1}^p \frac{1}{(p-q)!}K^{p-q} \mathcal{W}^{(q)}
		\end{align*}
		for $p=1,\dots,n$.
	\end{defn}
	Likewise, one can show that
	\begin{prop}\label{kgenseq}
		Every (general) $K$-sequence $\{ \mathcal{O}^{(p)}\}_{p=0}^n$ of an invariant preobservable $\mathcal{O}^{(0)}$ is a descendant sequence.
	\end{prop}
	
	Let $\{ \mathcal{O}^{(p)}\}_{p=0}^n$ be such that $\mathcal{O}^{(p)}=Q\mathcal{\rho}^{(p)} + d \mathcal{\rho}^{(p-1)}$ for $p>0$ and $\mathcal{O}^{(0)}=Q\mathcal{\rho}^{(0)}$, where $ \mathcal{\rho}^{(p)}$ is an arbitrary invariant local form of degree $(p,n-p-1)$. Then, $\{ \mathcal{O}^{(p)} \}_{p=0}^n$ is a solution to (\ref{deq}). Such a sequence  is called an exact sequence.  Obviously, $\{ \mathcal{O}^{(p)}\}_{p=0}^n$ is an exact sequence if and only if $\mathcal{O}=\sum_{p=0}^n \mathcal{O}^{(p)}$ is $(Q-d_h)$-exact.
	
	\begin{thm}\label{mthmphy}
		Every descendant sequence of an invariant preobservable $\mathcal{O}^{(0)}$ is locally a $K$-sequence up to an exact sequence.
	\end{thm}
	
	\begin{proof}
		Consider the ``Mathai-Quillen automorphism'' $j:=\exp(\widetilde{K})$ of $\Omega_{\circ loc}$, where $\widetilde{K}$ is defined by setting
	    $
	    \widetilde{K} \omega = (-1)^{d_{tot}(\omega) - n} K \omega
	    $ 
	    for $\omega \in \Omega_{\circ loc}$, $d_{tot}(\omega)$ is the total degree of $\omega$. 
		Note that the expression $\exp(\widetilde{K})$ is well-defined because $K$ is nilpotent. It is not hard to show that
		$
		[Q, \widetilde{K}]:= Q \widetilde{K}  - \widetilde{K} Q = \widetilde{d_{h,inv}},
		$
		where $\widetilde{d_{h,inv}}$ is defined in a similar manner as $\widetilde{K}$. It follows that
		\[
			j \circ Q \circ j^{-1} = \exp(\widetilde{K}) \left([Q,\exp(-\widetilde{K})] + \exp(-\widetilde{K}Q)\right) =-\exp(\widetilde{K})\widetilde{d_{h,inv}} \exp(-\widetilde{K}) + Q = Q - \widetilde{d_{h,inv}},
		\]
		where we use $\widetilde{d_{h,inv}} \widetilde{K} = \widetilde{K} \widetilde{d_{h,inv}}$. The proof is complete by observing that $d_{h,inv}=d_h$ when restricted to $\Omega_{\circ loc}$ and the total degree of $\mathcal{O}^{(p)}$ is $n$ for $p=0,\cdots,n$.
	\end{proof}
	
	
	\subsection{Cohomological Lagrangian gauge field theory}
	
	Let $(M,g)$ be a Riemannian manifold. Let $P$ be a principal $G$-bundle over $M$. Let $Y$ be a gauge natural bundle over $M$ associated to $P$. 
	
	\begin{defn}\label{CohLGFT}
		A cohomological Lagrangian gauge field theory (CohLGFT) is a LFT $(M,Y,\mathcal{L})$ such that
		\begin{enumerate}
			\item $M \times \Gamma(Y)$ is equipped with a (deformed) $QKG^{\star}$-structure;
			\item $Q$ is a Noether symmetry of the LFT;
			\item $I_{\lambda}$ and $\delta_{\lambda}$ are Noether symmetries of the LFT for all $\lambda \in \Gamma(\mathrm{ad} P)$.
		\end{enumerate}
		A CohLFT is called supersymmetric if $K_X$ is a Noether symmetry of the LFT for all $X$ in (a nontrivial subalgebra of) $\mathfrak{iso}(M) \cap \mathfrak{X}_P(M)$.
	\end{defn}
	
	\begin{rmk}
		Let's consider $\mathcal{L}$ of the form
		\begin{align}
			\mathcal{L} = \mathcal{L}_{top} + Q(\mathcal{V}_g),
		\end{align}
		where $\mathcal{L}_{top}$ is an invariant local form\footnote{In this paper, we do not consider gravity theories. That is to say, $g$ is a background field and we do not include the bundle $\mathrm{Met}(M)$ of Riemannian metrics as part of the configuration bundle $Y$. Therefore, any invariant local form is automatically independent of $g$.} of degree $(n,0,0)$, and $\mathcal{V}_g$ is a local form of degree $(n,0,-1)$ that is dependent on $g$ and is basic respect to the $\mathrm{Gau}(P)^{\star}$-action. The action functional of the CohLFT takes the form
		\begin{align*}
			S = S_{top} + Q(\Psi_g)
		\end{align*}
		where $S_{top} = \int_M \mathcal{L}_{top}$ and $\Psi_g=\int_M \mathcal{V}$. In other words, the triple $(\Gamma(Y),Q,S_{top})$ is a BRST system which can be gauge fixed to a BRST system $(\Gamma(Y),Q,S_{top})$ via the gauge fixing fermion $\Psi_g$.
	\end{rmk}
	
	\begin{defn}
		A preobservable $\mathcal{O}$ in the CohLGFT $(M, Y,\mathcal{L})$ is an (invariant) local form over $M \times \Gamma(Y)$ of vertical form degree $0$ such that
		\begin{enumerate}
			\item $\mathcal{O}$ is $Q$-closed up to a $d_h$-exact term;
			\item $\mathcal{O}$ is gauge invariant, i.e., it is basic with respect to the $\mathrm{Gau}(P)^{\star}$-action.
		\end{enumerate}
	\end{defn}
	Obviously, the integral $O[\gamma]:=\int_{\gamma} \mathcal{O}$ is a $Q$-closed function over $\Gamma(Y)$ which is basic with respect to the $\mathrm{Gau}(P)^{\star}$-action. Likewise, one can prove	
	\begin{prop}\label{autpqexO}
		The action $S$ and $O[\gamma]$ are $\mathrm{Aut}(P)$-invariant up to $Q$-exact terms.
	\end{prop} 
	
	\begin{defn}
		An observable of the CohLGFT $(M, Y,\mathcal{L})$ is an element in $\mathrm{Loc}(\Gamma(Y))$ which is $Q$-closed and $\mathrm{Aut}(P)$-invariant up to a $Q$-exact term.
	\end{defn}   
	
	The descendant sequence and the standard $K$-sequence of a preobservable in a CohLGFT can be defined in a similar manner as in the case of a CohLFT.
	\begin{lem}\label{shk2}
		Let $\omega$ be an element in $\Omega_{\circ loc}$ which is basic with respect to the $\mathrm{Gau}(P)^{\star}$-action. $K \omega$ is also basic.
	\end{lem}
	\begin{proof}
		This follows directly from Lemma \ref{shk}.
	\end{proof} 
	
	By Lemma \ref{shk2}, the standard $K$-sequence of a gauge invariant preobservable is again gauge invariant. We have
	\begin{prop}
		In a CohLGFT, every standard $K$-sequence $\{ \mathcal{O}^{(p)}\}_{p=0}^n$ of an invariant preobservable $\mathcal{O}^{(0)}$ is a descendant sequence.
	\end{prop}	
	Likewise, one can show that
	\begin{thm}\label{mthmg}
		In a CohLGFT, every gauge invariant descendant sequence is locally a $K$-sequence up to an exact sequence.
	\end{thm}
	\begin{proof}
		The proof is essentially the same as the proof of Theorem \ref{mthmphy}.
	\end{proof}

	\subsection{Examples}
	
	In this subsection, we will take $Y$ to be of the form
	\begin{align}
		Y=V[1]Y' \rightarrow Y'_0 \rightarrow M,
	\end{align}
	where $Y' \rightarrow Y'_0 \rightarrow M$ is a graded (gauge) natural bundle over $M$. In the case of gauge theories, $Y'_0$ needs to include the affine bundle $C$ whose sections can be identified with connection $1$-forms $A$. In simple terms, the fields can be divided into two groups $\Phi$ and $\Psi$ where $d(\Psi) = d(\Phi) + 1$. The canonical $QKG^{\star}$-structure is then given by
	\begin{align*}
		Q \Phi = \Psi, \quad Q \Psi = 0, \quad K_{\mu} \Phi =0, \quad K_{\mu} \Psi = \mathrm{Lie}_{\xi_{\partial_{\mu}}} \Phi, \quad I_{\lambda} \Phi = 0, \quad I_{\lambda} \Psi = \delta_{\lambda} \Phi.
	\end{align*}
	
	\subsubsection{N=2 supersymmetric quantum mechanics}
	
	Let $M=\mathbb{R}$. Let $(N,h)$ be a Riemannian manifold. We take $Y'$ to be the trivial bundle
	\begin{align*}
		Y'=T^*[-1]N \times M \rightarrow Y_0 = N \times M \rightarrow M.
	\end{align*} 
	A coordinate chart of $N$ induces a local coordinate system
	\begin{align*}
		(x^{\mu},~\chi^{\mu},~\psi^{\mu},~b^{\mu})
	\end{align*}
	for $Y$ with degrees $0, -1, 1, 0$, respectively. Let $t$ be the standard coordinate function over $M$. We adopt the conventional notation $\overline{Q}$ to denote the vertical local vector field $K_{\frac{d}{d t}}$. Let $\Phi$ be a local section of $Y$, we also use $\dot{\Phi}$ to denote the jet coordinates associated to $\frac{d \Phi}{d t}$. The $QK^{\star}$-structure of the theory is the canonical one, defined by\footnote{Note that any vertical local vector field is determined by its action on the zero jets.}
	\begin{align*}
		&Q x^{\mu} = \psi^{\mu}, \quad Q \psi^{\mu} = 0 \quad Q \chi^{\mu} = b^{\mu}, \quad Q b^{\mu} = 0, \\
		&\overline{Q} x^{\mu} = 0, \quad \overline{Q} \psi^{\mu} = \dot{x}^{\mu}, \quad \overline{Q} \chi^{\mu} = 0, \quad \overline{Q} b^{\mu} = \dot{\chi}^{\mu}.
	\end{align*}
	It is straightforward to verify that
	\begin{align*}
		Q^2 = 0, \quad Q \overline{Q} + \overline{Q} Q = \xi_{\frac{d}{dt}}, \quad \overline{Q}^2=0.
	\end{align*}
	Let $\Gamma=\Gamma^{\rho}_{\mu \nu}\partial_{\rho}\otimes dx^{\mu} \otimes dx^{\nu}$ be the Levi-Civita connection of $N$. Apply the following change of coordinates
	\begin{align*}
		b^{\mu} \rightarrow b^{\mu} - \Gamma ^{\mu}_{\rho \nu} \psi^{\rho}\chi^{\nu}.
	\end{align*}
	We obtain
	\begin{align*}
		&Q x^{\mu} = \psi^{\mu},
		\quad 
		Q \psi^{\mu}=0,
		\quad 
		Q \chi^{\mu}=b^{\mu} - \Gamma^{\mu}_{\rho \nu} \psi^{\rho} \chi^{\nu}, 
		\quad 
		Q b^{\mu} = -\Gamma^{\mu}_{\rho \nu}\psi^{\rho}b^{\nu} + \frac{1}{2}R^{\mu}_{\nu \rho \sigma} \psi^{\rho}\psi^{\sigma} \chi^{\nu} \\
		&\overline{Q} x^{\mu} = 0, 
		\quad 
		\overline{Q} \psi^{\mu} = \dot{x}^{\mu}, 
		\quad 
		\overline{Q} \chi^{\mu} = 0, 
		\quad 
		\overline{Q} b^{\mu} = \nabla_{\frac{d}{dt}} \chi^{\mu},  
	\end{align*}
	where $\nabla_{\frac{d}{dt}} \chi^{\mu} = \dot{\chi}^{\mu} + \Gamma^{\mu}_{\rho \nu} \dot{x}^{\rho}\chi^{\nu}$, and $R=R^{\nu}_{\mu \rho \sigma} \partial_{\nu} \otimes dx^{\mu} \otimes dx^{\rho} \otimes dx^{\sigma}$ is the Riemannian curvature tensor of $N$.
	
	The Lagrangian of the theory is then given by
	\begin{align}\label{lsqm1}
		&\mathcal{L} = Q(\chi_{\mu}(\dot{x}^{\mu} - b^{\mu})) dt.
	\end{align}
	where $\chi_{\mu} = h_{\mu\nu}\chi^{\nu}$. A straightforward computation shows that
	\begin{align*}
		\mathcal{L} = \left(b_{\mu} (\dot{x}^{\mu} - b^{\mu}) - \chi_{\mu}\nabla_{\frac{d}{dt}} \psi^{\mu} + \frac{1}{2}R_{\mu \beta \rho \sigma} \chi^{\mu} \chi^{\beta} \psi^{\rho}\psi^{\sigma} \right) dt,
	\end{align*}
	where $R_{\mu \beta \rho \sigma} = h_{\mu\nu}R^{\nu}_{\beta \rho \sigma}$.	
	
	By definition, $Q$ is a symmetry of $\mathcal{L}$, but $\overline{Q}$ is not. This problem can be solved by considering the following deformation of the canonical $QK^{\star}$-structure
	\begin{align}
		&Q x^{\mu} = \psi^{\mu}, \quad Q \psi^{\mu} = 0 \quad Q \chi^{\mu} = b^{\mu}, \quad Q b^{\mu} = 0, \label{f01} \\
		&\overline{Q} x^{\mu} = r \chi^{\mu}, \quad \overline{Q} \psi^{\mu} = \dot{x}^{\mu} - r b^{\mu}, \quad \overline{Q} \chi^{\mu} = 0, \quad \overline{Q} b^{\mu} = \dot{\chi}^{\mu}.\label{f02}
	\end{align}
	For our purpose, we set $r=1$ and apply again the change of coordinates
	$
	b^{\mu} \rightarrow b^{\mu} - \Gamma ^{\mu}_{\rho \nu} \psi^{\rho}\chi^{\nu}.
	$
	We then obtain
	\begin{align*}
		&Q x^{\mu} = \psi^{\mu}, \quad Q \psi^{\mu} = 0, \quad Q \chi^{\mu} = b^{\mu} - \Gamma ^{\mu}_{\rho \nu} \psi^{\rho}\chi^{\nu}, \quad Q b^{\mu} = -\Gamma^{\mu}_{\rho \nu}\psi^{\rho}b^{\nu} + \frac{1}{2}R^{\mu}_{\nu \rho \sigma} \psi^{\rho}\psi^{\sigma} \chi^{\nu} ,\\
		&\overline{Q} x^{\mu} = \chi^{\mu}, \quad \overline{Q} \psi^{\mu} = \dot{x}^{\mu} - b^{\mu} + \Gamma ^{\mu}_{\rho \nu} \psi^{\rho}\chi^{\nu}, \quad \overline{Q} \chi^{\mu} = 0, \quad \overline{Q} b^{\mu} = \nabla_{\frac{d}{dt}}{\chi}^{\mu}  -  \Gamma^{\mu}_{\rho \nu}\chi^{\rho} b^{\nu}+  \frac{1}{2}R^{\mu}_{\nu\rho\sigma}\chi^{\rho} \psi^{\sigma}\chi^{\nu}. 
	\end{align*}
	It is not hard to verify that $\overline{Q}$ is a Noether symmetry of the theory. In fact, we have
	\begin{align*}
		\overline{Q}(\chi_{\mu}(\dot{x}^{\mu} - b^{\mu})) = (\partial_{\sigma}h_{\mu\nu} + \Gamma_{\sigma\mu\nu})\chi^{\sigma}\chi^{\mu}(\dot{x}^{\nu} - b^{\nu}) + \frac{1}{2}R_{\mu\nu\rho\sigma}\chi^{\mu}\chi^{\nu}\chi^{\rho}\psi^{\sigma} = 0,
	\end{align*}
	where $\Gamma_{\sigma\mu\nu} = h_{\sigma\rho}\Gamma^{\rho}_{\mu\nu}$. We use the fact that $\partial_{\sigma}h_{\mu\nu} + \Gamma_{\sigma\mu\nu}$ is anti-symmetric in $\sigma$ and $\mu$, and the first Bianchi identity $R_{\mu\nu\rho\sigma} + R_{\nu\rho\mu\sigma} + R_{\rho\mu\nu\sigma}=0$. It follows that
	\begin{align*}
		\overline{Q} \mathcal{L} = \mathrm{Lie}_{\xi_{\frac{d}{dt}}}(\chi_{\mu}(\dot{x}^{\mu} - b^{\mu})) dt = d_h (\chi_{\mu}(\dot{x}^{\mu} - b^{\mu})) =: d_h \alpha_{\overline{Q}}.
	\end{align*}
	Let's compute the Noether currents of $Q$ and $\overline{Q}$. Observe that the boundary form $\gamma$ is of the form
	\begin{align*}
		\gamma = b_{\mu} \delta x^{\mu} + \chi_{\mu} (\nabla_{ver} \psi^{\mu}),
	\end{align*}
	where $\nabla_{ver} \psi^{\mu} = \delta \psi^{\mu} + \Gamma^{\mu}_{\rho\nu} \delta x^{\rho} \psi^{\nu}$. The Noether current $\mathcal{Q}$
	of $Q$ can be computed as
	\begin{align*}
		\mathcal{Q} = - \iota_Q \gamma = -b_{\mu}\psi^{\mu} - \chi_{\mu}\Gamma^{\mu}_{\rho\nu}\psi^{\rho}\psi^{\nu} = -b_{\mu}\psi^{\mu},
	\end{align*}
	where we use the torsion free condition $\Gamma^{\mu}_{\rho\nu} = \Gamma^{\mu}_{\nu\rho}$. Likewise, the Noether current $\overline{\mathcal{Q}}$ of $\overline{Q}$ can be computed as
	\begin{align*}
		\overline{\mathcal{Q}} = \alpha_{\overline{Q}} - \iota_{\overline{Q}} \gamma = \chi_{\mu}(\dot{x}^{\mu} - b^{\mu}) - b_{\mu}\chi^{\mu} - \chi_{\mu}(\dot{x}^{\mu}- b^{\mu} + \Gamma^{\mu}_{\rho\nu}\psi^{\rho}\chi^{\nu}) - \chi_{\mu} \Gamma^{\mu}_{\rho\nu} \chi^{\rho} \psi^{\nu} = -b_{\mu}\chi^{\mu}.
	\end{align*}
	
	Let $f$ be a Morse function over $N$. Let's consider the following change of coordinates
	\begin{align}\label{ccmorb}
		b^{\mu} \rightarrow b^{\mu} - \mathrm{grad} f^{\mu} - \Gamma^{\mu}_{\rho\nu}\psi^{\rho}\chi^{\mu},
	\end{align}
	where $\mathrm{grad}  f^{\mu} = h^{\mu\nu}\partial_{\nu} f$ is the gradient of $f$. \eqref{f01} and \eqref{f02} can be then generalized as follows.
	\begin{align*}
		&Q x^{\mu} = \psi^{\mu}, \quad Q \psi^{\mu} = 0,\\
		&Q \chi^{\mu} = b^{\mu}  - \mathrm{grad}  f^{\mu} - \Gamma ^{\mu}_{\rho \nu} \psi^{\rho}\chi^{\nu}, \quad Q b^{\mu} = -\Gamma^{\mu}_{\rho \nu}\psi^{\rho}b^{\nu} + \nabla_{\rho} \mathrm{grad} f^{\mu} \psi^{\rho} + \frac{1}{2}R^{\mu}_{\nu \rho \sigma} \psi^{\rho}\psi^{\sigma} \chi^{\nu} , \\
		&\overline{Q} x^{\mu} = \chi^{\mu}, \quad \overline{Q} \psi^{\mu} = \dot{x}^{\mu} - b^{\mu}  + \mathrm{grad}  f^{\mu}  + \Gamma ^{\mu}_{\rho \nu} \psi^{\rho}\chi^{\nu}, \\
		&\overline{Q} \chi^{\mu} = 0, \quad \overline{Q} b^{\mu} = \nabla_{\frac{d}{dt}}{\chi}^{\mu}  -  \Gamma^{\mu}_{\rho \nu}\chi^{\rho} b^{\nu} + \nabla_{\rho} \mathrm{grad} f^{\mu} \chi^{\rho} +  \frac{1}{2}R^{\mu}_{\nu\rho\sigma}\chi^{\rho} \psi^{\sigma}\chi^{\nu}.
	\end{align*}
	We consider the following Lagrangian as a generalization of \eqref{lsqm1} \cite{Witten1982}.
	\begin{align*}
		\mathcal{L} = \mathcal{L}_{top} + Q(\mathcal{V}),
	\end{align*}
	where 
	\begin{align*}
		\mathcal{L}_{top} = d_h f = \partial_{\mu} f \dot{x}^{\mu} dt, \quad \mathcal{V} = \chi_{\mu}(\dot{x}^{\mu}-b^{\mu}) dt.
	\end{align*}
	It is straightforward to show that 
	\begin{align*}
		\mathcal{L} 
		= \left(b_{\mu} (\dot{x}^{\mu} + \mathrm{grad} f^{\mu} - b^{\mu}) - \chi_{\mu}\nabla_{\frac{d}{dt}} \psi^{\mu} + \mathrm{Hess}f_{\mu\nu}  \chi^{\mu} \psi^{\nu} + \frac{1}{2}R_{\mu \beta \rho \sigma} \chi^{\mu} \chi^{\beta} \psi^{\rho}\psi^{\sigma} \right) dt,
	\end{align*}
	where $\mathrm{Hess} f = \nabla df$ is the Hessian of $f$. We have
	\begin{align*}
		Q \mathcal{L} = d_h (\partial_{\mu}f \psi^{\mu}) =: d_h \alpha_Q.
	\end{align*}
	On the other hand, note that
	\begin{align*}
		\overline{Q} \mathcal{V} = \left(\chi_{\mu} \nabla_{\rho} \mathrm{grad} f^{\mu} \chi^{\rho}\right) dt =  \left(\chi^{\nu} \mathrm{Hess}f_{\mu\nu} \chi^{\mu} \right) dt = 0,
	\end{align*}
	where we use the identity $h_{\rho\nu} \nabla_{\mu} \mathrm{grad} f^{\rho} = \mathrm{Hess} f_{\mu\nu}$ and the symmetric property of
	$\mathrm{Hess} f$. We then have
	\begin{align*}
		\overline{Q} \mathcal{L} = d_h (\chi_{\mu}(\dot{x}^{\mu} + \mathrm{grad}f^{\mu} - b^{\mu})) =: d_h \alpha_{\overline{Q}}.
	\end{align*}
	Note expression of the boundary form $\gamma$ remain unchanged. The Noether currents of $Q$ and $\overline{Q}$ are then
	\begin{align*}
		\mathcal{Q} = (\partial_{\mu}f - b_{\mu})\psi^{\mu}, \quad \overline{\mathcal{Q}} = -b_{\mu}\chi^{\mu}.
	\end{align*}
	\begin{rmk}
		The Morse function $f$ does not show up in the expression of $\overline{\mathcal{Q}}$ because we only deform $\overline{Q}$ in \eqref{f01} and \eqref{f02}. More generally, one can consider the deformation
		\begin{align*}
			&Q x^{\mu} = s \psi^{\mu}, \quad Q \psi^{\mu} = 0 \quad Q \chi^{\mu} = s b^{\mu} - \dot{x}^{\mu}, \quad Q b^{\mu} = \dot{\psi}^{\mu}, \\
			&\overline{Q} x^{\mu} = r \chi^{\mu}, \quad \overline{Q} \psi^{\mu} = \dot{x}^{\mu} - r b^{\mu}, \quad \overline{Q} \chi^{\mu} = 0, \quad \overline{Q} b^{\mu} = \dot{\chi}^{\mu}.
		\end{align*}
		with $r$ and $s$ satisfying $s-r=1$. By properly choosing $s$ and $r$, one can get symmetric expressions for $\mathcal{Q}$ and $\overline{\mathcal{Q}}$. We leave the details to the reader.
	\end{rmk}

	\subsubsection{Donaldson-Witten theory}
	
	
	Let $G=\mathrm{SU}(2)$. Let $\mathrm{Tr}$ be the Killing form on $\mathfrak{su}(2)$. Let $P$ be a principal $G$-bundle over an $4$-dimensional Riemannian manifold $(M,g)$. Let $\mathrm{ad} P$ denote the adjoint bundle of $P$. Let $\mathcal{A}$ denote the affine space of connection $1$-forms on $P$. Recall that $\mathcal{A}$ can be identified with $\Gamma(C)$ where $C$ is an affine bundle over $M$. Let $W$ be an associated vector bundle to $P$. For our purpose, we choose $W=\mathrm{ad}P \otimes \Lambda^2_- T^* M$, where $ \Lambda^2_- T^* M$ denotes the anti-sell-dual part of $\Lambda^2 T^* M$ with respect to the Hodge star operator $\star$ on $M$. We take $Y'$ to be 
	\begin{align*}
		Y'=\mathrm{ad} P[1] \times_M C \times_M W[-1]  \rightarrow Y_0 = C \rightarrow M
	\end{align*}
	A bundle chart of $P$ induces a local coordinate system
	\begin{align*}
		(x^{\mu},~\theta^a,~A^a_{\mu},~\chi^a_{\mu\nu},~ \phi^a,~ \upsilon^a_{\mu},~ b^a_{\mu\nu})
	\end{align*}
	for $Y$ with degrees $0,1,0,-1,2,1,0$, respectively. We use the Greek indices to denote the components of differential forms and the Roman indices to denote the components of elements in the Lie algebra $\mathfrak{g}$. The $QKG^{\star}$-structure is given by the canonical one.	\begin{align*}
		&Q \theta^a = \phi^a, \quad Q \phi^a = 0, \quad Q A^a_{\mu} = \upsilon^a_{\mu}, \quad Q \upsilon^a_{\mu}=0, \quad Q \chi^a_{\mu\nu} = b^a_{\mu\nu}, \quad Q b^a_{\mu\nu} = 0,\\
		&K_{\mu} \theta^a = 0, \quad K_{\mu} \phi^a = \theta^a_{\mu}, \quad K_{\mu} A^a_{\nu} =0, \quad K_{\mu} \upsilon^a_{\nu}= A^a_{\nu,\mu}, \quad K_{\mu} \chi^a_{\nu\rho} = 0, \quad K_{\mu} b^a_{\nu\rho} = \chi^a_{\nu\rho,\mu},\\
		&I_{\lambda} \theta^a = 0, ~ I_{\lambda} \phi^a = -f^a_{bc}\lambda^b\theta^a, ~ I_{\lambda} A^a_{\mu} = 0 , ~ I_{\lambda} \upsilon^a_{\mu}= \partial_{\mu} \lambda^a + f^a_{bc}A^b \lambda^c, ~ I_{\lambda} \chi^a_{\mu\nu} = 0, ~ I_{\lambda} b^a_{\mu\nu} = -f^a_{bc}\lambda^b \chi^c_{\mu\nu},
	\end{align*}
	where $f^a_{bc}$ are the structure constants of $\mathfrak{g}$, $\lambda = \lambda^a \xi_a$ is a local section of $\mathrm{ad} P$, and we use $\theta^a_{\mu}$ to denote the jet coordinates of associated to $\partial_{\mu}\theta^a$. The above expressions can be rewritten in physicists' notation as follows.
	\begin{align*}
		&Q \theta = \phi, \quad Q \phi = 0, \quad Q A = \upsilon, \quad Q \upsilon=0, \quad Q \chi= b, \quad Q b = 0,\\
		&K_{\mu} \theta= 0, \quad K_{\mu} \phi = \partial_{\mu} \theta, \quad K_{\mu} A =0, \quad K_{\mu} \upsilon = \partial_{\mu}A, \quad K_{\mu} \chi = 0, \quad K_{\mu} b = \partial_{\mu} \chi,\\
		&I_{\lambda} \theta = 0, \quad I_{\lambda} \phi = -[\lambda,\theta], \quad I_{\lambda} A = 0, \quad  I_{\lambda} \upsilon = d_A \lambda, \quad I_{\lambda} \chi = 0, \quad I_{\lambda} b =-[\lambda,\chi].
	\end{align*}
	One can deform the canonical $QKG^{\star}$-structure by
	\begin{align*}
		I_{\lambda} \theta = r \lambda.
	\end{align*}
	For our purpose, we set $r=1$. Note that this deformation is only vertically compatible with the canonical $QKG^{\star}$-structure.
	
	Let's apply the following change of coordinates
	\begin{align}\label{mqdw}
		\phi \rightarrow \phi - \frac{1}{2}[\theta,\theta], \quad \upsilon \rightarrow \upsilon + d_A \theta, \quad b \rightarrow b - [\theta, \chi].
	\end{align}
	We obtain
	\begin{align}
		& Q \theta = \phi - \frac{1}{2}[\theta,\theta], \quad Q \phi = -[\theta, \phi], \label{q1}\\
		& Q A = \upsilon + d_A \theta, \quad Q \upsilon = - [\theta, \upsilon] - d_A \phi \label{q2}\\
		& Q \chi = b - [\theta, \chi], \quad Q b = - [\theta, b] + [\phi, \chi], \label{q3}
	\end{align}
	and
	\begin{align*}
		I_{\lambda} \theta = \lambda, \quad I_{\lambda} \phi = I_{\lambda} A = I_{\lambda} \upsilon = I_{\lambda} \chi = I_{\lambda} b =0.
	\end{align*}
	The expression for the vertical local vector fields $K_{\mu}$ remain unchanged. It becomes evident that $Q$, $I_{\lambda}$, and $\delta_{\lambda} = [Q, I_{\lambda}]$ define an infinite dimensional BRST model for the equivariant cohomology of the $\mathrm{Gau}(P)$-manifold $\Gamma(Y')$.
	
	For the Lagrangian, we consider \cite{Witten1988}
	\begin{align*}
		\mathcal{L} =  \mathcal{L}_{top} + Q (\mathcal{V}),
	\end{align*}
	where 
	\begin{align*}
		\mathcal{L}_{top} = \mathrm{Tr}(F \wedge F)/2, \quad \mathcal{V} = \mathrm{Tr} \left( \chi \wedge (F + b) \right),
	\end{align*}
	and $F=d_A + \frac{1}{2}[A,A]$ is the curvature $2$-form of $A$. It is straightforward to show that
	\begin{align*}
		\mathcal{L} = \mathrm{Tr} \left( F \wedge F)/2 + b \wedge (F + b) - \chi \wedge d_A \upsilon  - \chi \wedge [\phi,\chi] \right).
	\end{align*}
	By definition, $Q$ and $\delta_{\lambda}$ are Noether symmetries of $\mathcal{L}$. Moreover, we have $I_{\lambda} \mathcal{L} = 0$ because $\mathcal{L}$ does not depend on $\theta$. 
	\begin{rmk}
		The equations of motion of $b$ are $b=-F_-/2$.  After integrating out $b$, the Lagrangian becomes
		\begin{align*}
			\mathcal{L} = \mathrm{Tr} \left( F_+ \wedge F_+)/2  - \chi \wedge d_A \upsilon  - \chi \wedge [\phi,\chi] \right).
		\end{align*}
	\end{rmk}
	
	For the preobservables, we consider 
	\begin{align}
		\mathcal{O}^{(0)}=\mathrm{Tr}(\phi^2), ~
		\mathcal{O}^{(1)}=-2\mathrm{Tr}(\phi \upsilon), ~
		\mathcal{O}^{(2)}=\mathrm{Tr}(\upsilon \wedge \upsilon - 2 \phi F), ~
		\mathcal{O}^{(3)}=2\mathrm{Tr}(\upsilon \wedge F),~
		\mathcal{O}^{(4)}=\mathrm{Tr}(F \wedge F).\label{preobdw}
	\end{align}
	It is straightforward to verify that $\{\mathcal{O}^{(p)}\}_{p=0}^n$ satisfy the descent equations \eqref{deq}. Moreover, by Theorem \ref{mthmphy}, it can be locally expressed as the sum of a $K$-sequence and an exact sequence. For example, we have
	\begin{align*}
		\mathcal{O}^{(1)} = Q(\mathrm{Tr}(\phi A)) - K \mathcal{O}^{(0)}. 
	\end{align*}
	It is not manifest that $\mathcal{O}^{(p)}$ for $p >1$ still can be expressed in the above form. Let's take $M$ to be $\mathbb{R}^{n}$. The canonical $K_{\mu}$ and $I_{\lambda}$ can be deformed by setting
	\begin{align*}
		K_{\mu} \theta = r A_{\mu},  \quad K_{\mu} \phi = \partial_{\mu} \theta - r \upsilon_{\mu}, \quad I_{\lambda} \theta = r \lambda.
	\end{align*}
	This deformation is indeed compatible with the canonical $QKG^{\star}$-structure. For our purposes, we set $r=1$.  Applying again the change of coordinates \eqref{mqdw}, we obtain
	\begin{align*}
		&K_{\mu} \theta= A_{\mu},  \quad K_{\mu} \phi = - \upsilon_{\mu}, \quad K_{\mu} A =  0, \quad K_{\mu} \upsilon = F_{\mu\nu}dx^{\nu}, \quad K_{\mu} \chi = 0, \quad K_{\mu} b = d_{A\mu} \chi, \\
		&I_{\lambda} \theta = \lambda, \quad I_{\lambda} \phi = I_{\lambda} A = I_{\lambda} \upsilon = I_{\lambda} \chi = I_{\lambda} b =0,
	\end{align*}
	where $d_{A\mu} \chi = \partial_{\mu} \chi + [A_{\mu}, \chi]$. 
	
	The operator $K=dx^{\mu} \wedge \mathrm{Lie}_{K_{\mu}}$ is now a globally defined. We have
	\begin{align*}
		K \theta= A  \quad K \phi = - \upsilon, \quad K A = 0, \quad K \upsilon = 2F, \quad K \chi = 0, \quad K b = d_{A} \chi.
	\end{align*}
	The descendant sequence $\{\mathcal{O}^{(p)}\}_{p=0}^n$ \eqref{preobdw} is now just the standard $K$-sequence of $\mathcal{O}^{(0)}$ with respect to the deformed $QKG^{\star}$-structure.
	
	The Donaldson-Witten theory also admits a supersymmetric quantum mechanical interpretation. Let's take $M=\mathbb{R} \times \Sigma$, where $\Sigma$ is a $3$-dimensional manifold. Let $g$ be the product metric. Since $G=\mathrm{SU}(2)$ is simply connected, the principle $G$-bundle $P$ is trivial over $M$. Let $\star_3$ denote the Hodge star operator on $\Sigma$. Let $a$ be a one form over $\Sigma$, $a$ can be identified as an anti-self dual $2$-form over $M$ via the map \cite{Donaldson2002}
	\begin{align*}
		a \mapsto a \wedge dt + \star_3 a.
	\end{align*}
	It follows that the bundle $W$ can be identified as $T^* \Sigma \otimes \mathfrak{g}$. The affine bundle $C$ can be identified as the vector bundle $(\underline{\mathbb{R}} \oplus T^*\Sigma) \otimes \mathfrak{g}$, where $\underline{\mathbb{R}}$ is the trivial line bundle over $M$. Every connection $A$ can be written as
	\begin{align*}
		A = A_0 dt + \sum_{\mu=1}^3 A_{\mu}(x,t) dx^{\mu}
	\end{align*}
	For our purpose, we need to choose the temporal gauge $A_0=0$. Or equivalently, we take $C$ to be the bundle $T^*\Sigma \otimes \mathfrak{g}$ instead of $T^* M \otimes \mathfrak{g}$. Since $P$ is a trivial bundle, 
	any gauge transformation $g$ is just a $G$-valued function $g(t,x)$ over $M=\mathbb{R} \times \Sigma$. The gauge transformation of $A$ under $g$ is given by $A \mapsto g A g^{-1} + g dg^{-1}$. Since we require $A_0$ to be $0$, $g$ must not depend on $t$. In other words, $g \in \mathrm{Gau}(P_{\Sigma})$, where $P_{\Sigma}$ is the trivial principle $G$-bundle over $\Sigma$.
	On the other hand, let $f \in \mathrm{Diff}(M)$. The pull back of $A$ under $f$ is locally given by $A \mapsto A_{\mu}(f(t,x)) \frac{\partial f^{\mu}(x,t)}{\partial x^{\nu}} dx^{\nu} + A_{\mu}(f(t,x)) \frac{\partial f^{\mu}(x,t)}{\partial t} dt$. Therefore, we also need to require $f^{\mu}$ to be independent of $t$. In other words, $f$ can be viewed as an automorphism of the trivial line bundle $\mathbb{R}_{\Sigma}$ over $\Sigma$. To conclude, we take $Y'$ to be the graded gauge natural bundle
	\begin{align*}
		Y' = \underline{\mathfrak{g}}[1] \times_M (T^*\Sigma \otimes \mathfrak{g}) \times_M (T^*[-1]\Sigma \otimes \mathfrak{g}) \rightarrow Y_0 = T^*\Sigma \otimes \mathfrak{g} \rightarrow M=\mathbb{R} \times \Sigma.
	\end{align*}
	And $M \times \Gamma(Y)$ is equipped with the $(\mathrm{Aut}(\mathbb{R}_{\Sigma}) \times \mathrm{Gau}(P_{\Sigma}))^{\star}$-action instead of the full $\mathrm{Aut}(P)^{\star}$-action. 
	
	A coordinate chart of $\Sigma$ induces a local coordinate system
	\begin{align*}
		(x^{\mu},~\theta^a,~A^a_{\mu},~\chi^a_{\mu},~ \phi^a,~ \upsilon^a_{\mu},~ b^a_{\mu})
	\end{align*}
	for $Y$ with degrees $0,1,0,-1,2,1,0$, respectively. The canonical $QKG^{\star}$-structure is
	\begin{align*}
		&Q \theta^a = \phi^a, \quad Q \phi^a = 0, \quad Q A^a_{\mu} = \upsilon^a_{\mu}, \quad Q \upsilon^a_{\mu}=0, \quad Q \chi^a_{\mu} = b^a_{\mu}, \quad Q b^a_{\mu} = 0,\\
		&K_{\frac{\partial}{\partial t}} \theta^a = 0, \quad K_{\frac{\partial}{\partial t}} \phi^a = \dot{\theta}^a, \quad K_{\frac{\partial}{\partial t}} A^a_{\mu} =0, \quad K_{\frac{\partial}{\partial t}} \upsilon^a_{\mu}= \dot{A}^a_{\mu}, \quad K_{\frac{\partial}{\partial t}} \chi^a_{\mu} = 0, \quad K_{\frac{\partial}{\partial t}} b^a_{\mu} = \dot{\chi}^a_{\mu},\\
		&K_{\mu} \theta^a = 0, \quad K_{\mu} \phi^a = \theta^a_{\mu}, \quad K_{\mu} A^a_{\nu} =0, \quad K_{\mu} \upsilon^a_{\nu}= A^a_{\nu,\mu}, \quad K_{\mu} \chi^a_{\nu} = 0, \quad K_{\mu} b^a_{\nu} = \chi^a_{\nu,\mu},\\
		&I_{\lambda} \theta^a = 0, ~ I_{\lambda} \phi^a = -f^a_{bc}\lambda^b\theta^a, ~ I_{\lambda} A^a_{\mu} = 0 , ~ I_{\lambda} \upsilon^a_{\mu}= \partial_{\mu} \lambda^a + f^a_{bc}A^b \lambda^c, ~ I_{\lambda} \chi^a_{\mu} = 0, ~ I_{\lambda} b^a_{\mu} = -f^a_{bc}\lambda^b \chi^c_{\mu}.
	\end{align*}
	Just like the case of supersymmetric quantum mechanics, we use $\overline{Q}$ to denote $K_{\frac{\partial}{\partial t}}$. Using physicists' notation, we rewrite the $QKG^{\star}$-structure as
	\begin{align*}
		&Q \theta = \phi, \quad Q \phi = 0, \quad Q A = \upsilon, \quad Q \upsilon=0, \quad Q \chi= b, \quad Q b = 0,\\
		&\overline{Q} \theta = 0, \quad \overline{Q} \phi = \dot{\theta}, \quad \overline{Q} A = 0, \quad Q \upsilon= \dot{A}, \quad \overline{Q} \chi= b, \quad \overline{Q} b = \dot{\chi},\\
		&K_{\mu} \theta= 0, \quad K_{\mu} \phi = \partial_{\mu} \theta, \quad K_{\mu} A =0, \quad K_{\mu} \upsilon = \partial_{\mu}A, \quad K_{\mu} \chi = 0, \quad K_{\mu} b = \partial_{\mu} \chi,\\
		&I_{\lambda} \theta = 0, \quad I_{\lambda} \phi = -[\lambda,\theta], \quad I_{\lambda} A = 0, \quad  I_{\lambda} \upsilon = d_A \lambda, \quad I_{\lambda} \chi = 0, \quad I_{\lambda} b =-[\lambda,\chi].
	\end{align*}
	In this case, one can consider the following deformation of $\overline{Q}$.
	\begin{align*}
		\overline{Q} \theta= 0, \quad \overline{Q} \phi = \dot{\theta}, \quad \overline{Q} A = s \chi, \quad \overline{Q} \upsilon = \dot{A}- s b, \quad \overline{Q} \chi = 0, \quad \overline{Q} b = \dot{\chi}.
	\end{align*}
	For simplicity, we set $s=1$. Applying the following change of coordinates as an analogue of \eqref{ccmorb}
	\begin{align*}
		\phi \rightarrow \phi - \frac{1}{2}[\theta,\theta], \quad \upsilon \rightarrow \upsilon + d_A \theta, \quad b \rightarrow b - \star_3 F - [\theta, \chi],
	\end{align*}
	we obtain
	\begin{align*}
		& Q \theta = \phi - \frac{1}{2}[\theta,\theta], \quad Q \phi = -[\theta, \phi], \\
		& Q A = \upsilon + d_A \theta, \quad Q \upsilon = - [\theta, \upsilon] - d_A \phi\\
		& Q \chi = b - \star_3 F - [\theta, \chi], \quad Q b = \star_3 d_A \upsilon - [\theta, b] + [\phi, \chi],
	\end{align*}
	and 
	\begin{align*}
		\overline{Q} \theta = 0, \quad  \overline{Q} \phi = \dot{\theta}, \quad
		\overline{Q} A = \chi, \quad  \overline{Q} \upsilon =\dot{A} - b + \star_3 F + [\theta,\chi], \quad
		\overline{Q} \chi = 0, \quad \overline{Q} b = \dot{\chi} + \star_3 d_A \chi.
	\end{align*}
	The expressions for $K_{\mu}$ and $I_{\lambda}$ are irrelevant. For the Lagrangian, we consider
	\begin{align*}
		&\mathcal{L} = \frac{d}{dt} \left( \int_{\Sigma}  \mathrm{CS}(A) \right)dt + Q\left( \int_{\Sigma} d \mathrm{vol}_{\Sigma} \mathrm{Tr}(\chi^{\mu}(\dot{A}_{\mu} - b_{\mu})) \right)dt
	\end{align*}
	where $\mathrm{vol}_{\Sigma}$ is the volume form on $\Sigma$, and $\mathrm{CS}(A)$ is the three dimensional Chern-Simons Lagrangian
	\begin{align*}
		\mathrm{CS}(A) = \mathrm{Tr}(\frac{1}{2} A \wedge dA + \frac{1}{6} A \wedge [A,A]).
	\end{align*}
	It is straightforward to show that
	\begin{align*}
		\mathcal{L} 
		&= \int_{\Sigma} d \mathrm{vol}_{\Sigma} \left( b^{\mu}(\dot{A}_{\mu} + (\star_3 F)_{\mu} - b_{\mu}) - \chi^{\mu}(\dot{\upsilon}_{\mu} + (d_A \dot{\theta})_{\mu} - (\star_3 d_A \upsilon)_{\mu} - [\phi, \chi_{\mu}])\right)dt.
	\end{align*}
	By definition, $Q$ is a Noether symmetry of $\mathcal{L}$. We have
	\begin{align*}
		Q \mathcal{L} =\frac{d}{dt} \left(\int_{\Sigma} \mathrm{Tr}(\upsilon \wedge F)\right) dt. 
	\end{align*}
	On the other hand, we have
	\begin{align*}
		\overline{Q} \left(\int_{\Sigma} d \mathrm{vol}_{\Sigma}  \mathrm{Tr}(\chi^{\mu}(\dot{A}_{\mu} - b_{\mu}))\right)= \int_{\Sigma} d \mathrm{vol}_{\Sigma}  \mathrm{Tr} (\chi^{\mu} (\star_3 d_A \chi)_{\mu})  = 2 \int_{\Sigma} d \mathrm{Tr}(\chi \wedge \chi) =0.
	\end{align*}
	It follows that
	\begin{align*}
		\overline{Q} \mathcal{L} &= \frac{d}{dt} \left(\int_{\Sigma} \mathrm{Tr}(\chi \wedge F) + \int_{\Sigma} d \mathrm{vol}_{\Sigma}  \mathrm{Tr}(\chi^{\mu}(\dot{A}_{\mu} - b_{\mu})) \right) dt \\
		&= \frac{d}{dt} \left(\int_{\Sigma} d \mathrm{vol}_{\Sigma}  \mathrm{Tr}(\chi^{\mu}(\dot{A}_{\mu} + (\star_3 F)_{\mu}- b_{\mu})) \right) dt.
	\end{align*}
	Hence, $\overline{Q}$ is also a symmetry of the theory. 
	The boundary term of the Lagrangian is
	\begin{align*}
		\gamma = \int_{\Sigma} d\mathrm{vol}_{\Sigma} \mathrm{Tr}(b^{\mu} \delta A_{\mu} + \chi^{\mu}(\delta \upsilon_{\mu} + (d_A \delta \theta)_{\mu})).
	\end{align*}
	It is not hard to show that the Nother currents of $Q$ and $\overline{Q}$ are
	\begin{align*}
		\mathcal{Q} = -\int_{\Sigma} d\mathrm{vol}_{\Sigma} \mathrm{Tr}((b^{\mu} - (\star_3 F)^{\mu} -[\theta,\chi^{\mu}])(\upsilon_{\mu} + (d_A \theta)_{\mu}))
	\end{align*}
	and 
	\begin{align*}
		\mathcal{\overline{Q}} = -\int_{\Sigma} d\mathrm{vol}_{\Sigma} \mathrm{Tr}((b^{\mu} -[\theta,\chi^{\mu}])\chi_{\mu}).
	\end{align*}
	
	\section{CohLFTs in the extended BV-BFV formalism}
	
	
	
	
	
	\subsection{Extended BV-BFV formalism in the variational bicomplex setting}
	
	Let $Y$ be a graded gauge natural bundle over an $n$-manifold $M$. For our purposes, we assume that the degree $0$ component $Y_0$ of $Y$ is an affine bundle. Let $\omega_i$, $i=1,2$, be two local form over $M \times \Gamma(Y)$. We say that $\omega_1$ is equivalent to $\omega_2$ if they are equal up to an $d_h$-exact term. We follow \cite{Sharapov2015} use ``$\simeq$'' to denote this equivalence relation.
	
	\subsubsection{BV Lagrangian field theory}
	
	\begin{defn}
		A presymplectic local structure/form of degree $q$ on $M \times \Gamma(Y)$ is a local form $\omega \in \Omega_{loc}^{p,2,q}(M \times \Gamma(Y))$ such that $d_v \omega \simeq 0$. 
		(In the case of $p=n$, we also require $\omega$ to be a functional form.)
		$\omega$ is called a symplectic local structure/form if it is nondegenerate with respect to vertical local vertical fields $\Xi$ over $M \times \Gamma(Y)$, i.e., if $\iota_{\Xi} \omega \simeq 0$ implies that $\Xi=0$. 
		An odd symplectic local form of degree $-1$ in $\Omega_{loc}^{n,2,-1}$ is called a BV symplectic local form.
	\end{defn}
	We will need the following lemma. 
	\begin{lem}\cite[Proposition A.1]{Sharapov2015}\label{trivdvdh}
		The cohomology groups $H^{p,q}(\Omega_{loc}; d_v/d_h)$ of the following cochain complex
		\begin{align*}
			\Omega_{loc}^{p,0}(M \times \Gamma(Y))/ d_h \Omega_{loc}^{p-1,0}(M \times \Gamma(Y)) \xlongrightarrow{d_v} \Omega_{loc}^{p,1}(M \times \Gamma(Y))/ d_h \Omega_{loc}^{p-1,1}(M \times \Gamma(Y)) \xlongrightarrow{d_v} \cdots
		\end{align*}
		is trivial for $q \geq 1$. While for $q=0$, 
		\begin{align}\label{p0m}
			H^{p,0}(\Omega_{loc}; d_v/d_h) \cong \Omega^p(M)/d\Omega^{p-1}(M).
		\end{align}
	\end{lem}
	
	Let $\omega$ be a (pre)symplectic local form. 
	It follows from Lemma \ref{trivdvdh} that one can always find 
	a local form $\gamma$ such that $\omega \simeq d_v \gamma$. $\gamma$ is called the (pre)symplectic local potential of $\omega$.
	
	\begin{defn}
		Let $\omega$ be a (pre)symplectic local form. A vertical local vector field $\Xi$ is called symplectic with respect to $\omega$ if $\mathrm{Lie}_{\Xi} \omega \simeq 0$.  $\Xi$ is called Hamiltonian with respect to $\omega$ if there exists a local form $\mathcal{F}$ of form degree $(n,0)$ such that $\iota_{\Xi} \omega - d_v \mathcal{F} \simeq 0$. 
	\end{defn}
	
	By Lemma \ref{trivdvdh}, every symplectic $\Xi$ is Hamiltonian. 
	
	\begin{defn}
		A BV Lagrangian field theory is a LFT $(M,Y,\mathcal{L})$ together with a Noether symmetry $Q$ of degree $1$ and a BV symplectic local form $\omega$ such that 
		\begin{enumerate}
			\item $Q$ is a cohomological vector field, i.e., $Q^2=0$;
			\item $Q$ is Hamiltonian with respect to $\omega$ and $\iota_Q \omega \simeq d_v \mathcal{L} $.
		\end{enumerate}
	\end{defn} 
	
	Let $\omega$ be a BV symplectic local form. Let $\omega_M:=\int_M \omega$. By definition, $\omega$ is a presymplectic form over $\Gamma(Y)$ in the usual sense, i.e.,
	\begin{align*}
		\delta \omega_M = \int_M d_v \omega = 0.
	\end{align*} 
	Moreover, $\omega_M$ is nondegenerate with respect to local vector fields over $\Gamma(Y)$ since $\omega$ is symplectic. Therefore, if $F$ is a local functional over $\Gamma(Y)$, one can always find a local vector field $\Xi_F$ such that $\delta F = \iota_{\Xi_F} \omega_M$. It follows that there exists a well-defined Poisson bracket $\{\cdot,\cdot\}$ between local functionals, given by 
	\begin{align*}
		\{F_1, F_2\} := \iota_{\Xi_{F_1}} \iota_{\Xi_{F_2}} \omega_M.
	\end{align*}
	In particular, we can take $F$ to be the action $S=\int_M \mathcal{L}$ and $\Xi_F$ to be the cohomological vector field $Q$. By definition, we have $\{S,S\} = Q(S)=0$. In this way, one can derive a classical BV theory from a BV Lagrangian field theory.
	
	\begin{rmk}
		If $\Phi \in C_{\mathcal{L}}$ is a critical point of $S$, then it is also in the zero locus of the cohomological vector field $Q$. In fact, from the LFT point of view, we have
		\begin{align*}
			\iota_Q \omega \simeq d_v \mathcal{L} \simeq EL,
		\end{align*}
		where $EL$ is the Euler-Lagrange form of the BV LFT. It follows that
		\begin{align*}
			(\iota_{\Xi} \iota_Q \omega)(x, \Phi)  \simeq \iota_{\Xi} EL (x, \Phi) = 0
		\end{align*}
		vanishes over $M$ for all vertical local vector field $\Xi$. Since $\omega$ is symplectic, we must have $Q(\cdot,\Phi) \equiv 0$.
	\end{rmk}
	
	\subsubsection{Extended BV-BFV Lagrangian field theory}
	
	Let $\omega^{(0)}$ be a BV symplectic local form. Let $Q$ be a cohomological vector field which is symplectic with respect to $\omega$, i.e., $\mathrm{Lie}_Q \omega^{(0)} + d_h \omega^{(1)} = 0$ for some $\omega^{(1)}$ of degree $(n-1,2,0)$. Note that
	\begin{align*}
		d_v(d_h \omega^{(1)})= -\mathrm{Lie}_Q d_v \omega^{(0)} \simeq 0.
	\end{align*}
	By Lemma \ref{flv}, we have $d_v \omega^{(1)} \simeq 0$, i.e., it is also a presymplectic local form. Moreover, note that
	\begin{align*}
		d_h (\mathrm{Lie}_Q \omega^{(1)}) =- \mathrm{Lie}_{Q^2} \omega^{(0)} = 0.
	\end{align*}
	By Lemma \ref{flv}, one can also find a local form $\omega^{(2)}$ of degree $(n-2,2,1)$ satisfying $\mathrm{Lie}_Q \omega^{(1)} + d_h \omega^{(2)} = 0$ and $d_v \omega^{(2)} \simeq 0$. Repeating this process, one can find a sequence of presymplectic local forms $\{\omega^{(p)}\}_{p=0}^n$ satisfying 
	\begin{align}\label{aspresym}
		(\mathrm{Lie}_Q + d_h) \sum_{p=0}^n \omega^{(p)} = 0.
	\end{align}
	\begin{defn}
		Let $\alpha^{(0)}$ be a local form of degree $(n,q,r)$. An ascendant sequence of $\alpha^{(0)}$ is a sequence $\{\alpha^{(p)}\}_{p=0}^n$ of local forms of degrees $(n-p,q,p+r)$ satisfying
		\begin{align}\label{as}
			(\mathrm{Lie}_Q - (-1)^{q+r} d_h) \sum_{p=0}^n \alpha^{(p)} = 0.
		\end{align}
		\eqref{as} is called the ascent equations.
	\end{defn}
	
	Let $\omega^{(0)} = d_v \gamma^{(0)}$ be a BV symplectic local form. Let $\{\gamma^{(p)}\}_{p=0}^n$ be an ascendant sequence of $\gamma^{(0)}$. Let $\omega^{(p)}:= d_v \gamma^{(p)}$. We have
	\begin{align*}
		(\mathrm{Lie}_Q + d_h) \sum_{p=0}^n \omega^{(p)} = (\mathrm{Lie}_Q + d_h) d_v \sum_{p=0}^n \gamma^{(p)} = -d_v (\mathrm{Lie}_Q - d_h) \sum_{p=0}^n \gamma^{(p)} =0.
	\end{align*} 
	In other words, $\{\omega^{(p)}\}_{p=0}^n$ is an ascendant sequence of $\omega^{(0)}$.
	
	\begin{defn}\label{bvbfvlft}
		Let $m$ be an integer, $0 \leq m \leq n$. An $m$-extended BV-BFV Lagrangian field theory consists of the following data: 
		\begin{enumerate}
			\item a LFT $(M,Y,\mathcal{L})$ together with a Noether symmetry $Q$, where $Q$ is a cohomological vector field;
			\item a BV symplectic local form $\omega^{(0)}=d_v \gamma^{(0)}$ over $M\times \Gamma(Y)$;
			\item an ascendant sequence $\{\omega^{(p)}\}_{p=0}^n$  of $\omega^{(0)}$ such that $\omega^{(p)}=d_v \gamma^{(p)}$ for some presymplectic potential $\gamma^{(p)}$ when $p \leq m$ and $\omega^{(p)}=0$ when $p>m$;
			\item a sequence of local forms $\{\mathcal{L}^{(p)}\}_{p=0}^n$ with $\mathcal{L}^{(0)}=\mathcal{L}$ such that
			\begin{align}\label{shuniu}
				\iota_Q \omega^{(p)} = d_v \mathcal{L}^{(p)} + d_h \gamma^{(p+1)} 
			\end{align}
			for $p=0,\cdots,n-1$, and $\iota_Q \omega^{(n)} = d_v \mathcal{L}^{(n)}$.
		\end{enumerate}
		The theory is said to be fully extended if $m=n$. An $1$-extended BV-BFV LFT is simply called a BV-BFV LFT. 
	\end{defn}
	\begin{rmk}
		The requirement that $\{\omega^{(p)}\}_{p=0}^n$ is an ascendant sequence is actually redundant and can be derived from \eqref{shuniu}.
	\end{rmk}
	For $m=0$, Definition \ref{bvbfvlft} reduces to the definition of a BV Lagrangian field theory. 
	
	\begin{prop}
		$Q$ is a symmetry of $\mathcal{L}^{(p)}$, i.e., $\mathrm{Lie}_Q \mathcal{L}^{(p)} \simeq 0$.
	\end{prop}
	\begin{proof}
		On one hand,
		\begin{align*}
			\mathrm{Lie}_Q (\iota_Q \omega^{(p)}) = \iota_Q (\mathrm{Lie}_Q \omega^{(p)})  = -d_h(\iota_Q \omega^{(p+1)}) =-d_h(d_v \mathcal{L}^{(p+1)}), 
		\end{align*}
		where we use the ascent equations of $\omega^{(p)}$ and \eqref{shuniu}. On the other hand,
		\begin{align*}
			&\mathrm{Lie}_Q (\iota_Q \omega^{(p)}) = \mathrm{Lie}_Q(d_v \mathcal{L}^{(p)} + d_h \gamma^{(p+1)}) = -d_v \mathrm{Lie}_Q \mathcal{L}^{(p)} + d_h ((\iota_Q d_v -d_v \iota_Q)\gamma^{(p+1)}) \\
			&= -d_v \mathrm{Lie}_Q \mathcal{L}^{(p)} + d_h(d_v \mathcal{L}^{(p+1)}) - d_h(d_v(\iota_Q \gamma^{(p+1)})),
		\end{align*}
		where we use $\mathrm{Lie}_Q=[\iota_Q, d_v]$ and \eqref{shuniu}. It follows that
		\begin{align*}
			d_v \left(\mathrm{Lie}_Q \mathcal{L}^{(p)} - d_h(2 \mathcal{L}^{(p+1)} - \iota_Q \gamma^{(p+1)} )\right)=0.
		\end{align*}
		Note that $\mathrm{Lie}_Q \mathcal{L}^{(p)}$ is of vertical form degree $0$ and ghost number degree $p+1>0$. By \eqref{p0m}, we must have
		\begin{align}\label{ncbvbfv}
			\mathrm{Lie}_Q \mathcal{L}^{(p)} = d_h(\mathcal{L}_{CMR}^{(p+1)}),
		\end{align}
		where $\mathcal{L}_{CMR}^{(p)}:=2 \mathcal{L}^{(p)} - \iota_Q \gamma^{(p)}$ is known as the modified Lagrangian of the extended BV-BFV LFT \cite{Mnev2019}. 
	\end{proof}   
	\begin{rmk}
		Noting that $\mathrm{Lie}_Q \mathcal{L}^{(p)} = \iota_Q d_v \mathcal{L}^{(p)} = \iota_Q^2 \omega^{(p)} - d_h(\iota_Q \gamma^{(p+1)})$, \eqref{ncbvbfv} is equivalent to
		\begin{align*}
			\iota_Q^2 \omega^{(p)} = 2 d_h \mathcal{L}^{(p+1)}.
		\end{align*}
	\end{rmk}
	Let $\mathbb{\Delta}^{(p)}:=\mathcal{L}^{(p)} - \iota_Q \gamma^{(p)}$ be the difference between the modified Lagrangian $\mathcal{L}^{(p)}_{CMR}$ and $\mathcal{L}^{(p)}$. $\mathbb{\Delta}^{(p)}$ is known as the BV-BFV difference \cite{Mnev2019}. By definition,
	\begin{align*}
		(\mathrm{Lie}_Q - d_h)\sum_{p=0}^n \mathcal{L}^{(p)} = d_h \sum_{p=0}^n \mathbb{\Delta}^{(p)},
        \quad
		(\mathrm{Lie}_Q - d_h)\sum_{p=0}^n \gamma^{(p)} = d_v \sum_{p=0}^n  \mathbb{\Delta}^{(p)}. 
	\end{align*}
	It follows that 
	\begin{align*}
		0=(\mathrm{Lie}_Q + d_h)(\mathrm{Lie}_Q - d_h)\sum_{p=0}^n \gamma^{(p)} = (\mathrm{Lie}_Q + d_h)d_v \sum_{p=0}^n  \mathbb{\Delta}^{(p)} = -d_v\left((\mathrm{Lie}_Q - d_h) \sum_{p=0}^n  \mathbb{\Delta}^{(p)}\right). 
	\end{align*}
	By Lemma \ref{trivdvdh}, we have 
	\begin{align}\label{asdelta}
		(\mathrm{Lie}_Q - d_h)\sum_{p=0}^n \mathbb{\Delta}^{(p)} = 0.
	\end{align}
	In other words, $\{\mathbb{\Delta}^{(p)}\}_{p=0}^n$ satisfies the ascent equations and measures the failure of $\{\mathcal{L}^{(p)}\}_{p=0}^n$ and $\{\gamma^{(p)}\}_{p=0}^n$ to be ascendant sequences. In the case of a vanishing $\{\mathbb{\Delta}^{(p)}\}_{p=0}^n$, the extended BV-BFV LFT is fully determined by the presymplectic local potentials $\gamma^{(p)}$ and the cohomological vector field $Q$.
	
	\begin{defn}
		An $f$-transformation of an extended BV-BFV LFT is the map
		\begin{align*}
			P_f: (\sum_{p=0}^n \mathcal{L}^{(p)}, \sum_{p=0}^n \gamma^{(p)}) \mapsto (\sum_{p=0}^n \mathcal{L}^{(p)} + d_h \sum_{p=0}^n f^{(p)}, \sum_{p=0}^n \gamma^{(p)} - d_v \sum_{p=0}^n f^{(p)}),
		\end{align*}
		where $f^{(p)}$ is a local form of degree $(n-p,0,p-1)$.
	\end{defn}
	
	It is easy to observe that  \eqref{shuniu} is preserved by $P_f$. Therefore, $f$-transformations are well-defined over the space of extended BV-BFV LFTs. Moreover, one can check that
	\begin{align*}
		P_f (\mathbb{\Delta}^{(p)}) - \mathbb{\Delta}^{(p)} = \mathrm{Lie}_Q f^{(p)} + d_h f^{(p+1)}.
	\end{align*}
	Or equivalently,
	\begin{align}\label{pfex}
		P_f(\sum_{p=0}^n \mathbb{\Delta}^{(p)}) = \sum_{p=0}^n \mathbb{\Delta}^{(p)} + (\mathrm{Lie}_Q + d_h) \sum_{p=0}^n  f^{(p)}.
	\end{align}
	In other words, the $(\mathrm{Lie}_Q -d_h)$-cohomology class of the BV-BFV difference $\sum_{p=0}^n \mathbb{\Delta}^{(p)}$ is preserved under $f$-transformations.
	
	\begin{lem}\label{bvex}
		Let $\mathcal{L}^{(0)}$ be the Lagrangian of an extended BV-BFV LFT. Let $\gamma$ be the (canonical) boundary form of $\mathcal{L}^{(0)}$. We have $\gamma \simeq -\gamma^{(1)}$ if $\omega^{(0)} = d_v \gamma^{(0)}$ takes the form
		\begin{align}\label{canbvform}
			\omega^{(0)} = \omega_{ab} \delta \Phi^a \wedge \delta \Phi^b \wedge \nu
		\end{align}
		where $\nu$ is a volume form on $M$.
	\end{lem}
	\begin{proof}
		By Corollary \ref{eleq}, we have
		\begin{align*}
			\iota_Q \omega^{(0)} = d_v \mathcal{L}^{(0)} + d_h \gamma^{(1)} = EL + d_h (\gamma+\gamma^{(1)}),
		\end{align*}
		where $EL$ is the Euler-Lagrange form of the LFT. On the other hand, note that 
		\begin{align*}
			\iota_Q \omega^{(0)} = \omega_{ab} (Q(\Phi^a) \delta \Phi^b + \delta \Phi^a  Q(\Phi^b)) \wedge \nu 
		\end{align*}
		is a source form. By Lemma \ref{flv}, we must have $\iota_Q \omega^{(0)} = EL$ and $d_h(\gamma + \gamma^{(1)})=0$. It follows again from Lemma \ref{flv} that $\gamma \simeq -\gamma^{(1)}$.
	\end{proof}
	
	\subsubsection{K-sequences}
	
	Supposing that there exists a $QKG^{\star}$-structure on $M \times \Gamma(Y)$, one can then locally define two homotopy operators $K_0:=dx^{\mu} \wedge \iota_{\xi_{\partial_{\mu}}}$ and $K:=dx^{\mu} \wedge \mathrm{Lie}_{K_{\mu}}$ on $\Omega_{\circ loc}$ as before. Let $\gamma^{(n)}$, $\mathbb{\Delta}^{(n)}$, and $\mathcal{L}$ be local forms over $M \times \Gamma(Y)$ of degrees $(0,1,n-1)$, $(0,0,n)$, and $(0,0,n)$, respectively, such that
	\begin{align}\label{inin}
		\mathrm{Lie}_Q \gamma^{(n)} = d_v \mathbb{\Delta}^{(n)}, \quad \mathcal{L}^{(n)} = \mathbb{\Delta}^{(n)} + \iota_Q \gamma^{(n)}.
	\end{align}
	By Lemma \eqref{trivdvdh}, $\mathbb{\Delta}^{(n)}$ is $Q$-closed. $\mathrm{Lie}_Q \mathcal{L}^{(n)} = \mathrm{Lie}_Q \iota_Q \gamma^{(n)} = \iota_Q d_v \mathbb{\Delta}^{(n)} = 0$, i.e., $\mathcal{L}^{(n)}$ is also $Q$-closed.
	
	Using the homotopy operator $K$, one can construct a fully extended BV-BFV LFT from \eqref{inin}.
	Let
	\[
	\gamma_K^{(n-p)} :=\frac{K^p}{p!} \gamma^{(n)}, \quad \mathbb{\Delta}_K^{(n-p)} := \frac{K^p}{p!} \mathbb{\Delta}^{(n)}, \quad \mathcal{L}_K^{(n-p)}:=\frac{K^p}{p!} \mathcal{L}^{(n)}.
	\]
	Or equivalently, 
	\[
	\sum_{p=0}^n \gamma_K^{(p)} := \exp(K) \gamma^{(n)}, \quad \sum_{p=0}^n \mathbb{\Delta}_K^{(p)} := \exp(K) \mathbb{\Delta}^{(n)}, \quad \sum_{p=0}^n \mathcal{L}_K^{(p)} := \exp(K) \mathcal{L}^{(n)}.
	\]
	Let $\widetilde{D} \omega:= (-1)^{i_D(d_{tot}(\omega)-n)} D \omega$ for a derivation $D$ of $\Omega_{\circ loc}$ of horizontal form degree $i_D$. We have
	\begin{align*}
		(\mathrm{Lie}_Q - d_{h}) \sum_{p=0}^n \gamma_K^{(p)} =\left(\mathrm{Lie}_Q - \widetilde{d_{h,inv}}\right) \exp(\widetilde{K})\gamma^{(n)} = \exp(\widetilde{K}) \mathrm{Lie}_Q \gamma^{(n)} = d_v \exp(\widetilde{K}) \mathbb{\Delta}^{(n)}= d_v \sum_{p=0}^n \mathbb{\Delta}_K^{(p)},
	\end{align*}
	where we use $\exp(\widetilde{K}) \mathrm{Lie}_Q = (\mathrm{Lie}_Q- \widetilde{d_{h,inv}}) \exp(\widetilde{K})$. Let's assume that both $\{\gamma_K^{(p)}\}_{p=0}^n$ and $\{\mathbb{\Delta}_K^{(p)}\}_{p=0}^n$ are globally well-defined. They then determine a fully extended BV-BFV LFT, whose Lagrangians are given by
	\begin{align}
		\sum_{p=0}^n \mathcal{L}^{(p)} &= \sum_{p=0}^n \mathbb{\Delta}_K^{(p)} + \iota_Q \sum_{p=0}^n \gamma_K^{(p)} = \sum_{p=0}^n \mathbb{\Delta}_K^{(p)} + \left([\iota_Q, \exp(\widetilde{K})] + \exp(\widetilde{K}) \iota_Q \right) \gamma^{(n)} \notag \\
		&= \sum_{p=0}^n \mathbb{\Delta}_K^{(p)}  - \widetilde{K_0} \exp(\widetilde{K}) \gamma^{(n)} + \exp(\widetilde{K})\left(\mathcal{L}^{(n)}-\mathbb{\Delta}^{(n)}\right) = \sum_{p=0}^n \mathcal{L}_K^{(p)} - K_0 \sum_{p=0}^n \gamma_K^{(p)}, \label{lk}
	\end{align}
	where we use \eqref{KiQ} to pass to the second line. 
	
    \begin{rmk}
    	The above discussion can be easily generalized to the case of $m$-extended BV-BFV LFTs, $m=1,\cdots,n$.
    \end{rmk}
	
	\subsection{Cotangent lift of CohLGFTs}
	
	\subsubsection{Cotangent lift of BRST theories}
	
	Let $Y$ be a graded gauge natural bundle over an $n$-dimensional Riemannian manifold $(M,g)$. Let $Y_{ct}:= V^*[-1]Y$. There is a canonical BV symplectic local form $\omega_{ct}$ over $M \times \Gamma(Y_{ct})$. Let $(x^{\mu}, \Phi^a)$ be a local coordinate system of $Y$, which induces a local coordinate system $(x^{\mu}, \Phi^a, \Phi^+_a)$ of $Y_{ct}$, which again induces a local coordinate system of $M \times \Gamma(Y_{ct})$. $\omega_{ct}$ is then given by
	\begin{align*}
		\omega_{ct} = \delta \Phi^+_a \wedge \delta \Phi^a \wedge d\mathrm{vol}_g = d_v(\Phi^+_a \delta \Phi^a \wedge d\mathrm{vol}_g) =: d_v \gamma_{ct}.
	\end{align*}
	Every vertical local vector field $\Xi =  \Xi^a \frac{\partial}{\partial \Phi^a} + \widehat{\partial_I}(\Xi^a) \frac{\partial}{\partial \Phi^a_I}$ over $M \times \Gamma(Y)$ can be lifted to a local form over $M \times \Gamma(Y_{ct})$ of horizontal degree $n$ defined by the formula
	\begin{align*}
		\widetilde{\Xi} = \Phi^+_a \Xi^a d\mathrm{vol}_g.
	\end{align*}
	Recall that $d_v \widetilde{\Xi}$ can be composed as
	\begin{align*}
		d_v \widetilde{\Xi}  = \mathcal{I}(d_v \widetilde{\Xi}) + d_h\text{-exact term},
	\end{align*}
	where $\mathcal{I}$ is the interior Euler operator and $\mathcal{I}(d_v \widetilde{\Xi})$ is a source form. There exists a unique vertical local vector field $\Xi_{cl}$ over $M \times \Gamma(Y_{ct})$ such that $\iota_{\Xi_{cl}} \omega_{ct} = \mathcal{I}(d_v \widetilde{\Xi})$. One can verify that the map $\Xi \mapsto \Xi_{cl}$ defines a homomorphism of graded Lie superalgebras, and that $\Xi_{cl}$ can be written as
	\begin{align}
		\Xi_{cl} = \Xi^a \frac{\partial}{\partial \Phi^a} + \text{terms involving }\frac{\partial}{\partial \Phi^a_{I}}, \frac{\partial}{\partial \Phi^+_{a}}, \text{ and }\frac{\partial}{\partial \Phi^+_{a,I}},\label{xiv1}
	\end{align}
	when $\Xi$ is odd, and
	\begin{align}
		\Xi_{cl} = (-1)^{|\Phi^a|}\Xi^{a} \frac{\partial}{\partial \Phi^a} + \text{terms involving }\frac{\partial}{\partial \Phi^a_{I}}, \frac{\partial}{\partial \Phi^+_{a}}, \text{ and }\frac{\partial}{\partial \Phi^+_{a,I}},\label{xiv2}
	\end{align}
	when $\Xi$ is even. From \eqref{xiv1} and \eqref{xiv2}, one can easily see that $\widetilde{\Xi} = \iota_{\Xi_{cl}} \gamma_{ct}$.
	
	Let $(M, Y, \mathcal{L})$ be any LFT such that $\Xi$ is a Noether symmetry of $\mathcal{L}$. $\mathcal{L}$ can be canonically viewed as a local form over $M \times \Gamma(Y_{ct})$. By \eqref{xiv1} and \eqref{xiv2}, we have
	\begin{align*}
		\mathrm{Lie}_{\Xi_{cl}} \mathcal{L} = \mathrm{Lie}_{\Xi} \mathcal{L} \simeq 0.
	\end{align*}
	Let $Q_{\mathcal{L}}$ denote the vector field associated to $\mathcal{L}$ via $\omega_{ct}$. By definition, $Q_{\mathcal{L}}^2 = 0$ and satisfy the equation
	\begin{align*}
		\iota_{Q_{\mathcal{L}}} \omega_{ct} = EL,
	\end{align*}
	where $EL$ is the Euler-Lagrange form of the LFT. Note that
	\begin{align*}
		\iota_{[\Xi_{cl}, Q_{\mathcal{L}}]} \omega_{ct} = (\mathrm{Lie}_{\Xi_{cl}} \iota_{Q_{\mathcal{L}}} - (-1)^{|\Xi|}\iota_{Q_{\mathcal{L}}} \mathrm{Lie}_{\Xi_{cl}}) \omega_{ct} \simeq  \mathrm{Lie}_{\Xi_{cl}} (d_v \mathcal{L}) \simeq 0,
	\end{align*}
	where we use $[d_h, \iota_{Q_{\mathcal{L}}}]=0$, $\mathrm{Lie}_{\Xi_{cl}} \mathcal{L} \simeq 0$, and $\mathrm{Lie}_{\Xi_{cl}} \omega_{ct} \simeq 0$. Since $\omega_{ct}$ is symplectic, we conclude that
	\begin{align*}
		[\Xi_{cl}, Q_{\mathcal{L}}] = 0.
	\end{align*}
	In particular, one can choose $\Xi$ to be a cohomological vector field $Q$ and define 
	\begin{align*}
		Q_{BV}:= Q_{ct} + Q_{\mathcal{L}},
	\end{align*}
	which is again a cohomological vector field. By definition, it is also the Hamiltonian vector field associated to
	\begin{align}
		\mathcal{L}_{BV}:= \mathcal{L} + \widetilde{Q} = \mathcal{L} + \iota_{Q_{BV}} \gamma_{ct}.
	\end{align} 
	$(M, Y_{ct}, \mathcal{L}_{BV})$ together with $Q_{BV}$ and $\omega_{ct}$ defines a BV Lagrangian field theory, which is called the cotangent lift of $(M,Y,\mathcal{L})$.
	
	\begin{defn}
		An extended BV-BFV LFT $(M, Y_{ct}, \mathcal{L}_{BV})$ is said to be of BRST type if
		\begin{enumerate}
			\item $\gamma^{(0)}=\gamma_{ct}$;
			\item $\mathcal{L}^{(0)}=\mathcal{L}_{BV}=\mathcal{L}_{BRST} + \iota_{Q_{BV}} \gamma_{ct}$,
		\end{enumerate}
		where $Q_{BV}=(Q_{BRST})_{cl} + Q_{\mathcal{L}_{BRST}}$ is the cohomological vector field of the extended BV-BFV LFT, $(M,Y,\mathcal{L}_{BRST})$ is a LFT with $Q_{BRST}$ as a Noether symmetry.
	\end{defn}
	
	By definition, we have $\mathbb{\Delta}^{(0)}=\mathcal{L}^{(0)}-\iota_{Q_{BV}} \gamma^{(0)} = \mathcal{L}_{BRST}$. By \eqref{asdelta}, we have
	\begin{align*}
		\mathrm{Lie}_{Q_{BRST}} \mathbb{\Delta}^{(0)} = \mathrm{Lie}_{Q_{BV}} \mathbb{\Delta}^{(0)} = d_h \mathbb{\Delta}^{(1)} 
	\end{align*}
	It follows that
	\begin{align*}
		(\mathrm{Lie}_{Q_{BRST}}-d_h) \sum_{p=0}^n \mathbb{\Delta}^{(p)} = 0.
	\end{align*}
	Therefore, $f$-transformations of extended BV-BFV LFTs of BRST type should be only defined for local forms $f^{(p)}$ over $M \times \Gamma(Y)$.
	
	\subsubsection{Cotangent lift of CohLGFTs}
	
	Let $(M, Y, \mathcal{L})$ be a CohLGFT. Recall that there is a cohomological vector field $Q$ over $M \times \Gamma(Y)$ and a family of vertical local vertical fields $K_{\Xi}$ parameterized by $G$-invariant vector fields $\Xi$ over the principal $G$-bundle $P$ defining $Y$, satisfying
	\begin{align*}
		[K_{\Xi_1}, K_{\Xi_2}] = 0, \quad [\Xi_1, K_{\Xi_2}]= K_{[\Xi_1, \Xi_2]}, \quad [Q, K_{\Xi}] = \Xi,
	\end{align*}
	where we identify $\Xi$ as a vertical local vector field of degree $0$ over $M \times \Gamma(Y)$ via taking Lie derivatives. A $G$-invariant vertical vector field $\Xi$ over $P$ can be identified with a section $\lambda$ of $\mathrm{ad} P$. We write $I_{\lambda}$ to denote such $K_{\Xi}$ and $\delta_{\lambda}$ to denote the corresponding gauge transformations. By definition, $Q$, $I_{\lambda}$, and $\delta_{\lambda}$ are Noether symmetries of the Lagrangian. Therefore, $Q_{BV}$, $(I_{\lambda})_{cl}$, and $(\delta_{\lambda})_{cl}$ satisfy
	\begin{align*}
		[Q_{BV}, (I_{\lambda})_{cl}] = (\delta_{\lambda})_{cl}.
	\end{align*} 
	In other words, they define a vertical local $\mathrm{Gau}(P)^{\star}$-action on $M \times \Gamma(Y_{ct})$. Note that $\mathcal{L}_{BV}=\mathcal{L} + \iota_{Q_{BV}} \gamma_{ct}$ is not basic with respect to this $\mathrm{Gau}(P)^{\star}$-action. We have
	\begin{align*}
		(I_{\lambda})_{cl} \mathcal{L}_{BV} = \widetilde{\delta_{\lambda}},
	\end{align*}
	which is not $d_h$-exact. 
	
	\begin{rmk}
		Just like the finite dimensional case, one can always choose a $\mathrm{Gau}(P)^{\star}$-invariant Lagrangian submanifold $\mathcal{L}(\Gamma(Y_{ct}))$ of $\Gamma(Y_{ct})$ such that $\widetilde{\delta_{\lambda}}$ vanishes over $\mathcal{L}(\Gamma(Y_{ct}))$. (Such $\mathcal{L}(\Gamma(Y_{ct}))$ are usually chosen to be the space of sections of some subbundle of $Y_{ct}$, e.g., $\Gamma(Y)$.) The restriction of $\mathcal{L}_{BV}|_{M \times \mathcal{L}(\Gamma(Y_{ct}))}$ is then basic with respect to this $\mathrm{Gau}(P)^{\star}$-action. 
	\end{rmk}
	\begin{rmk}
		One can also consider the cotangent lift $(K_{\xi_X})_{cl}$ of $K_{\xi_X}$, $X \in \mathfrak{X}_P(M)$. However, $[Q_{BV}, (K_{\xi_X})_{cl}] = (\xi_{X})_{cl}$ does not hold unless $X$ is a Noether symmetry of $\mathcal{L}$. By definition, such $X$ exists only if the CohLGFT is supersymmetric.
	\end{rmk}
	
	\begin{defn}
		An extended BV-BFV LFT $(M, Y_{ct}, \mathcal{L}_{BV})$ is said to be of CohLGFT type if it is of BRST type, $(M,Y,\mathcal{L}_{BRST})$ is a CohLGFT, and the BV-BFV difference $\sum_{p=0}^n \mathbb{\Delta}^{(p)}$ is basic with respect to the $\mathrm{Gau}(P)^{\star}$-action on $M \times \Gamma(Y_{ct})$, i.e.,
		\begin{align*}
			(\iota_{\lambda})_{cl} \mathbb{\Delta}^{(p)} \simeq 0, \quad (\delta_{\lambda})_{cl} \mathbb{\Delta}^{(p)} \simeq 0.
		\end{align*}
	\end{defn}
	
	$f$-transformations of extended BV-BFV LFTs of CohLGFT type should be only defined for local forms $f^{(p)}$ over $M \times \Gamma(Y)$ that are basic with respect to the $\mathrm{Gau}(P)^{\star}$-action.
	
	\begin{thm}
		The BV-BFV difference of an $m$-extended BV-BFV LFT of CohLGFT type is uniquely determined by the (general) $K$-sequence of $\mathbb{\Delta}^{(m)}$ up to an $f$-transformation.
	\end{thm}
	
	\begin{proof}
		This follows directly from  \eqref{pfex} and Theorem \ref{mthmg}.
	\end{proof}
	\subsubsection{Cotangent lift of Donaldson-Witten theory}
	
	Recall that the configuration bundle $Y$ of Donaldson-Witten theory is given by $Y=V[1]Y'$, where 
	\begin{align*}
		Y'=\mathrm{ad} P[1] \times_M C \times_M W[-1].
	\end{align*}
	For our purpose, we will choose $W$ to be $\mathrm{ad} P \otimes \Lambda^2 T^*M$ instead of $\mathrm{ad} P \otimes \Lambda^2_- T^*M$. This is justifiable because within the BV formalism, one can apply a gauge fixing to eliminate the self-dual component of $W$. Let
	\begin{align*}
		(x,~\theta,~A,~\chi,~\phi,~\upsilon,~b)
	\end{align*}
	be a coordinate system of $Y$. (We are adopting physicists' notation again for convenience.) It induces a coordinate system
	\begin{align*}
		(x,~\theta,~A,~\chi,~\phi,~\upsilon,~b,~\theta^+,~A^+,~\chi^+,~\phi^+,~\upsilon^+,~b^+),
	\end{align*}
	of
	\begin{align*}
		Y_{ct}=V^*[-1]Y \cong Y \times_M \mathrm{ad}P[-2] \times_M (\mathrm{ad} P \otimes \Lambda^3 T^*M)[-1] \times_M (\mathrm{ad} P \otimes \Lambda^2 T^*M) \\
		\times_M \mathrm{ad}P[-3] \times_M (\mathrm{ad} P \otimes \Lambda^3 T^*M)[-2] \times_M (\mathrm{ad} P \otimes \Lambda^2 T^*M)[-1],
	\end{align*} 
	where we use the Killing form $\mathrm{Tr}$ of $\mathfrak{g}$, the Riemannian metric $g$, and the Hodge star operator $\star$ to make the identifications $\mathrm{ad} P^* \cong \mathrm{ad} P$ and $\Lambda^p TM \cong \Lambda^{4-p} T^*M$. For brevity, we use $\Phi$ to denote the fields $(\theta,~A,~\chi,~\phi,~\upsilon,~b)$ and $\Phi^+$ to denote the anti-fields $(\theta^+,~A^+,~\chi^+,~\phi^+,~\upsilon^+,~b^+)$. The canonical symplectic local form $\omega_{ct}$ is of the form
	\begin{align*}
		\omega_{ct} = \mathrm{Tr}(\delta \Phi^+ \wedge \delta \Phi) = d_v \mathrm{Tr}(\Phi^+ \wedge  \delta \Phi) =: d_v \gamma_{ct}.
	\end{align*}
	The local form $\widetilde{Q}$ associated to the cohomological vector field $Q$ defined by \eqref{q1} to \eqref{q3} is given by
	\begin{align*}
		\widetilde{Q} = &\mathrm{Tr}(\theta^+(\phi - \frac{1}{2}[\theta,\theta]) + \phi^+(-[\theta, \phi]) + A^+\wedge (\upsilon + d_A \theta) + \upsilon^+\wedge (- [\theta, \upsilon] - d_A \phi ) \\
		&+ \chi^+\wedge (b - [\theta, \chi]) + b^+\wedge (- [\theta, b] + [\phi, \chi])).
	\end{align*}
	We then have
	\begin{align}
		&d_v \widetilde{Q} = \mathrm{Tr}(\delta \Phi^+ \wedge Q(\Phi) ) + \mathrm{Tr}(\theta^+(\delta \phi + [\theta,\delta \theta]) - \phi^+(-[\delta \theta, \phi] + [\theta, \delta \phi]) \notag \\
		&- A^+\wedge (\delta \upsilon + [\delta A, \theta] + d_A \delta \theta) + \upsilon^+\wedge (- [\delta \theta, \upsilon] + [\theta, \delta \upsilon]- [\delta A, \phi] - d_A \delta \phi )  \notag \\
		&+ \chi^+\wedge (\delta b - [\delta \theta, \chi] + [\theta, \delta \chi]) - b^+\wedge (- [\delta \theta, b] + [\theta, \delta b] + [\delta \phi, \chi] + [\phi, \delta \chi])) d\mathrm{vol}_g. \notag \\
		&= \mathrm{Tr}(\delta \Phi^+ \wedge Q(\Phi)) + \mathrm{Tr}(\theta^+ \wedge \delta \phi + [\theta^+,\theta] \wedge \delta \theta-  [\phi^+,\phi] \wedge \delta \theta - [\phi^+, \theta] \wedge \delta \phi  \notag \\
		&- A^+ \wedge  \delta \upsilon -[A^+, \theta] \wedge \delta A - d_A  A^+ \wedge \delta \theta + [\upsilon^+, \upsilon] \wedge \delta \theta + [\upsilon^+, \theta] \wedge \delta \upsilon + [\upsilon^+, \phi] \wedge \delta A - d_A  \upsilon^+ \wedge \delta \phi )  \notag \\
		&+ \chi^+ \wedge \delta b + [\chi^+, \chi] \wedge \delta \theta + [\chi^+, \theta] \wedge \delta \chi - [b^+, b] \wedge \delta \theta -  [b^+, \theta] \wedge \delta b -  [b^+, \chi] \wedge \delta \phi - [b^+, \phi] \wedge \delta \chi)  \notag \\
		&+ d_h \mathrm{Tr}(A^+ \wedge \delta \theta + \upsilon^+ \wedge \delta \phi)\label{bd1}
	\end{align}
	The cotangent lift $Q_{cl}$ of the cohomological vector field $Q$ defined by \eqref{q1} to \eqref{q3} is given by
	\begin{align*}
		& Q_{cl} \theta = \phi - \frac{1}{2}[\theta,\theta], \quad Q_{cl} \phi = -[\theta, \phi], \\
		& Q_{cl} A  = \upsilon + d_A \theta, \quad Q _{cl}\upsilon  = - [\theta, \upsilon] - d_A \phi \\
		& Q_{cl} \chi = b - [\theta, \chi], \quad Q_{cl} b  = - [\theta, b] + [\phi, \chi], \\
		& Q_{cl} \theta^+ =  -[\theta, \theta^+] + [\phi, \phi^+] - d_A A^+ +[\upsilon,\upsilon^+] - [\chi,\chi^+] +[b, b^+], \\
		&Q_{cl} \phi^+  = \theta^+ - [\theta,\phi^+] - d_A \upsilon^+ - [\chi, b^+],\\
		& Q_{cl} A^+  = - [\theta, A^+] - [\phi, \upsilon^+], \quad Q _{cl}\upsilon^+  =  -A^+ - [\theta, \upsilon^+], \\
		& Q_{cl} \chi^+ = - [\theta, \chi^+] + [\phi, b^+],  \quad Q_{cl} b^+  =  \chi^+ - [\theta, b^+],
	\end{align*}
	\begin{rmk}
		Applying the change of coordinates $A^+ \mapsto - A^+$, the cohomological vector field $Q_{cl}$ restricted to $(A^+,\upsilon^+,\chi^+,b^+)$ becomes the Kalkman differential in the BRST model of equivariant cohomology.
	\end{rmk}
	One can also check that the cotangent lifts $(I_{\lambda})_{cl}$, $(\delta_{\lambda})_{cl}$ of $I_{\lambda}$ and $\delta_{\lambda}$ are given by
	\begin{align*}
		(I_{\lambda})_{cl} \theta = \lambda,  \quad I_{\lambda} \phi = I_{\lambda} A = I_{\lambda} \upsilon = I_{\lambda} \chi = I_{\lambda} b =0, \quad I_{\lambda} \Phi^+ =0, 
	\end{align*} 
	and
	\begin{align*}
		(\delta_{\lambda})_{cl} \Phi = \delta_{\lambda} \Phi, \quad	(\delta_{\lambda})_{cl} \Phi^+ = \delta_{\lambda} \Phi^+.
	\end{align*}
	Therefore, we will omit the subscript ``$cl$'' and simply use $I_{\lambda}$ and $\delta_{\lambda}$ to denote the corresponding vector fields.
	
	The BRST Lagrangian of Donaldson-Witten theory is given by
	\begin{align}\label{brstldw}
		\mathcal{L}_{BRST} &= \mathrm{Tr}(F \wedge F)/2 + Q \mathrm{Tr}(\chi \wedge (F+b/2)) \notag \\
		&= \mathrm{Tr} \left( (F+b) \wedge (F+b)/2 - \chi \wedge d_A \upsilon  - \chi \wedge [\phi,\chi]/2 \right).
	\end{align}
	One can easily show that
	\begin{align}
		d_v \mathcal{L}_{BRST} 
		&= \mathrm{Tr}((F + b) \wedge \delta b - (d_A \upsilon + [\phi,\chi]) \wedge \delta \chi - (d_A b + [\chi,\upsilon] ) \wedge \delta A  - d_A  \chi \wedge \delta \upsilon + [\chi,\chi] \wedge \delta \phi/2) \notag \\
		& + d_h \mathrm{Tr}((b+F) \wedge \delta A + \chi \wedge \delta \upsilon).\label{bd2}
	\end{align}
	It follows that $Q_{\mathcal{L}_{BRST}} \Phi=0$ and
	\begin{align*}
		& Q_{\mathcal{L}_{BRST} } \theta^+ =  0, \quad Q_{\mathcal{L}_{BRST} } \phi^+  = [\chi,\chi]/2, \\
		& Q_{\mathcal{L}_{BRST} } A^+  = -d_A b - [\chi,\upsilon], \quad Q _{\mathcal{L}_{BRST} }\upsilon^+  = - d_A \chi, \\
		& Q_{\mathcal{L}_{BRST} } \chi^+ = - d_A \upsilon - [\phi,\chi],  \quad Q_{\mathcal{L}_{BRST} } b^+  =  F + b,
	\end{align*}
	Summing up, we have
	\begin{align*}
		& Q_{BV} \theta = \phi - \frac{1}{2}[\theta,\theta], \quad Q_{BV} \phi = -[\theta, \phi], \\
		& Q_{BV} A  = \upsilon + d_A \theta, \quad Q_{BV} \upsilon  = - [\theta, \upsilon] - d_A \phi \\
		& Q_{BV} \chi = b - [\theta, \chi], \quad Q_{BV} b  = - [\theta, b] + [\phi, \chi], \\
		& Q_{BV} \theta^+ =  -[\theta, \theta^+] + [\phi, \phi^+] - d_A A^+ +[\upsilon,\upsilon^+] - [\chi,\chi^+] +[b, b^+], \\
		&Q_{BV} \phi^+  = \theta^+ - [\theta,\phi^+] - d_A \upsilon^+ - [\chi, b^+ -\chi/2],\\
		& Q_{BV} A^+  = -d_A b - [\chi,\upsilon] - [\theta, A^+] - [\phi, \upsilon^+], \quad Q_{BV} \upsilon^+  =  -d_A \chi -A^+ - [\theta, \upsilon^+], \\
		& Q_{BV} \chi^+ = - d_A \upsilon - [\phi,\chi]- [\theta, \chi^+] + [\phi, b^+],  \quad Q_{BV} b^+  =  F + b + \chi^+ - [\theta, b^+],
	\end{align*}
	The cotangent lift of the BRST Lagrangian \eqref{brstldw} is given by
	\begin{align*}
		&\mathcal{L}_{BV}= \mathrm{Tr} ((F+b) \wedge (F+ b)/2 - \chi \wedge d_A \upsilon  - \chi \wedge [\phi,\chi]/2 + \theta^+(\phi - \frac{1}{2}[\theta,\theta]) - \phi^+[\theta, \phi]  \\
		&+ A^+\wedge (\upsilon + d_A \theta) - \upsilon^+\wedge ( [\theta, \upsilon] + d_A \phi ) 
		+ \chi^+\wedge (b - [\theta, \chi]) - b^+\wedge ([\theta, b] - [\phi, \chi])).
	\end{align*}
	Let $\gamma$ denote the boundary term of $\mathcal{L}_{BV}$. $\gamma$ is the sum of the boundaries terms in \eqref{bd1} and \eqref{bd2}.
	\begin{align*}
		\gamma = \mathrm{Tr}(A^+ \wedge \delta \theta + \upsilon^+ \wedge \delta \phi + (b+F) \wedge \delta A + \chi \wedge \delta \upsilon).
	\end{align*}
	$(\mathcal{L}_{BV}, \gamma)$ admit a BV-BFV extension. Consider the following change of coordinates.
	\begin{align*}
		&\widetilde{\phi}=\phi - \frac{1}{2}[\theta,\theta], \quad \widetilde{\upsilon} =  \upsilon + d_A \theta,\quad \tilde{b}= b - [\theta, \chi],\\
		&\widetilde{\theta^+}=\theta^+ - [\theta,\phi^+] - d_A \upsilon^+ - [\chi, b^+ -\chi/2], \quad
		\widetilde{A^+}=-d_A \chi -A^+ - [\theta, \upsilon^+], \quad
		\widetilde{\chi^+}=F + b + \chi^+ - [\theta, b^+].
	\end{align*}
	The cohomological vector field $Q_{BV}$ takes a simplified form in the new coordinates.
	\begin{align*}
		&Q_{BV} \theta = \widetilde{\phi}, \quad Q_{BV} A = \widetilde{\upsilon}, \quad Q_{BV} \chi = \widetilde{b}, \quad 
		&&Q_{BV} \widetilde{\phi} = 0, \quad Q_{BV} \widetilde{\upsilon} = 0, \quad Q_{BV} \widetilde{b} = 0,\\
		&Q_{BV} \phi^+ = \widetilde{\theta^+}, \quad Q_{BV} \upsilon^+ = \widetilde{A^+}, \quad Q_{BV} b^+= \widetilde{\chi^+}, \quad 
		&&Q_{BV} \widetilde{\theta^+} = 0, \quad Q_{BV} \widetilde{A^+} = 0, \quad Q_{BV} \widetilde{\chi^+} = 0.
	\end{align*}
    It is then straightforward to write down an expression for the homotopy operator $K$.
	\begin{align*}
		&K \theta = A, \quad K A = s \chi + s' b^+, \quad K \chi = t \upsilon^+, \quad 
		&&K \widetilde{\phi} = d\theta - \widetilde{\upsilon}, \quad K \widetilde{\upsilon} = dA - s \widetilde{b} - s' \widetilde{\chi^+}, \quad K \widetilde{b} = d\chi - t \widetilde{A^+},\\
		&K \phi^+ = 0, \quad K \upsilon^+ = u \phi^+, \quad K b^+= w \upsilon^+, \quad 
		&&K \widetilde{\theta^+} = 0, \quad K \widetilde{A^+} = d\upsilon^+ - u \widetilde{\theta^+}, \quad K \widetilde{\chi^+} = db^+ - w \widetilde{A^+},
	\end{align*}
	where $s, s' , t, u, w$ are real numbers. Reverting to the original coordinates, we have
	\begin{align*}
		&K \theta = A, \quad K A = s \chi + s' b^+, \quad K \chi = t \upsilon^+, \\
		&K \phi = -\upsilon, \quad K \upsilon = (2-s')F - (s+s') b - s' \chi^+, \quad K b = (1+t)d_A \chi + t A^+,\\
		&K \phi^+ = 0, \quad K \upsilon^+ = u \phi^+, \quad K b^+= w \upsilon^+, \\
		&K \theta^+ = 0, \quad K A^+ = (u/2-s)[\chi,\chi] - (s'+u)[\chi,b^+] + (t-1-u) d_A \upsilon^+ + u\theta^+, \\
		&K \chi^+ = (s'+1) d_Ab^+ +(s+w-1-t)d_A \chi + (w-t)A^+.	
	\end{align*}
	Let 
	\begin{align*}
		\gamma^{(4)} :=  \mathrm{Tr}(\phi \wedge \delta \theta), \quad \mathbb{\Delta}^{(4)}:= -\mathrm{Tr}(\phi^2)/2, \quad \mathcal{L}^{(4)}:= \mathbb{\Delta}^{(4)} + \iota_Q \gamma^{(4)} = \mathrm{Tr}(\phi^2 - \phi[\theta,\theta])/2.
	\end{align*}
	One can easily verify that $\mathrm{Lie}_Q \gamma^{(4)} = d_v \mathbb{\Delta}^{(4)}$. Therefore, $\gamma_K:= \exp(K) \gamma^{(4)}$ together with $\mathbb{\Delta}_K:= \exp(K) \mathbb{\Delta}^{(4)}$ define a fully extended BV-BFV LFT. Let 
	\begin{align*}
		\theta_K:=\exp(\widetilde{K})\theta = \exp(-K)\theta, \quad
		\phi_K:=\exp(\widetilde{K})\phi = \exp(K)\theta.
	\end{align*}
	We have
	\begin{align*}
		\gamma_K = -\mathrm{Tr}(\phi_K \wedge \delta \theta_K), \quad \mathbb{\Delta}_K= \mathrm{Tr}(\phi_K^2)/2, \quad \mathcal{L}_K = \mathrm{Tr}(\phi_K \wedge \phi_K - \phi_K \wedge [\theta_K, \theta_K])/2.
	\end{align*}
	By \eqref{lk}, the Lagrangian $\mathcal{L}=\sum_{p=0}^n \mathcal{L}^{p}$ of the theory takes the following form
	\begin{align}\label{bfbb}
		\mathcal{L} = \mathcal{L}_K - K_0 \gamma_K = \mathrm{Tr}(\phi_K \wedge \phi_K /2 - \phi_K \wedge F_K), 
	\end{align}
	where $F_K:= d\theta_K + \frac{1}{2}[\theta_K, \theta_K]$. $\theta_K$, $\phi_K$, and \eqref{bfbb} can be interpreted as the superfields and the AKSZ Lagrangian of the ``$BF+B^2$'' theory \cite[Section 7.4]{Cattaneo2014}.
	
	More concretely, let's set $w=s'=0, s=-2, t=-3, u=-4$. We have
	\begin{align*}
		&K \theta = A, \quad K A = -2\chi, \quad K b^+ = 0, \quad K \phi = -\upsilon, \quad K \upsilon = 2F+2b , \quad K \chi^+ = d_A b^+ + 3A^+, \\
		&K b = -2d_A\chi - 3A^+, \quad K A^+ = 4[\chi, b^+]-4\theta^+, \quad K \theta^+ =0, \quad K \chi = -3\upsilon^+, \quad K \upsilon^+ = -4\phi^+, \quad K \phi^+ =0.
	\end{align*}
	and
	\begin{align*}
		&\theta_K =  \theta - A - \chi - \upsilon^+ - \phi^+, \quad
		\phi_K  =  \phi - \upsilon - F - b + A^+ - \theta^+ + [\chi,b^+].
	\end{align*}
	It follows that
	\begin{align*}
		&\gamma_K^{(4)} = -\mathrm{Tr}(\phi \wedge \delta \theta), \quad 
		\gamma_K^{(3)} = \mathrm{Tr}(\upsilon \wedge \delta \theta + \phi \wedge \delta A), \quad 
		\gamma_K^{(2)} = \mathrm{Tr}((F+b) \wedge \delta \theta - \upsilon \wedge \delta A + \phi \wedge \delta \chi), \\
		&\gamma_K^{(1)} = -\mathrm{Tr}(A^+ \wedge \delta \theta + (F+b) \wedge \delta A + \upsilon \wedge \delta \chi - \phi \wedge \delta \upsilon^+), \\
		&\gamma_K^{(0)} = \mathrm{Tr}((\theta^+ - [\chi,b^+]) \wedge \delta \theta + A^+ \wedge \delta A - (F+b) \wedge \delta \chi - \upsilon \wedge \delta \upsilon^+ + \phi \wedge \delta \phi^+),
	\end{align*}
	and
	\begin{align*}
		&\mathbb{\Delta}_K^{(4)} = \mathrm{Tr}(\phi^2)/2, \quad 
		\mathbb{\Delta}_K^{(3)} = -\mathrm{Tr}(\phi \wedge \upsilon), \quad 
		\mathbb{\Delta}_K^{(2)} = \mathrm{Tr}(\upsilon \wedge \upsilon/2 - \phi \wedge (F+b)), \\
		&\mathbb{\Delta}_K^{(1)} =\mathrm{Tr}(\phi \wedge A^+ + \upsilon \wedge (F+b)), \quad
		\mathbb{\Delta}_K^{(0)} = \mathrm{Tr}(\phi \wedge [\chi, b^+] - \phi \wedge \theta^+ - \upsilon \wedge A^+ + (F+b)\wedge(F+b)/2).
	\end{align*}
	The Lagrangian $\mathcal{L}^{(0)}$ takes the form
	\begin{align*}
		&\mathcal{L}^{(0)}=\iota_Q\gamma_K^{(0)} + \mathbb{\Delta}_K^{(0)} \\
		&= \mathrm{Tr}\left((F+b)\wedge(F+b)/2 + \upsilon \wedge d_A \chi  + \phi \wedge [\chi, \chi]/2 + \theta^+\wedge (\phi-[\theta,\theta]/2) - \phi^+ \wedge [\theta, \phi] \right. \\
		&\left. + A^+ \wedge (\upsilon + d_A \theta) - \upsilon^+ \wedge [\theta, \upsilon] - \phi \wedge d_A \upsilon^++b^+\wedge ([\phi, \chi] - [\theta,[\theta,\chi]]) - (F+b) \wedge (b - [\theta,\chi]) \right).
	\end{align*}
	\begin{rmk}
		The subspace $\Gamma(Y)_{red}$ of $\Gamma(Y)$ defined by the equations 
		\begin{align*}
			F+ \chi^+ + b =0, \quad b^+= 0.
		\end{align*}
		is preserved by the action of $Q_{BV}$ and $K$. It is not hard to see that $(\mathcal{L}^{(0)},\gamma_K^{(1)})|_{M \times \Gamma_{red}}$ is equal to $(\mathcal{L}_{BV}, -\gamma)|_{M \times \Gamma_{red}}$ up to an $f$-transformation.
	    Moreover, $Q_{BV}|_{\Gamma_{red}}$, $\mathcal{L}_{BV}|_{M \times \Gamma_{red}}$, and $\gamma|_{M \times \Gamma_{red}}$  are just the cohomological vector field, the Lagrangian, and the boundary term in the AKSZ construction of the Donaldson-Witten theory \cite{Bonechi2020}.
	\end{rmk}
	
	\appendix
	
	\addcontentsline{toc}{section}{Sign conventions}
	\section*{Sign conventions}
	
	Let $A = \bigoplus_{i,j,k} A^{i,j,k}$ be a trigraded associative algebra. $A$ is said to be commutative if
	\begin{align*}
		ab = (-1)^{i_ai_b + (j_a+k_a)(j_b+k_b)}ba,
	\end{align*}
	where $a \in A^{i_a,j_a,k_a}$ and $b \in A^{i_b,j_b,k_b}$. In our specific case, consider the algebra $\Omega_{loc}= \bigoplus_{i,j,k} \Omega_{loc}^{i,j,k}$ of local forms over $M \times \Gamma(Y)$ with the wedge product $\wedge$ as the algebraic product. Here, $Y$ is a graded fiber bundle, and $i$, $j$, and $k$ denote the horizontal form degree, vertical form degree, and ghost number degree, respectively. A derivation $D$ (of degree $(i_D, j_D, k_D)$) of $A$ is an element in $\mathrm{End}(A)^{i_D, j_D, k_D}$ such that
	\begin{align*}
		D(ab) = D(a)b +  (-1)^{i_ai_D + (j_a+k_a)(j_D+k_D)}a(Db).
	\end{align*}
	For $A=\Omega_{loca}$, consider, for example, $D=d_h$, $d_v$, $\iota_Q$, and $\mathrm{Lie}_Q$, where $Q$ is a cohomological vector field. We have
	\begin{align*}
		&d_h(a \wedge b) = d_h a \wedge b + (-1)^{i_a} a \wedge d_h b, \\
		&d_v(a \wedge b) = d_v a \wedge b + (-1)^{j_a+k_a} a \wedge d_v b, \\
		&\mathrm{Lie}_Q(a \wedge b) = \mathrm{Lie}_Q a \wedge b + (-1)^{j_a + k_a} a \wedge \mathrm{Lie}_Q b,  \\
		&\iota_Q (a \wedge b) = \iota_Q  a \wedge b + a \wedge \iota_Q  b.
	\end{align*}
	$\mathrm{Lie}_Q^2=d_h^2=d_v^2=0$. One can combine any two of them to make $\Omega_{loc}$ into a double cochain complex. 
	For example, one can consider $B = \bigoplus_{i,j} B^{i,j}$ with $B^{i,j} := \bigoplus_{i'+k'=i} \Omega_{loc}^{i',j,k'}$. The two differentials on $B$ are given by
	\begin{align*}
	d_v: B^{i,j} \rightarrow B^{i,j+1}, \quad
	(\mathrm{Lie}_Q - \widetilde{d_h}): B^{i,j} \rightarrow B^{i+1,j},
	\end{align*}
	where $\widetilde{D}:=(-1)^{i_D(i+j-\mathrm{dim}(M))}D$ for a derivation $D$ of horizontal form degree $i_D$. It is easy to see that $(\mathrm{Lie}_Q - \widetilde{d_h})$ squares to zero and anti-commutes with $d_v$.


\addcontentsline{toc}{section}{References}
	
	\begin{bibsection}
		\begin{biblist}
			\bib{Alexandrov1997}{article}{
				title={The geometry of the master equation and topological quantum field theory},
				author={Alexandrov, M.},
				author={Schwarz, A.},
				author={Zaboronsky, O.},
				author={Kontsevich, M.},
				journal={Int. J. Mod. Phys. A},
				volume={12},
				number={07},
				pages={1405--1429},
				date={1997},
				publisher={World Scientific}
			}
			\bib{Atiyah1990}{article}{
				title = {Topological Lagrangians and cohomology},
				journal = {J. Geom. Phys.},
				volume = {7},
				number = {1},
				pages = {119-136},
				date = {1990},
				author = {Atiyah, M. F.},
				author = {Jeffrey, L.},
			}
			\bib{Baulieu1988}{article}{
				author = {Baulieu, L.},
				author = {Singer, I. M.},
				title = {Topological Yang-Mills symmetry},
				journal = {Nucl. Phys. B Proc. Suppl.},
				volume = {5},
				pages = {12--19},
				date = {1988}
			}
			\bib{Baulieu1989}{article}{
				title={The topological sigma model},
				author={Baulieu, L.},
				author={Singer, I. M.},
				journal={Comm. Math. Phys.},
				volume={125},
				number={2},
				pages={227--237},
				date={1989},
				publisher={Springer}
			}
			\bib{Baulieu2005}{article}{
				title={Topological vector symmetry of BRSTQFT topological gauge fixing of BRSTQFT and construction of maximal supersymmetry},
				author={Baulieu, L.},
				author={Bossard, G.},
				author={Tanzini, A.},
				journal={J. High Energy Phys.},
				volume={2005},
				number={08},
				pages={037},
				date={2005},
				publisher={IOP Publishing}
			}
			\bib{Birmingham1991}{article}{
				title={Topological field theory},
				author={Birmingham, D.},
				author={Blau, M.},
				author={Rakowski, M.},
				author={Thompson, G.},
				journal={Phys. Rep.},
				volume={209},
				number={4-5},
				pages={129--340},
				date={1991},
				publisher={Elsevier}
			}
			\bib{Blau1993}{article}{
				title={The Mathai-Quillen formalism and topological field theory},
				author={Blau, M.},
				journal={J. Geom. Phys.},
				volume={11},
				number={1-4},
				pages={95--127},
				date={1993},
				publisher={Elsevier}
			}
			\bib{Blohmann14}{article}{
				title={Lagrangian Field Theory},
				author={Blohmann, Christian},
				journal={Unpublished manuscript, version 23.0},
				note = {Available at \url{https://people.mpim-bonn.mpg.de/blohmann/Lagrangian_Field_Theory.pdf}},
				date={2023},
			}
			\bib{Blohmann2023}{article}{
				title={The homotopy momentum map of general relativity},
				author={Blohmann, Christian},
				journal={Int. Math. Res. Not.},
				volume={2023},
				number={10},
				pages={8212--8250},
				date={2023},
				publisher={Oxford University Press}
			}
			\bib{Bonechi2020}{article}{
				title={Equivariant Batalin--Vilkovisky formalism},
				author={Bonechi, Francesco},
				author={Cattaneo, Alberto S},
				author={Qiu, Jian},
				author={Zabzine, Maxim},
				journal={J. Geom. Phys.},
				volume={154},
				pages={103720},
				date={2020},
				publisher={Elsevier}
			}
			\bib{Cattaneo2014}{article}{
				title={Classical BV theories on manifolds with boundary},
				author={Cattaneo, Alberto S.},
				author={Mnev, Pavel},
				author={Reshetikhin, Nicolai},
				journal={Comm. Math. Phys.},
				volume={332},
				pages={535--603},
				date={2014},
				publisher={Springer}
			} 
			\bib{Deligne1999}{incollection}{
				title={Classical field theory},
				author={Deligne, P.},
				author={Freed, Daniel S.}
				booktitle={Quantum Fields and Strings: A Course for Mathematicians},
				volume={1},
				pages={99-135},
				date={1999},
				publisher={American Mathematical Society}
			}
			\bib{Donaldson2002}{book}{,
				title={Floer Homology Groups in Yang-Mills Theory},
				author={Donaldson, Simon Kirwan},
				volume={147},
				date={2002},
				publisher={Cambridge University Press}
			}
			\bib{Fatibene2003}{book}{
				title={Natural and Gauge Natural Formalism for Classical Field Theories: A Geometric Perspective including Spinors and Gauge Theories},
				author={Fatibene, Lorenzo},
				author={Francaviglia, Mauro},
				date={2003},
				publisher={Springer Science \& Business Media}
			}
			\bib{Guillemin2013}{book}{
				title={Supersymmetry and Equivariant de Rham Theory},
				author={Guillemin, V. W.},
				author={Sternberg, S.},
				date={2013},
				publisher={Springer Science \& Business Media}
			}
			\bib{Jiang2023a}{article}{,
				title={Mathematical structures of cohomological field theories},
				author={Jiang, S.},
				journal={J. Geom. Phys.},
				volume={185},
				pages={104744},
				date={2023}
			}
			\bib{Kalkman1993}{article}{
				title={BRST model for equivariant cohomology and representatives for the equivariant Thom class},
				author={Kalkman, J.},
				journal={Comm. Math. Phys.},
				volume={153},
				number={3},
				pages={447--463},
				date={1993},
				publisher={Springer}
			}
			\bib{Leites1980}{article}{
				title={Introduction to the theory of supermanifolds},
				author={Leites, D. A.},
				journal={Russ. Math. Surv.},
				volume={35},
				number={1},
				pages={1--64},
				date={1980},
				publisher={IOP Publishing}
			}
			
			\bib{Mathai1986}{article}{
				title = {Superconnections, Thom classes, and equivariant differential forms},
				journal = {Topology},
				volume = {25},
				number = {1},
				pages = {85-110},
				date = {1986},
				author = {Mathai, V.},
				author = {Quillen, D.}
			}
			\bib{Mnev2019}{article}{
				author = {Mnev, Pavel},
				author ={Schiavina, Michele},
				author = {Wernli, Konstantin},
				date = {2019},
				pages = {993-1044},
				title = {Towards Holography in the BV-BFV Setting},
				volume = {21},
				journal = {Ann. Henri Poincaré}
			}
			\bib{Ouvry1989}{article}{
				title={On the algebraic characterization of Witten's topological Yang-Mills theory},
				author={Ouvry, S.},
				author={Stora, R.},
				author={Van Baal, P.},
				journal={Phys. Lett. B},
				volume={220},
				number={1-2},
				pages={159--163},
				date={1989},
				publisher={Elsevier}
			}
			\bib{Piguet2008}{book}{
				title={Algebraic Renormalization: Perturbative Renormalization, Symmetries and Anomalies},
				author={Piguet, O.},
				author = {Sorella, S. P.},
				volume={28},
				date={2008},
				publisher={Springer Science \& Business Media}
			}
			\bib{Sardanashvily2005}{article}{
				title={The variational bicomplex on graded manifolds and its cohomology},
				author={Sardanashvily, G},
				journal={arXiv preprint math/0504529},
				date={2005}
			}
			\bib{Schwarz1993}{article}{
				title={Geometry of batalin-vilkovisky quantization},
				author={Schwarz, Albert},
				journal={Comm. Math. Phys.},
				volume={155},
				number={2},
				pages={249--260},
				date={1993},
				publisher={Springer}
			}
			\bib{Sharapov2015}{article}{
				title={Variational tricomplex of a local gauge system, Lagrange structure and weak Poisson bracket},
				author={Sharapov, Alexey A.},
				journal={Int. J. Mod. Phys. A},
				volume={30},
				number={25},
				pages={1550152},
				date={2015},
				publisher={World Scientific}
			}
			\bib{Sorella1998}{article}{
				title={Algebraic characterization of vector supersymmetry in topological field theories},
				author={Sorella, S. P.},
				author={Vilar, L. C. Q.},
				author={Ventura, O. S.},
				author={Sasaki, C. A. G.},
				journal={J. Math. Phys.},
				volume={39},
				number={2},
				pages={848--866},
				date={1998},
				publisher={American Institute of Physics}
			}
			\bib{Witten1982}{article}{
				title={Supersymmetry and Morse theory},
				author={Witten, E.},
				journal={J. Differ. Geom.},
				volume={17},
				number={4},
				pages={661--692},
				date={1982},
				publisher={Lehigh University}
			}
			\bib{Witten1988}{article}{
				author = {Witten, E.},
				journal = {Comm. Math. Phys.},
				number = {3},
				pages = {353--386},
				title = {Topological quantum field theory},
				volume = {117},
				date = {1988}
			}
			\bib{Witten1991}{article}{
				title={Introduction to cohomological field theories},
				author={Witten, E.},
				journal={Int. J. Mod. Phys. A},
				volume={6},
				number={16},
				pages={2775--2792},
				date={1991},
				publisher={World Scientific}
			}
			\bib{Zinn1975}{article}{
				title={Trends in elementary particle theory},
				author={Zinn-Justin, J.},
				journal={Lect. Notes Phys.},
				volume={37},
				pages={34--35},
				date={1975},
				publisher={Springer Berlin}
			}
			\bib{Zuckerman1987}{incollection}{
				title={Action principles and global geometry},
				author={Zuckerman, G. J.},
				booktitle={Mathematical Aspects of String Theory},
				pages={259--284},
				date={1987},
				publisher={World Scientific}
			}
		\end{biblist}
	\end{bibsection}
\end{document}